\documentclass[epj]{svjour}

\usepackage{microtype}
\usepackage[pdfborder={0 0 0}]{hyperref} 
\usepackage{mathtools}
\usepackage{amstext}
\usepackage{amsmath}
\usepackage{amssymb}
\usepackage{amsfonts}
\usepackage{graphicx}
\usepackage{color}
\usepackage{cite}
\usepackage[normalem]{ulem} 

\usepackage[version=4]{mhchem}

\renewcommand{\vec}[1]{\boldsymbol{#1}}

\definecolor{Red}{rgb}{0.9,0.0,0.1}
\definecolor{Lila}{rgb}{0.7,0.0,0.9}
\definecolor{Darkblue}{rgb}{0.22,0.33,0.64}
\definecolor{Darkgray}{rgb}{0.4,0.4,0.4}
\definecolor{Blue}{rgb}{0.1,0.0,0.9}

\begin{document}

\title{From diffusive mass transfer 
in Stokes flow  to low Reynolds number  Marangoni boats}
  \author{Hendrik Ender \and Jan Kierfeld}
\institute{                    
   Department of Physics, Technische Universit\"{a}t Dortmund, 
  44221 Dortmund, Germany
}
\date{Received: date / Revised version: date}

\authorrunning{H. Ender and J. Kierfeld}
\titlerunning {Low Reynolds number  Marangoni boats}

\abstract{
  We present a theory for the  self-propulsion of
    symmetric, half spherical
  Marangoni boats (soap or camphor boats) at low Reynolds numbers. 
  Propulsion is generated by release (diffusive emission or dissolution)
  of   water-soluble surfactant 
    molecules, which  modulate the
    air-water interfacial tension.
    Propulsion either requires asymmetric release or
    spontaneous symmetry breaking by coupling to advection 
    for a perfectly symmetrical swimmer.
   We study the diffusion-advection
   problem for a sphere in Stokes flow  analytically and numerically
   both for constant concentration
   and constant flux boundary conditions. We derive novel results
   for concentration profiles under
   constant flux boundary conditions and for the Nusselt number
   (the dimensionless ratio of total emitted flux  and 
    diffusive flux). 
   Based on these results, we analyze the Marangoni boat for small
   Marangoni propulsion (low Peclet number) and
    show that two swimming regimes exist, a diffusive regime at low
    velocities and an advection-dominated regime at high
    swimmer  velocities. 
    We  describe both the limit of large Marangoni propulsion
    (high Peclet number)
     and the effects
  from evaporation  by approximative analytical theories.
 The swimming velocity is determined by force balance, and we
 obtain a general expression for the Marangoni forces, which 
  comprises both
  direct Marangoni forces from the surface tension gradient along the
  air-water-swimmer  contact line and  Marangoni   flow forces.
  We unravel whether the Marangoni flow contribution is exerting a
  forward or backward force during propulsion.
  Our main result is the relation between Peclet number  and
  swimming velocity. 
Spontaneous symmetry breaking and, thus, swimming occur
  for a perfectly symmetrical swimmer above a critical Peclet number,
  which becomes small for large system sizes. 
  We find a supercritical swimming bifurcation for a symmetric swimmer
  and an avoided bifurcation in the presence of an asymmetry.
} 

\maketitle

\section{Introduction}

Swimming on the microscale is governed by low Reynolds numbers
and requires special propulsion
mechanisms which are effective in the presence of dominating viscous
forces.
An important  class of low Reynolds number swimming strategies generates 
interfacial fluid slip-velocities
at the swimmer surface, which then lead to self-propulsion
because the swimmer must be force-free. 
This class of swimming strategies comprises phoretic and
Marangoni mechanisms.
Phoretic
mechanisms  self-create gradients in concentration
(self-diffusiophoresis) or temperature (self-thermo\-pho\-resis)
\cite{ebbens2010,Illien2017}
which, in turn, give rise to interfacial fluid flow
in a thin interaction layer
\cite{Anderson1989}.

  There are two types of swimmers based on the Marangoni effect
  \cite{Scriven1960}:
  droplet swimmers with liquid interfaces, which can operate in the
  bulk  and solid Marangoni boats or surfers operating
  at a liquid-air interface.
The  liquid droplet swimmer is  fully immersed in a
 liquid that carries  surfactant. 
 Propulsion is generated by the Marangoni effect, which
 creates a slip velocity from a surfactant concentration
 gradient along the entire 
  liquid-liquid interface between swimmer and surrounding liquid.
  One typical mechanism to maintain such a surfactant gradient
  is that more surfactant is adsorbed  at the
 front (in swimming direction)
 of the swimmer, which depresses the interfacial tension
 in the front \cite{Yoshinaga2012,Herminghaus2014,Schmitt2016}.
 In Ref.\ \cite{Izri2014}, an auto-diffusiophoretic  mechanism
 coupled to advection
 \cite{michelin2013,Michelin2014}  
 has been proposed to maintain the surfactant concentration gradient.
This propulsion mechanism based on the Marangoni effect
is utilized in different liquid Marangoni swimmers, for example,
active liquid droplets or active emulsions \cite{Herminghaus2014},
such as  pure water droplets in an
 oil-surfactant medium (squalane and monoolein) \cite{Izri2014}
 or liquid crystal droplets in surfactant solutions \cite{Herminghaus2014}.
 Many liquid Marangoni swimmers are spherically symmetric initially and
   swimming spontaneously breaks this symmetry.
   Beyond the instability, 
   advection and/or preferred adsorption can
   produce sufficiently strong  surfactant concentration gradients
    and swimming velocities
   to maintain advection and/or preferred adsorption
   \cite{Yoshinaga2012,michelin2013,Schmitt2016}.
  Also asymmetric
   shape changes can give rise to concentration gradients and
   sufficient swimming velocities to 
   maintain asymmetric shapes \cite{Nagai2005,Yoshinaga2014}.

 Here, we consider 
 Marangoni boats or surfers, which employ a different
 propulsion mechanism.
Important examples are soap or camphor boats which
have a long history \cite{Tomlinson1864}.
The crucial  difference to liquid Marangoni swimmers is
that these boats or surfers operate at a liquid-air interface
rather than in the bulk of the liquid.
Propulsion is not caused by 
surfactants that are anisotropically
distributed along the swimmer-liquid interface but
by the  anisotropic distribution of surfactant at 
the  liquid-air interface along which the
soap boat propels \cite{Nakata2015}.
The surfactant molecules  at 
the  liquid-air interface are emitted or dissolved
   from the swimmer; this can be achieved by     depositing them  on
 the floating swimmer initially \cite{Renney2013},
  by soaking  the swimmer in surfactant
   \cite{Hayashima2001,Nagayama2004,
     Soh2008,Akella2018,Boniface2019,Sur2019}, or by using a
   swimmer body
    made from dissolving surfactant \cite{Loffler2019}. 
 There are many examples based on DMF (dimethylformamide) \cite{Wang2016},
 alcohol \cite{Renney2013,Sur2019}, soap
 \cite{Sur2019}, camphor
 \cite{Hayashima2001,Nagayama2004,Soh2008,Suematsu2014,Akella2018,Boniface2019}
 or camphene \cite{Loffler2019}
 that have also been investigated quantitatively.
 In a companion paper \cite{Ender2020}, we discuss alginate capsules
 as versatile interfacial Marangoni swimmers working with
 many surface tensions
 reducing ``fuels'' in detail,  in particular,   polyethylene glycol
 (PEG)-loaded alginate capsules.

 So far, Marangoni boats can be produced down to radii $a\sim 150\,
 \mathrm{\mu  m}$
 \cite{Ender2020}, and quantitative results are available down to
 $a\sim 1500\,\mathrm{\mu m}$ with Reynolds numbers
 ${\rm Re} \sim 60$, which is
 still  above the low Reynolds number regime.
 Miniaturization is approaching the low Reynolds number
 regime,  which is the regime we address in detail
 in the present paper.
 In a companion paper \cite{Ender2020}, we discussed low Reynolds
 number results more briefly and aimed to generalize to
 high Reynolds numbers using the concept of the Nusselt number in
 order to describe  experiments on PEG-alginate capsule swimmers
 and camphor boats quantitatively.
 There is a related system of thermal Marangoni surfers propelled
 by the thermal Marangoni effect, which was successfully
 realized only recently 
 \cite{Dietrich2020}. Its theoretical description
 is equivalent to surfactant-driven
 Marangoni boats with thermal advection-diffusion replacing
 surfactant advection-diffusion.
 Because thermal diffusion coefficients are much higher and
 swimmer radii reach down to micrometers, this system operates
 at low Reynolds numbers.

 The surfactant molecules are emitted or dissolved from the Marangoni boat,
 diffuse and advect to fluid flow 
 in the water phase and adsorb to the air-water interface,
eventually in interplay with evaporation for volatile surfactants.
This  creates  surface tension
gradients and  Marangoni stresses on the fluid.
Surface tension gradients give rise to a direct net propulsion force
(\emph{direct Marangoni force} in the following).
Marangoni stresses on the fluid give rise to 
 symmetry-broken Marangoni flows, which also  contribute to (or impede)
 propulsion via hydrodynamic drag onto the swimmer (\emph{Marangoni flow
   forces} in the following).
Direct Marangoni forces 
propel into the direction of
\emph{higher} surface tension, i.e., \emph{lower} surfactant
concentration along the air-water-swimmer contact line.
We note that this is \emph{opposite} to the propulsion in the
direction of higher surfactant concentration for the
liquid  Marangoni swimmers operating in the bulk
\cite{Yoshinaga2012,Herminghaus2014,Schmitt2016,Izri2014,michelin2013}.

\begin{figure}
  \centerline{\includegraphics[width=0.99\linewidth]{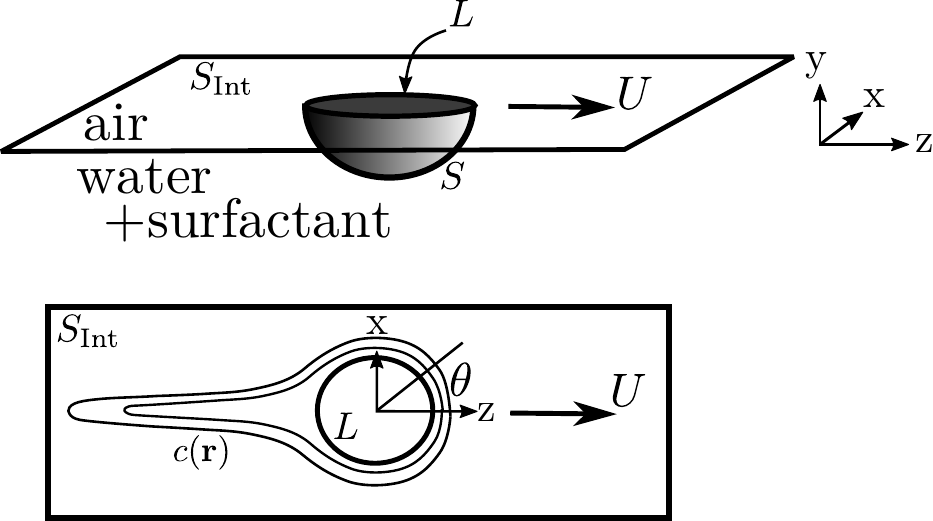}}
  \caption{Side view (top) and top view (bottom) of
    the  half-spherical Marangoni swimmer geometry with surfactant
     concentration field $c(\vec{r})$ and coordinates. 
     }
\label{fig:design}
\end{figure}

A full quantitative theory of Marangoni boats including hydrodynamics,
surfactant advection, direct Marangoni forces and
Marangoni flows  is still elusive but  numerical approaches
exist \cite{Soh2008,Nakata2015,Gidituri2019,JafariKang2020}.
 Theoretical  approaches ignore
 the advection of the surfactant concentration field
 \cite{Lauga2012,Wurger2014,Vandadi2017},
or  even ignore   hydrodynamic flow fields
\cite{Hayashima2001,Nagayama2004,Iida2014,Suematsu2014} 
or approximate it by uniform flow \cite{Boniface2019}, which clearly
oversimplifies the description of surfactant transport.
Here, we focus on low Reynolds numbers as in  Refs.\
\cite{Lauga2012,Wurger2014,Vandadi2017,Gidituri2019}
and consider a  half-spherical
swimmer geometry (see Fig.\ \ref{fig:design}),
which can simplify the theoretical treatment because axial
symmetry can be exploited in certain limits.
Experimentally, half-spherical swimmers can be 
fabricated using the PEG-alginate system \cite{Ender2020}. 
In the limit of weak Marangoni flows, for example, the fluid flow reduces
to  the
well-known Stokes flow  around a sphere.
We fully include advection of the surfactant concentration
  field into our analysis as opposed to
  Refs.\ \cite{Lauga2012,Wurger2014,Vandadi2017},
  where 
disks and spheres propelled by the soap boat mechanism
have been considered previously.
  If advection is ignored,
  the formation of a concentration boundary layer at higher
  velocities, as it is well known
  from the related problem of mass transfer  from a sphere in
  laminar Stokes flow \cite{Acrivos1960,Acrivos1962,Acrivos1965,Leal},
  will be missed. This happens for  velocities $U\gg  a/D$, where $a$ is the
  sphere radius and $D$ the surfactant diffusion constant, and
  is essential for the resulting swimming velocity, which is
  the  quantity of main interest in  this paper.

  In order to obtain the Marangoni forces onto the swimmer, we have
  to calculate the surfactant concentration profile and 
  have to revisit the problem of  mass transfer  from a sphere in
  laminar Stokes flow
    both for constant concentration
   and constant flux boundary conditions.
 In particular,  analytical results on the relevant flux
  boundary conditions are missing in the literature.
  We fill this gap and derive results
  for concentration profiles and for the angular dependence of the
  Nusselt number
  both  for isotropic and anisotropic emission. 
  This allows us to calculate Marangoni propulsion forces
  both in the diffusive and in the advection-dominated regime
  and provide  analytical results.
  There is a direct Marangoni force,
  which is propelling the swimmer into the direction of
  higher surface tension
  and a Marangoni flow force, as has been worked out by
    Masoud and Stone \cite{Masoud2014}. We will unravel, under which
  conditions the flow force contribution is propelling or
  dragging the swimmer.
  We can extend our analysis to
  large Marangoni propulsion (high Peclet number)
    and  include  effects
    from evaporation  by approximative analytical theories.
    This part of the analysis is also the focus of a companion paper
    \cite{Ender2020}.
    Therefore, this discussion is shortened in this paper.
    All analytical results are corroborated by  numerical finite element
 calculations employing a novel iterative approach.

As opposed to previous work
\cite{Soh2008,Lauga2012,Wurger2014,Nakata2015,Vandadi2017,
  Gidituri2019,JafariKang2020}, 
we  consider here  completely symmetric Marangoni boats with isotropic
surfactant emission as motivated by the experiments in
Ref.\ \cite{Ender2020}. Swimming is established in a symmetry-breaking
bifurcation. Similar swimming bifurcations have been analyzed by Michelin and
coworkers \cite{Yoshinaga2012,michelin2013,Izri2014,Michelin2014,Schmitt2016},
but for a different
type of swimmer, namely 
liquid droplet bulk Marangoni swimmers. For symmetric surface swimmers
propelled by thermal Marangoni forces, a related symmetry-breaking
effect has been observed in Ref.\ \cite{Girot2016}, where
symmetric microbeads spontaneously circle around a heating laser beam.
    Our analysis
    allows us to obtain the swimming velocity of Marangoni boats
    as a function of Marangoni
  propulsion strength (Peclet number) and to analyze in detail 
   the nature of the symmetry-breaking
   swimming bifurcation.

\section{Model}

We introduce coordinates such that the origin $r=0$
is at the center 
of the planar surface of the
half-sphere,  the liquid-air interface is at
   $y=0$ (with $y<0$ being the liquid phase), and
   $\vec{e}_z$ will coincide with
the spontaneously selected swimming direction, see Fig.\ \ref{fig:design}.
We also use spherical coordinates such that $\theta = 0$ is the swimming
direction and
the interfacial plane is located at $\phi = 0,\pi$ ($y=0$).
The half-sphere has  radius $a$ such that the  contact line
is at $r=a$  and $\phi = 0,\pi$ (and parameterized by $\theta$).
We denote the half-spherical surface of the swimmer by $S$,  the
circular  air-water-swimmer  contact line by $L$,
and the liquid-air interface
outside the swimmer  as $S_{\rm Int}$.

 The general strategy to determine
 the swimming speed has been  outlined in Ref.\ \cite{Ender2020}. 
 We first prescribe a stationary
velocity $\vec{U} = U \vec{e}_z$
of the swimmer and  analyze
the following three coupled problems for its stationary state:
\begin{itemize}
  \item[(i)] Surface tension reduction by surfactant adsorption
  at the air-water interface;
  depending on the volatility of the surfactant,
  we also need to include evaporation.
\item[(ii)]
   Low Reynolds number fluid flow including both 
   Stokes flow around the half-sphere 
   and additional surfactant-induced fluid Marangoni flow.
\item[(iii)]  
  Diffusive surfactant release or surfactant dissolution
   from the swimmer and subsequent 
  diffusion and advection.
\end{itemize}
We will start from the diffusion-advection problem (iii) in the
presence of Stokes flow around a sphere and neglecting Marangoni flow.
This will also give new results for the mass transfer from spheres in
laminar Stokes flow. 
We later examine the additional effects of Marangoni flow  and evaporation.
The fully coupled problems can also be treated numerically. 

Solving these three coupled problems, we can obtain the
Marangoni forces as a function  of the prescribed velocity $U$
from the surfactant concentration profile by employing the reciprocal
theorem. 
Finally, the actual swimming velocity $U=U_{\rm swim}$
is determined from the condition of a force-free swimmer,
 i.e., the force equilibrium between Stokes  drag force,
 direct propelling Marangoni forces from the surface tension gradient
 along the
 air-water-swimmer   contact line and  Marangoni   flow forces.

We begin with a short recapitulation
  of the governing equations \cite{Ender2020}.

\subsection{Coupled adsorption, fluid flow, and diffusion-advection problems}

Regarding sub-problem (i),
  we use a local and linear relationship for the
  surface tension reduction
\begin{align}
  \Delta \gamma(\vec{r})  &= -\kappa c(\vec{r})
  \label{eq:kappa}
\end{align}
($\vec{r}$ is an interfacial vector with $y=0$)
by the local surfactant concentration difference
$ c(\vec{r})$ with respect to a bulk concentration
background value $c_0$ (the concentration at $|\vec{r}|\to \infty$);
$\kappa$ is a coefficient characterizing the propulsion strength. 
In formulating Eq.\ (\ref{eq:kappa}) locally,
we  assumed  fast  on and off kinetics
of surfactant to the interface \cite{li1994} such that
the interfacial concentration $\Gamma(\vec{r})$
is slaved to the bulk
and only a passive  ``reporter''
of the bulk subsurface concentration $\left.c(\vec{r})\right|_{y=0}$.
This is appropriate for water-soluble surfactants but, for example,
  the opposite limit of what has been considered in Refs.\
  \cite{Lauga2012,Vandadi2017}, where surfactant is strictly confined
  to the interface.

Fast  on and off kinetics also implies that an
imbalance of flux to and from the interface into the bulk
can only arise from an additional evaporating flux from the
interface to the gas phase. 
In the general case including  surfactant evaporation
from the interface,  the balance of fluxes to and from the
interface gives 
\begin{equation}
  j_{\rm Int} =
  -\left. D \vec{\nabla} c(\vec{r})\cdot \vec{n}^{\rm out}\right|_{y=0}
  = -  j_{\rm ev} =  k\left. c(\vec{r})\right|_{y=0},
  \label{eq:jev}
\end{equation}
where $k$ is the rate constant for evaporation.
This provides the  boundary condition to the
diffusion-advection sub-problem (iii) in the bulk.

Regarding the low Reynolds number fluid flow sub-problem (ii),
we consider the rest frame of the  swimmer 
and linearly decompose the total fluid flow field into
a field $ \vec{v}(\vec{r})$, which is the flow field of a
half-sphere pulled with velocity $U\vec{e}_z$ through the liquid
and a correction $\vec{{v}}_{\rm M}(\vec{r})$ from Marangoni flows, 
$ \vec{v}_{\rm tot}(\vec{r})
              = \vec{v}(\vec{r}) +
              \vec{{v}}_{\rm M}(\vec{r})$.
For low Reynolds numbers, both $ \vec{v}(\vec{r})$ and $\vec{{v}}_{\rm
  M}(\vec{r})$ (and the associated pressure fields)
fulfill the incompressibility condition $\vec{\nabla}\cdot \vec{v} = 0$ and
the linear Stokes equation  $\mu \vec{\nabla}^2 \vec{v}
= \vec{\nabla} p$, where $\mu$ is the fluid viscosity.

The flow field  $\vec{v}(\vec{r})$ 
of an externally pulled half-sphere
is given by ``half'' ($y<0$)
the Stokes flow field around a sphere, which automatically
fulfills the boundary condition  $\left. v_y(\vec{r}) \right|_{y=0} = 0$
for symmetry reasons. In spherical coordinates, the axisymmetric
Stokes flow field is 
\begin{subequations}
\begin{align}
  \vec{v}(\vec{r}) &= \hat{u}(r,\theta) \vec{e}_r +
                      \hat{v}(r,\theta) \vec{e}_\theta
                    ~~\mbox{with} \nonumber\\
  \hat{u}(r,\theta) &=
          U \cos\theta \left[-\frac{1}{2} \left(\frac{a}{r}\right)^3 +
                \frac{3}{2} \frac{a}{r} - 1 \right]
         \equiv  U \cos\theta  u(r/a),
         \label{u}\\
  \hat{v}(r,\theta) &=
      U \sin\theta \left[-\frac{1}{4} \left(\frac{a}{r}\right)^3 -
               \frac{3}{4} \frac{a}{r} + 1 \right]
              \equiv  U \sin\theta  v(r/a).
               \label{v}
\end{align}
\end{subequations}

The total flow field  $\vec{v}_{\rm tot}(\vec{r})$ also  has
 no-slip boundary
 conditions on the surface of the sphere and 
 assumes $\vec{v}_{\rm tot}(\infty) =  - U\vec{e}_z$ at infinity,
 but is subject to Marangoni stresses
 at the liquid-air interface.
 Consequently, the difference
 $\vec{v}_{\rm M}(\vec{r}) =  \vec{v}_{\rm tot}(\vec{r})-
\vec{v}(\vec{r}) $ from Marangoni flows has no-slip boundary
 conditions on the surface of the sphere, has vanishing velocity 
 $\vec{v}_{\rm M}(\infty) = 0$ at infinity, and
 is subject  to Marangoni stresses
 at the liquid-air interface.
Moreover, for all three flow fields, there is no normal flow across the
liquid-air interface.
We will assume that the liquid-air interface remains flat,
even if the sphere moves.
This requires that typical viscous forces
remain small compared to interfacial stress,
$\mu U  \ll \gamma$, which is fulfilled with
$\mu U \sim  10^{-5}\, {\rm N/m}$ for generic Marangoni boats with 
$U\sim 1\, {\rm cm/s}$ and $\gamma \sim  0.07\, {\rm N/m}$ for the air-water
interface. We also neglect a possible  curvature of the interface from wetting
effects. 

The Marangoni flow is caused by  tangential Marangoni
stresses at the liquid-air interface $y=0$,
\begin{equation}
  \mu  \vec{n}^{\rm out}\cdot \vec{\nabla}
  \left. \vec{v}_{\rm M}(\vec{r})\right|_{y=0} =
  \mu \partial_y  \left. \vec{v}_{\rm M}(\vec{r})\right|_{y=0}
  = \vec{\nabla}_S \Delta \gamma(\vec{r}),
  \label{eq:Marangoni}
\end{equation}
which act both on $ \vec{v}_{\rm M}$ and $ \vec{v}_{\rm tot}$.

Surfactant diffusion and advection (iii) will play a
central role. 
Surfactant molecules are emitted from the
half-spherical surface $S$ and diffuse in the liquid phase.
At the same time, they are advected by the total fluid flow. 
In the stationary state, the bulk concentration field
is governed by the diffusion-advection equation
\begin{align}
  0=  \partial_t c &= D \vec{\nabla}^2 c -
      (\vec{v}(\vec{r})+ \vec{v}_{\rm M}(\vec{r})) \cdot \vec{\nabla} c.
          \label{eq:diffadv0}
\end{align}
We consider two types of boundary conditions which seem
  most important for applications: (A)
slow diffusional surfactant release on $S$
leading to a constant flux boundary condition
or (B) surfactant dissolution from the swimmer or surfactant  production
by some chemical reaction by the swimmer leading to a constant
concentration boundary condition,
\begin{align}
  \mbox{(A)~constant flux:}~~ &&  \left. \vec{j}\cdot\vec{n}\right|_S
  &=- D\left. \vec{\nabla} c\cdot\vec{n} \right|_S  =\alpha,
    \label{eq:constantflux}\\
  \mbox{(B)~constant conc.:}~~ &&
               \left.   c   \right|_S &= c_S
         \label{eq:constantconc}             
\end{align}
together with $c(\infty)=0$ and  the no-flux
boundary condition at the
interface $S_{\rm Int}$.
The surface flux $\alpha$ or the surface concentration $c_S$ is assumed
to be only  slowly changing  on the
time scales of the fluid flow and the surfactant diffusion
and approximated as a constant for the calculation of quasi-stationary
fluid flow and
concentration fields.

We non-dimensionalize sub-problems (i)-(iii)  by
measuring lengths in units of $a$, velocities in units of $D/a$,
concentrations in units of $\alpha a/D$
\begin{align}
  \vec{\rho} &\equiv \vec{r}/a,~~\bar{\vec{\nabla}} \equiv a \vec{\nabla}
                  = \vec{\nabla}_\rho,
               ~~\bar{\vec{v}}\equiv \vec{v} \frac{a}{D},~~
         \bar{U}  \equiv U \frac{a}{D},
 \nonumber \\
  \bar{c} &\equiv   c \frac{D}{\alpha a},~~ \bar{j} \equiv  j\frac{1}{\alpha},
          ~~\bar{p} \equiv p \frac{a^2}{D\mu}.
            \label{eq:nondim}
  \end{align}
The prescribed dimensionless velocity $\bar{U}$ of the swimmer is the first
control parameter of the problem,\footnote{In many publications on the
  diffusion-advection problem, such as Refs.\
  \cite{Acrivos1960,Acrivos1962,Acrivos1965} but also in Refs.\
  \cite{Lauga2012,Yariv2015,Vandadi2017,Boniface2019,JafariKang2020},
  $\bar{U}$ is called the Peclet-number.
  Here, we define the Peclet number as ${\rm Pe} \equiv \bar{U}_{\alpha}$,
  i.e.,  by the 
  characteristic velocity $\bar{U}_{\alpha} = {\kappa \alpha a}/{D \mu}$
  for constant flux boundary conditions.
    We define it as ${\rm Pe} \equiv \bar{U}_{c_s} = {\kappa c_S a}/{D \mu}$
    for constant concentration boundary conditions.
  $\bar{U}_{\alpha}$ and $\bar{U}_{c_s}$ are characteristic velocities,
  where a typical direct Marangoni force $F_{\rm M} \sim \kappa a^2
  \partial_rc(r=a)\sim \kappa a^2\alpha/D$ for constant flux boundary
  conditions or $F_{\rm M} \sim \kappa a
  c(r=a)\sim \kappa c_S a/D$ for constant concentration boundary
  conditions is balanced by the 
  typical Stokes drag force $F_D \sim \mu a U$.
  The Peclet number is a dimensionless measure of propulsion strength
  with these definitions.
}
which is related to  the Reynolds number, ${\rm Re} = 2\bar{U}/{\rm Sc}$,
via the Schmidt number ${\rm Sc} \equiv\mu /\rho D$. 
Low Reynolds numbers ${\rm Re}\ll 1$
are realized for $\bar{U} \ll {\rm Sc}/2$, which can still be much larger
than unity as typical Schmidt numbers for surfactants in aqueous
solutions are of the order of $1000$.
Therefore, we have to discuss \emph{both} the diffusive case $\bar{U}\ll 1$
\emph{and} the advective case $\bar{U}\gg 1$, even at low Reynolds numbers.

Our dimensionless set of equations for problems (i)-(iii) becomes 
\begin{subequations}
  \label{eq:i-iii}
\begin{align}
  {\rm (i)}&&& 
      -\left. \bar{\vec{\nabla}}  \bar{c}(\vec{\rho})\cdot \vec{n}^{\rm out}
                                     \right|_{\bar{y}=0}\approx 0\nonumber\\
    &&&\mbox{without evaporation,}   \label{eq:bcnoevap} \\
     &&&
  -\left.  \bar{\vec{\nabla}}  \bar{c}(\vec{\rho})\cdot \vec{n}^{\rm out}
                                     \right|_{\bar{y}=0}
               \approx \bar{k}\left. \bar{c}(\vec{\rho})
                            \right|_{\bar{y}=0}
                            \nonumber\\
     &&&\mbox{with evaporation,} \label{eq:bcevap} \\
{\rm (ii)}&& \bar{\vec{v}}_{\rm tot}(\vec{\rho})
                        &= \bar{\vec{v}}(\vec{\rho}) +
                                      \bar{\vec{v}}_{\rm M}(\vec{\rho}),
  \nonumber \\
{\rm (iia)}&&
  \bar{\vec{v}}(\rho,\theta) &= \bar{U}\cos\theta  u(\rho) \vec{e}_r +
                  \bar{U} \sin\theta  v(\rho) \vec{e}_\theta \nonumber\\
    &&&\mbox{Stokes~flow~field,}  \label{eq:Stokes}
  \\
{\rm (iib)}&&
  \bar{\vec{\nabla}}\cdot \bar{\vec{v}}_{\rm M} &= 0\nonumber\\
    &&&\mbox{Marangoni~flow~field,}
                             \nonumber \\
  &&\bar{\vec{\nabla}}^2 \bar{\vec{v}}_{\rm M} &=
               \bar{\vec{\nabla}} \bar{p}_{\rm M},
  \nonumber \\
  &&  \bar{\vec{v}}_{\rm M}(\infty) &=0,
                             \nonumber  \\
  &&  \left. \bar{\vec{v}}_{\rm M}(\vec{\rho}) \right|_{\rho=1}&=0,
  \nonumber \\
   &&\left. \bar{v}_{\rm M, y}(\vec{\rho}) \right|_{\bar{y}=0} &=0,
 \nonumber \\
     && \left. \partial_{\bar{y}}  \bar{\vec{v}}_{\rm M}(\vec{\rho})
        \right|_{\bar{y}=0}
              &= -{\rm Pe}
    \left.    \bar{\vec{\nabla}}_{S}  \bar{c}(\vec{\rho}) \right|_{\bar{y}=0},
                          \label{eq:Mflow}\\      
{\rm (iii)}&&  0 &=  \bar{\vec{\nabla}}^2 \bar{c} -
               (\bar{\vec{v}}(\vec{\rho})+ \bar{\vec{v}}_{\rm M}(\vec{\rho}))
      \cdot \bar{\vec{\nabla}} \bar{c},
  \label{Dadvdim} \\
 &&   \bar{c}(\infty)  &= 0,
   \nonumber  \\
 &&\mbox{(A)~const. flux:}&\nonumber\\
      &&  \left. \bar{\vec{j}}\cdot\vec{n}\right|_S
             &=-
               \left. \bar{\vec{\nabla}} \bar{c}\cdot\vec{n} \right|_S
               =1,
               \label{eq:boundconstflux}\\
&& \mbox{(B)~const. conc.:}&\nonumber\\
            &&      \left.   \bar{c}   \right|_S &= 1
                  \label{eq:boundconstc}
\end{align}
\end{subequations}
with the dimensionless Peclet number
\begin{align}
  \mbox{(A)~const. flux:}&&
         {\rm Pe} &\equiv {\rm Pe}_A \equiv \frac{\kappa \alpha a^2}{D^2 \mu}=
                    \frac{\kappa \dot{m}}{2\pi D^2 \mu},
  \nonumber\\
   \mbox{(B)~const. conc.:}&&
    {\rm Pe} &\equiv {\rm Pe}_B\equiv \frac{\kappa c_S a}{D \mu} ,              
            \label{eq:Pe}
 \end{align}
where $\dot{m}= 2\pi a^2\alpha$ is the mass loss per time
of the swimmer.
We also introduced the  dimensionless Biot number
\begin{equation}
   \bar{k} \equiv \frac{ak}{D}
   \label{eq:Biot}
 \end{equation}
governing possible evaporation.
From Eq.\ (\ref{eq:Mflow}), we see that the Peclet number ${\rm Pe}$
determines the velocity scale of the Marangoni flow field.
Therefore, we can also assign a Reynolds number
${\rm Re}_{\rm M} = {2{\rm Pe}}/{{\rm Sc}}= {\rm Re} {\rm Pe}/\bar{U}$
to the Marangoni flow.
In the following, we will address the low Reynolds number regime
implying that \emph{both} ${\rm Re}\ll 1$ \emph{and} ${\rm Re}_{\rm M}\ll 1$
such that both flow contributions fulfill the Stokes equation. 
Via the advection with $\bar{\vec{v}}(\vec{\rho})+ \bar{\vec{v}}_{\rm
  M}(\vec{\rho})$,  the concentration field $c(\vec{\rho})$ depends
both on the dimensionless velocity scale $\bar{U}$ of the Stokes field
and the dimensionless velocity scale ${\rm Pe}$ of the Marangoni
flow field, in general.
All dimensionless parameters are summarized in Table \ref{tab:nondimen}.

\begin{table*}
  \begin{center}
  \caption{\label{tab:nondimen}
    Dimensionless parameters.
    ${\rm Re}$ or $\bar{U}$,
    ${\rm Sc}$, ${\rm Pe}$, and $\bar{k}$ are control parameters
    of the problem.  ${\rm Re}_{\rm M}$ and ${\rm Nu}$
    cannot be independently  controlled but characterize the
    resulting solutions; the swimming velocity $\bar{U}_{\rm swim}$
    is determined by the
     force balance swimming condition.} 
   \begin{tabular}{ l @{\qquad\qquad} l @{\qquad\qquad}  l }
     \hline\noalign{\smallskip}
     Dimensionless parameter &      Formula      & Eqs.    
     \\
   \noalign{\smallskip}\hline\noalign{\smallskip}
   Reynolds number ${\rm Re}$ &  $ ={\rho U 2a}/{\mu}
                  = {2\bar{U}}/{{\rm Sc}}$  &   \\
   dimensionless velocity  $\bar{U}$  & $ = U{a}/{D}  $
      &\\
   Schmidt number ${\rm Sc}$  &  $ = {\mu}/{\rho D}$  &  \\
   Peclet number ${\rm Pe}$ & $ =  {\kappa \alpha a^2}/{D^2 \mu}$
                                  &    (\ref{eq:Pe})\\
     Biot number $\bar{k}$ & $= {ak}/{D}$ & (\ref{eq:Biot}) 
     \\
     \noalign{\smallskip}\hline\noalign{\smallskip}
   swimming velocity  $\bar{U}_{\rm swim}$  & $ = U_{\rm swim}{a}/{D}  $
                        & (\ref{eq:swimcond})  \\
   Marangoni Reynolds number ${\rm Re}_{\rm M}$ &
   $= {2{\rm Pe}}/{{\rm Sc}}$ &  
     \\
   Nusselt (or Sherwood) number ${\rm Nu}$ (${\rm Sh}$)
            &   $ ={-\partial_\rho \bar{c}_0(1)}/{\bar{c}_0(1)}$ &
                    (\ref{eq:Nu})  \\
\noalign{\smallskip}\hline
   \end{tabular}
   \end{center}
\end{table*}

\subsection{Marangoni forces, energy transduction and swimming condition}

The half-spherical swimmer moving at velocity $U$
must be force-free and 
is subject to three forces. First, there is
the drag force, which is given by
the standard Stokes drag for a half-sphere, $
\vec{F}_{\rm D} = F_{\rm D}\vec{e}_z$.
In dimensionless form using $\bar{F} \equiv F/{D\mu}$,
this is 
\begin{equation}
   \bar{F}_{\rm D}=  -3\pi  a \bar{U}.
   \label{FD}
 \end{equation}
 Second, there is the direct Marangoni propulsion
 force  $\vec{F}_{\rm M} = F_{\rm M} \vec{e}_z$ from
 integrating the  surface stress $\Delta\gamma(\vec{r})= -\kappa c(\vec{r})$
 along the air-water-swimmer  contact line $L$ around the swimmer
 at $y=0$,   
 \begin{align}
   \frac{\bar{F}_{\rm M}}{\rm Pe}
   &=  -   \oint_L d\bar{s}  (\vec{e}_n\cdot \vec{e}_z)
                  \bar{c}(\vec{\rho})
              \nonumber\\
    &=- 2  \int_0^\pi d\theta \cos\theta  \bar{c}(1,\theta)|_{\bar{y}=0},
                \label{FMdim}  
\end{align}
in dimensionless form.
For constant concentration boundary conditions (B), there is no direct
  Marangoni force $\bar{F}_{\rm M}=0$
  because there are no concentration and, thus, surface tension gradients
  along the  contact line $L$ by definition.

Third, there is the Marangoni flow force
$\vec{F}_{\rm M, fl} = F_{\rm M, fl} \vec{e}_z$, which is by definition
the force transmitted by fluid stresses of the Marangoni flow onto
the sphere, 
$F_{\rm M, fl} \equiv - \int_{S} da_i \sigma_{\rm M, iz}$.
For low Reynolds numbers, we can employ the reciprocal theorem
to calculate the Marangoni flow force without explicitly
calculating the Marangoni flow $\vec{v}_{\rm M}$ \cite{Masoud2014}.
In Appendix \ref{app:energy}, we discuss the reciprocal
theorem in terms of  energy transduction and find
the result (\ref{eq:transduction}), which states that 
the mutual power input by Marangoni stresses via the
 Stokes flow field is \emph{completely}
 transduced via the  Marangoni flow force
 onto the sphere,
 while the power input  by Marangoni stresses via the
 Marangoni flow field itself is completely dissipated.
This energy transduction statement (\ref{eq:transduction}) is equivalent
 to the result derived by Masoud and Stone  \cite{Masoud2014}
 for the Marangoni flow force directly from the reciprocal theorem.
 In the rest frame of the sphere, we obtain in dimensionless form
 \begin{align}
  \frac{\bar{F}_{\rm M,fl}}{\rm Pe} 
   &= - \int_{S_{\rm Int}} d\bar{S}
     \frac{\bar{\vec{v}}(\vec{\rho})+\bar{U}\vec{e}_z}{\bar{U}}
     \cdot \bar{\vec{\nabla}}_S
        \bar{c}(\vec{\rho}),
              \label{FMfl}
\end{align}
where $\bar{\vec{v}}(\vec{\rho})/\bar{U}$
is the dimensionless Stokes flow field
from (\ref{u}) and (\ref{v}) (in particular, this is independent of $\bar{U}$)
in the sphere frame.

The total
Marangoni force $\bar{F}_{\rm M,tot} = \bar{F}_{\rm M} +  \bar{F}_{\rm M,fl}$
is obtained by using Eqs.\ (\ref{FMdim}) and (\ref{FMfl})
and the Gauss theorem,
\begin{align}
  \frac{\bar{F}_{\rm M,tot}}{\rm Pe} 
  &=    \int_{S_{\rm Int}} d\bar{S}
    \left(\bar{\vec{\nabla}}_S\cdot
    \frac{\bar{\vec{v}}(\vec{\rho})}{\bar{U}}\right)
     \bar{c}(\vec{\rho})\nonumber\\
 &=-\frac{3\pi }{4}
      \int_1^\infty d\rho 
    \left(\frac{1}{\rho} - \frac{1}{\rho^{3}} \right)
  \bar{c}_M(\rho)\label{FMtotaldim}\\
  \mbox{with}~~\bar{c}_M(\rho) &\equiv  \frac{2}{\pi} \int_0^\pi
d\theta \cos\theta  \bar{c}(\rho,\theta)|_{y=0}.
    \nonumber
\end{align}
The total Marangoni driving force has to be 
 determined from the concentration field
 $\bar{c}(\rho,\theta)$ of surfactant molecules (at the interface $y=0$).
 Note that $\bar{\vec{\nabla}}_S\cdot \bar{\vec{v}}(\vec{\rho})$
 is the two-dimensional
 surface divergence of the 3D fluid velocity field; therefore,
 $\bar{\vec{\nabla}}_S\cdot \bar{\vec{v}}(\vec{\rho})\neq 0$
 in general, although
 $\bar{\vec{\nabla}}\cdot \bar{\vec{v}}(\vec{\rho})= 0$
 for the 3D divergence of
 the stationary velocity field.
The contribution from a constant velocity $\bar{U}\vec{e}_z$ of the whole 
fluid (if all the fluid would be dragged along by the particle) exactly 
cancels the direct Marangoni force in (\ref{FMtotaldim}),
and the velocity $\bar{\vec{v}}(\vec{\rho})$ in 
the sphere frame determines the total force.

The sign of the Marangoni flow force $\bar{F}_{\rm M,fl}$
determines 
  whether it increases or decreases the
  direct Marangoni force  into the direction
  of higher surface tension:
\begin{itemize}
  \item
  For  anisotropic pure 2D surface diffusion without advection, 
  $\bar{c}_{2D}(\rho,\theta) = {\rm const} + 2A_1 \rho^{-1} \cos\theta
  + ...$ ($A_1<0$),
 as in Refs.\ \cite{Lauga2012,Masoud2014}, we find
 $\bar{F}_{\rm M,tot}/{\rm Pe} = - \pi A_1 =
 \frac{1}{2}\bar{F}_{\rm M}/{\rm Pe}$,
   i.e., the total Marangoni force is half the direct Marangoni
   force if only the first $\cos\theta$-component is relevant.
   Here, Marangoni flow forces drag and \emph{decrease}
   the direct driving force ($\bar{F}_{\rm M,fl}<0$).
   This result will change as we (i) consider 3D diffusion and (ii)
   as symmetry breaking is only caused by
    advection, which can focus the concentration field and
   lead to higher Legendre components becoming
   relevant in $\bar{c}(\vec{\rho})$.   

  \item
Because $\rho^{-1} - \rho^{-3} >0$ for $\rho>1$, 
the total Marangoni force is always positive for concentration
profiles with $\bar{c}_M(\rho) <0$, 
which are increasing  toward the rear side.
 Vandadi {\it et al.} have shown that
  this can change in confinement,
  when the high of the fluid container becomes comparable to the
  sphere radius\cite{Vandadi2017}.

For constant concentration boundary conditions (B),
  this means that $\bar{F}_{\rm M,tot} = \bar{F}_{\rm M,fl}>0$
  because  there is no direct
  Marangoni force $\bar{F}_{\rm M}=0$ for these boundary conditions.

For constant flux boundary conditions (A),
the Marangoni flow contribution $\bar{F}_{\rm M,fl}$, however, can have
both signs. For $\bar{F}_{\rm M,fl}>0$, the flow force increases the
direct Marangoni force resulting in $\bar{F}_{\rm M,tot}> \bar{F}_{\rm M}$;
for  $\bar{F}_{\rm M,fl}<0$, the flow force is directed backward and
increases the
drag force  resulting in $\bar{F}_{\rm M,tot}< \bar{F}_{\rm M}$.
As opposed to Ref.\ \cite{Lauga2012}, we will find that
both cases are possible. A backward force is found
for steep radial gradients in the concentration $\bar{c}(\rho)$,
which is the case for high velocities $\bar{U}\gg 1$
  in the advection-dominated
  regime, and a forward force is found
at low velocities $\bar{U}\ll 1$ in the diffusive regime. 

\item 
Advection leads to  a tangential  $\vec{e}_\theta$-component of 
 $ \bar{\vec{\nabla}}_S  \bar{c}(\vec{\rho})$
 pointing from the front  to the rear corresponding to an increasing 
surfactant concentration toward the rear, which 
gives rise to a forward Marangoni flow $\sim - \vec{e}_\theta$.
Accordingly, this increases  the driving force ($\bar{F}_{\rm M,fl}>0$)
because $ -\vec{e}_z \cdot \bar{\vec{\nabla}}_S
\bar{c}(\vec{\rho})\sim -\vec{e}_z \cdot \vec{e}_\theta\sim \sin\theta >0$
in Eq.\ (\ref{FMfl}).
This effect dominates in the diffusive regime.

  A radial $\vec{e}_r$-component of
  $\bar{\vec{\nabla}}_S  \bar{c}(\vec{\rho})$
pointing inward corresponding to a radially decaying surfactant
concentration and, on the other hand, gives rise to 
to radially outward Marangoni flows.
Because
$ -\vec{e}_z \cdot \bar{\vec{\nabla}}_S  \bar{c}(\vec{\rho})\sim
\vec{e}_z \cdot\vec{e}_r  \propto \cos\theta$ in Eq.\ (\ref{FMfl}),
this increases
the direct force in the
front (around $\theta=0$) but decreases it in the back (around $\theta=\pi$).
Advection gives rise  to bigger surfactant concentrations in the back,
which lead to bigger radial
concentration gradients on the rear side (in some distance from the
sphere because constant flux boundary conditions assure uniform
radial gradients right at the surface of the sphere).
Overall, the radial Marangoni flows in the back
are stronger and decrease the  direct force
or   increase the  drag
($\bar{F}_{\rm M,fl}<0$).
This effect dominates in the advective regime for constant
flux boundary conditions (A) and
 is  rather subtle, as can be seen from the fact that
it is  absent for the constant concentration boundary conditions (B),
where $\bar{F}_{\rm M,tot} = \bar{F}_{\rm M,fl}>0$ always.
Then, the constant concentration at the surface of the sphere
leads to smaller  radial
concentration gradients on the rear side, because the 
concentration decay is stretched over a larger distance by advection.
Then,  radial Marangoni flows in the front are stronger and
increase the direct force.

\item 
 The last equality in (\ref{FMtotaldim}) shows that the effect of including the 
Marangoni flow contribution is that the  total Marangoni forces are 
dominated by the concentration profile $\bar{c}(\rho,\theta)$ around
$\rho\sim 2$,
where $\rho^{-1} - \rho^{-3}$ assumes its maximal value.
Concentration  boundary 
layer profiles concentrated around  $\rho \approx 1$, as we will
find for large swimmer velocities $\bar{U}> 1$ in the advection-dominated
  regime, give a small 
total Marangoni force (because 
$\rho^{-1} - \rho^{-3}\approx 2(\rho-1)$ is small), i.e.,  
Marangoni flows decrease the direct Marangoni driving force.

 \item
Long-range contributions as, for example, from a long advection tail 
can be important, even if they are limited to a small angular 
regime around $\theta\sim \pi$ as for high velocities.
The highest total force is obtained if a long-range $-\cos\theta$-component
is present in the concentration profile, as we will find for
small swimmer velocities; then Marangoni flows   increase 
 the direct Marangoni driving force.
This makes a Marangoni swimmer also susceptible to disturbances in its 
far-field as, for example, induced by other swimmers. 

\end{itemize}

These results for Marangoni forces as a function of $\bar{U}$
are inserted into  the force balance or swimming condition
\begin{equation}
  -\bar{F}_{\rm D} = 3\pi  \bar{U}_{\rm swim} =
  \bar{F}_{\rm M}({\rm Pe},\bar{U}_{\rm swim}) +
  \bar{F}_{\rm M,fl} ({\rm Pe},\bar{U}_{\rm swim}),
  \label{eq:swimcond}
\end{equation}
 in order to obtain an additional equation
whose solution determines the actual
swimmer velocity $\bar{U}=\bar{U}_{\rm swim}$
as a function of the remaining control parameters
${\rm Pe}$ (``fuel'' emission) and eventually $\bar{k}$ (evaporation).

\subsection{Control parameters and parameter regimes}

The non-dimensionalization reveals that the coupled problems (i)-(iii)
and the Marangoni forces 
depend on three dimensionless control parameters (see also Table
\ref{tab:nondimen}):
first, the prescribed dimensionless velocity of the swimmer $\bar{U}$;
second, the Peclet number $\rm Pe$ characterizing the strength $\alpha$
of the  surfactant emission, and third, the Biot number $\bar{k}$
characterizing the evaporation.
A suitable  Peclet number can be defined  for both constant flux boundary
conditions (A) and constant concentration boundary conditions (B). 
We also see that the Peclet number both controls the
strength of the Marangoni flow via Eq.\ (\ref{eq:Mflow}) and
the strength of all  Marangoni forces. 
We note, however,
that $\bar{F}_{\rm M}/{\rm Pe}$ and $\bar{F}_{\rm M,tot}/{\rm Pe}$
still depend on $\bar{U}$ and
${\rm Pe}$ via the dependence of  $\bar{c}(\vec{\rho})$ on these
parameters.

Another important finding from non-dimensionalization is that 
the diffusion-advection problem (iii) with boundary conditions (i)
decouples from the Marangoni flow
problem (iib) for ${\rm Pe} \ll \bar{U}$ or ${\rm Re}_{\rm M} \ll {\rm Re}$,
where $|\vec{v}_{\rm M}| \ll |\vec{v}|$,
and we can neglect   $\vec{v}_{\rm M}$ in the advection term. 
Then, the concentration profile is only determined
by a classic diffusion-advection problem for mass transfer from a sphere
in Stokes flow in the case of constant concentration boundary
conditions (B)  \cite{Acrivos1960,Acrivos1962,Acrivos1965,Leal},
but with unusual constant flux boundary conditions for case (A).
It becomes axisymmetric, and  only depends on $\bar{U}$.
In this limit,  
the Marangoni flow field need not to be calculated in order
to calculate the total Marangoni force for the swimming condition.
This limit will be the starting point of several analytical calculations. 

All in all, we have the following regimes for a
symmetric Marangoni boat at low Reynolds numbers:
 \begin{itemize}
 \item
   $\bar{U}<1$ and  ${\rm Pe}<1$:
   The concentration profile is
   governed by diffusion, which is slightly perturbed by
     advection and described by a linear response
     in diffusion-advection (iii) with respect to  $\bar{U}$
     and ${\rm Pe}$.
     Only the linear response in $\bar{U}$ is relevant for
     symmetry breaking; therefore, the Marangoni flow can be
     neglected for the swimming problem. 
      Only
      for ${\rm Pe}\ll \bar{U}$,  the Marangoni flow
      decouples from the advection problem 
          and strict analytical analysis is 
    possible.
     Swimming sets in (starting with $\bar{U}=0$) for a critical
     Peclet number ${\rm Pe} >{\rm Pe}_c$;
     if Marangoni flows forces are included, we find   ${\rm Pe}_c\ll 1$
     and  the symmetry-breaking bifurcation  takes place within
       this regime. 
\item
    $\bar{U}<1$ and   $1<{\rm Pe} <{\rm Sc}$:
    All fluid flows are still at low  Reynolds numbers,
    but Marangoni flows are relevant.  The concentration profile is
   governed by symmetric Marangoni advection, which is slightly perturbed 
      by a linear response
      in diffusion-advection (iii) with respect to  $\bar{U}$,
      which causes symmetry breaking and swimming. 

\item
    $1<\bar{U}<{\rm Sc}$ and   ${\rm Pe} <{\rm Sc}$:
    All fluid flows are still at low  Reynolds numbers,
    but the concentration profile is
    governed by advection by the swimming flow for $\bar{U}>1$.
    Advection leads to the formation of  a 
    concentration boundary layer  of width
    $\bar{U}^{-1/3}$ around the half-sphere for $\bar{U}>1$.
    Only
      for ${\rm Pe}\ll \bar{U}$,  the Marangoni flow
      decouples from the advection problem 
          and strict analytical analysis is 
    possible.
    For ${\rm Pe}> \bar{U}$, Marangoni flows are relevant to advection,
    in principle, but
    the surfactant is transported away by the swimming flow field
    via the concentration boundary layer before it can advect
    to the  Marangoni flow field. There are, however,
    Marangoni flows in the advection tail, which will become relevant then. 
\end{itemize}

\begin{figure*}
  \begin{center}
    \includegraphics[width=0.99\textwidth]{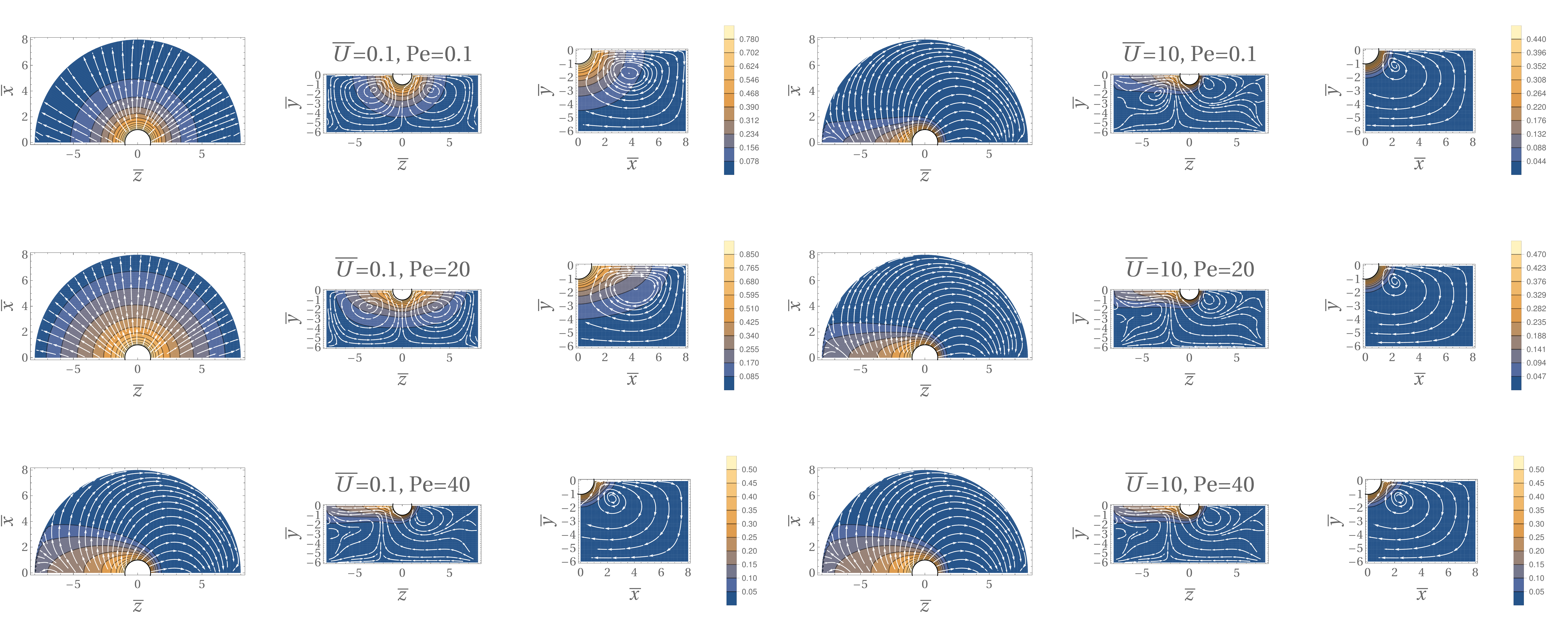}
 \caption{\label{fig:flows}
   Contour plots of the concentration $\bar{c}(\vec{\rho})$
   and the stream lines of the Marangoni flow field
   $\bar{\vec{v}}_{\rm M}(\vec{\rho})/{\rm Pe}$
   (in the comoving frame) for constant flux boundary conditions
  (A) and 
   $\bar{U}=0.1,10$ and  ${\rm Pe} = 0.1,20,40$
   from numerical iterative three-dimensional FEM result for
   a half-cylindrical region ($0<\rho<8$, $\bar{x}>0$,
     $-6 < \bar{y} <0$).
     After division by ${\rm Pe}$ the Marangoni flow field
     $\bar{\vec{v}}_{\rm M}(\vec{\rho})/{\rm Pe}$ is
     rather independent of ${\rm Pe}$ suggesting that
     $\bar{\vec{v}}_{\rm M}(\vec{\rho})/{\rm Pe}$ essentially depends on $\bar{U}$.
     The Marangoni flow forms a roll. 
   }
   \end{center}
   \end{figure*}

Figure \ref{fig:flows}
shows exemplary numerical finite element
results for the concentration field $\bar{c}(\vec{\rho})$
and the stream lines of the Marangoni flow
$\bar{\vec{v}}_{\rm M}(\vec{\rho})/{\rm Pe}$
for  different  parameter regimes for constant flux boundary conditions
  (A). 
At low velocities $\bar{U}\ll 1$, the Marangoni flow is
mostly radial  at the interface
because the radial concentration profile is only slightly
perturbed by advection at the interface; it  features a Marangoni roll (vortex)
around the swimmer with an upward flow directly around the particle.
For increasing ${\rm Pe}$,  the normalized Marangoni flow field
 $\bar{\vec{v}}_{\rm M}(\vec{\rho})/{\rm Pe}$
as plotted in Fig.\ \ref{fig:flows} seems unchanged indicating 
a Marangoni flow $\bar{\vec{v}}_{\rm M}(\vec{\rho})$ that
is simply proportional to ${\rm Pe}$ in strength but otherwise independent
of ${\rm Pe}$.

At high velocities $\bar{U}\gg 1$, the Marangoni flow pattern changes
because the concentration pattern develops the typical advection tail.
As a result, there forms a vortex pair within the interface plane, which
directs Marangoni flow  from
the tail to the front.
In front of the particle, the flow reaches beneath the particle
(around $\bar{z}=5$ in Fig.\ \ref{fig:flows})
and resurfaces behind the particle.
This leads to a slightly distorted Marangoni vortex roll
around the particle. Similar vortex patterns (with a vortex pair
within the interfacial plane next to the swimmer) have been observed
in Ref.\ \cite{Sur2019}, however, by  particle image velocimetry
(PIV) measurements at high Reynolds
numbers. 
Again, for increasing ${\rm Pe}$,  the normalized Marangoni flow field
 $\bar{\vec{v}}_{\rm M}(\vec{\rho})/{\rm Pe}$ seems more or less unchanged 
 in Fig.\ \ref{fig:flows}.

At small $\bar{U}\ll 1$,  the Marangoni flow force $\bar{F}_{\rm M,fl}>0$
will \emph{increase} the direct Marangoni force into positive z-direction
because there is a net forward tangential component of
$ \bar{\vec{\nabla}}_S  \bar{c}(\vec{\rho})$ from the symmetry-breaking
advection perturbation proportional to $\bar{U}$; the radial component  of
$ \bar{\vec{\nabla}}_S  \bar{c}(\vec{\rho})$
increases the drag  but is slowly decaying
at small $\bar{U}$ and weaker.

At high $\bar{U}\gg 1$, on the other hand, there is a strong radial
component in the concentration boundary layer around the swimmer,
which increases
the drag. This is created by the large radial component  of 
$ \bar{\vec{\nabla}}_S  \bar{c}(\vec{\rho})$
in the concentration boundary layer region of size
$\bar{U}^{-1/3}$ and 
 leads to a  Marangoni flow force $\bar{F}_{\rm M,fl}<0$
that  \emph{decreases} the direct Marangoni force.
This effect is counter-intuitive as the large vortex pair suggests
a strong forward Marangoni force on the  large scale picture. The strong radial
flows directly around the particle (which are stronger in the backward
direction and, thus, dragging the particle) are not clearly visible  on the
larger scale in  Fig.\ \ref{fig:flows}.
The remaining total Marangoni force mainly
comes from the net forward motion in the
horizontal vortex pairs but will be weaker than the direct force. 

   \subsection{Legendre decomposition
     for the decoupled limit ${\rm Pe} \ll \bar{U}$}

   In the decoupled limit  ${\rm Pe} \ll \bar{U}$,
   the
   diffusion-advection problem  becomes axisymmetric.
   Then, $\bar{c}=\bar{c}(\rho,\theta)$
   only depends on the radial coordinate and one angular coordinate,
   and we can also employ a decomposition of the
   concentration field into Legendre polynomials with respect to the
   angle $\theta$:
   $\bar{c}(\rho,\theta) = \sum_{n=0}^\infty \bar{c}_n(\rho) P_n(\cos \theta)$.
As derived in the Appendix \ref{app:Legendre},
  the diffusion-advection equation (\ref{Dadvdim}) only couples
  coefficients $\bar{c}_n(\rho)$ to coefficients $\bar{c}_{n\pm 1}(\rho)$
  because the Stokes  velocity field (\ref{u}) and (\ref{v})
can be written
in terms of $n=1$ polynomials only.
We find the  diffusion-advection equation (\ref{Dadvdim})
in Legendre representation,
\begin{align}
 & \left[ \frac{1}{\rho} \partial_\rho^2(\rho \bar{c}_n)
    - \frac{n(n+1)}{\rho^2}
  \bar{c}_n  \right] \nonumber\\
  &=  \bar{U} u(\rho) \left(   \frac{n}{2n-1}   
    \partial_\rho \bar{c}_{n-1}
    +  \frac{n+1}{2n+3} \partial_\rho \bar{c}_{n+1}
    \right)
    \nonumber\\
   & +\bar{U} \frac{v(\rho)}{\rho}
    \left(\frac{n(n-1)}{2n-1}   \bar{c}_{n-1}
     -  \frac{(n+1)(n+2)}{2n+3} \bar{c}_{n+1}  \right),
  \nonumber\\
  &\bar{c}_0(\infty)  = \bar{c}_\infty~,~~ \bar{c}_{n>0}(\infty)=0,
  \nonumber\\
 & \mbox{(A)~constant flux:}~~  
    \partial_\rho \bar{c}_0(1) = -1~,~~\partial_\rho \bar{c}_{n>0}(1) = 0,
    \nonumber\\
  &  \mbox{(B)~constant concentration:}~~
    \bar{c}_0(1) =  1~,~~\bar{c}_{n>0}(1) = 0
     \label{DadvLd}
\end{align}
 for $n=0,1,....$. 
For small $\bar{U}\ll 1$, the Legendre coefficients will scale as
$\bar{c}_n(\rho) \sim \bar{U}^n$ and truncation of Legendre
decomposition becomes an excellent approximation.
This is one strategy for analytical progress in the linear response
regime. 
In Appendix \ref{app:Legendre}, we also show how the Marangoni forces
are expressed by the Legendre coefficients of the concentration field.

Both types of boundary 
conditions are completely isotropic and only $n=0$ components
are nonzero. We can include traditional soap boats
into our description
by introducing  explicitly symmetry-breaking
anisotropic flux components $n>0$ into the boundary
conditions, such as
\begin{align}
  \mbox{(A)~constant flux:}~~&
    \partial_\rho\bar{c}_1(1)= \bar{\beta} >0,
                               \label{eq:beta}\\
  \mbox{(B)~constant concentration:}~~
        &
     \bar{c}_1(1) = \bar{c}_{S,1} >0
  \label{eq:c1S}                           
\end{align}
in the simplest generic case. 
Then, the soap boat emits
preferentially on the lower half $\theta > \pi/2$ in case (A) 
or produces  surfactant preferentially on the lower half of its surface
in case (B)
as in a standard asymmetric soap boat.
Such symmetry-breaking emission
will give rise to an avoided swimming bifurcation.

\section{Numerical methods}

\subsection{Full iterative FEM solution}

Numerically, we can consider the problems (i)-(iii) without further
approximations at low Reynolds numbers, i.e.,
solve the coupled diffusion-advection problem and the Marangoni flow problem
for a prescribed swimmer velocity $\bar{U}$.

For the coupled problems of 
three-dimensional coupled diffusion-advection and Marangoni flow,
we  use an iterative scheme of three-dimensional FEM solutions
to both problems, employing FEM-routines from Wolfram
MATHEMATICA in a finite cylindrical or rectangular domain. 
We  iteratively solve for the Marangoni flow field
(iib) starting from an  initial guess for the
concentration profile; then we solve the diffusion-advection 
equation (iii) with  the resulting total flow field, which gives
an improved approximation for the concentration profile.
With this improved approximation we go back into solving for the 
Marangoni flow field (iib) and start an iteration, which should converge
to the final Marangoni flow field and surfactant concentration field.
The iterative approach has the advantage that the Marangoni boundary
condition in the fluid flow problem (iib) is a fixed one
at each iterative step
and only adjusts over the iteration;
the coupling of the two problems is correctly established over the
iteration. 
Similar iterative numerical schemes for coupled problems have been
applied successfully in Refs.\ \cite{Boltz2015,Wischnewski2018}.

The FEM solution  of the stationary equations (iib) and (iii)
is obtained on a cylindrical or cubical irregular tetrahedral mesh.
We use  cubical (for example, with edge length 14
in $\bar{x}\bar{z}$-plane  and height 7 in $\bar{y}$-direction
in Fig.\ \ref{fig:FMtotalPe}) or cylindrical volumes
(for example, with
 radius 8 in $\bar{x}\bar{z}$-plane and height 4  in $\bar{y}$-direction
   in Fig.\ \ref{fig:flows}) for the FEM calculations. 
   The maximal volume of mesh elements is $0.2$, and  the mean volume is
   $0.01$.
    Mesh volumes are smaller ($<0.005$)
     in the region $-1<\bar{y}<0$ below the
     interface to capture Marangoni advection. 
Because of the mirror symmetry $\bar{x}\to -\bar{x}$, we only need to solve on
half-cubes and half-cylinders $\bar{x}>0$ and apply
Neumann boundary conditions $\left. \partial_{\bar{x}}
  \bar{c}\right|_{\bar{x}=0}=0$ and $\left. \partial_{\bar{x}}
  \bar{\vec{v}}_{\rm M}\right|_{\bar{x}=0}=0$ to  enforce the mirror symmetry. 
The boundary conditions at the outer boundaries are Dirichlet
conditions 
for the concentration $\bar{c}=0$ and the Marangoni flow
$\bar{\vec{v}}_{\rm M}=0$. For sufficiently large  cubes or cylinders,
these boundary conditions should not matter but we still have finite
size effects. In particular, at large Peclet numbers this can trigger
numerical instabilities if the Marangoni roll interferes with the
system boundary. 

We are interested in the resulting symmetry-breaking Marangoni forces
caused by a symmetry-breaking swimming motion as a function
of the  velocity $\bar{U}$.
At small $\bar{U}$, there is the problem that 
artificial symmetry breaking from lattice irregularities/defects
is often larger than symmetry breaking by swimming.
Therefore, we  average all measured quantities
over two simulations with $\bar{U}$ and $-\bar{U}$ to
cancel artificial  symmetry-breaking  effects.

\subsection{Two-dimensional  FEM solution and Legendre representation
  for the decoupled limit ${\rm Pe} \ll \bar{U}$}

For  ${\rm Pe} \ll \bar{U}$, we obtain the decoupled limit,
where Marangoni flow does not need to be calculated 
and the diffusion-advection
problem becomes axisymmetric. Then, $\bar{c}=\bar{c}(\rho,\theta)$
only depends on the radial coordinate and one angular coordinate.
We can 
solve the diffusion-advection problem in a two-dimensional
angular representation using finite element methods (FEM), i.e,
FEM-routines from Wolfram
MATHEMATICA.

For a given $\bar{U}$, we can also employ the
Legendre decomposition (\ref{DadvLd}) of the diffusion-advection
equation and calculate all functions $\bar{c}_n(\rho)$ by solving
the resulting coupled ordinary differential equation  boundary value
problem. We  use the MATLAB routine {\tt bvp4c}
for a domain $1\le \rho \le \bar{R}=300$ \cite{Khair2013,Michelin2011}
with  Legendre components up to $n=61$.
In this way, we obtain
all relevant coefficients $\bar{c}_n(\rho)$ to
calculate all Marangoni forces for the force balance.

\section{Diffusion-advection equation
  in the decoupled limit ${\rm Pe} \ll \bar{U}$ and mass transfer from
  a sphere in Stokes flow}

First, we will consider the  limit 
${\rm Pe} \ll \bar{U}$, where the diffusion-advection problem
for a half-sphere with prescribed velocity $U$
decouples from the Marangoni flow problem because
$\bar{\vec{v}}_{\rm M}$ can be neglected. 
We also neglect  evaporation in the beginning.
This problem is axisymmetric and
equivalent to mass transfer from a  full sphere
in laminar Stokes flow  \cite{Acrivos1960,Acrivos1962,Acrivos1965,Leal},
but with unusual constant flux boundary conditions for case (A).
  Therefore, we first derive
  new analytical results
  for concentration profiles and for the angular dependence of the
  Nusselt number for these boundary conditions, 
  both  for isotropic and anisotropic emission from the sphere. 
In the decoupled limit, the concentration profile only
depends on $\bar{U}$ and is independent of ${\rm Pe}$.
Thus,
the dimensionless Marangoni forces (\ref{FMdim}) and
(\ref{FMtotaldim}) only depend trivially linearly on  ${\rm Pe}$,
but $\bar{F}_{\rm M}/{\rm Pe}$ and  $\bar{F}_{\rm M,tot}/{\rm Pe}$
are independent of ${\rm Pe}$ as well. This will make
analysis of the swimming condition (\ref{eq:swimcond}) much easier.

\subsection{Nusselt number}

Diffusive release in an advecting flow can be  characterized by the
average Nusselt number (or Sherwood number {\rm Sh}),
\begin{equation}
  {\rm Nu} \equiv \frac{\int_S \vec{j}(\vec{r})\cdot\vec{n}\,dA}
  { (D/a) \int_S c(\vec{r})\,dA}
  = \frac{-\partial_\rho \bar{c}_0(\rho=1)}{\bar{c}_0(\rho=1)},
  \label{eq:Nu}
\end{equation}
which is the dimensionless ratio of the total emitted flux  and the
typical diffusive flux \cite{Leal}.
The average Nusselt number becomes ${\rm Nu}=1$ for a quiescent  fluid
($\bar{U}=0$), where the flow is purely diffusive
$\bar{c}_0(\rho) \propto 1/\rho$; as soon
as advection is present ($\bar{U}>0$) the current out of the sphere
is increased  resulting  in 
${\rm Nu}>1$. The  Nusselt number thus measures how much
the current  out of the sphere is increased by advection over
its purely diffusional value. It is an increasing function of the
fluid velocity $\bar{U}$.

The Nusselt number has been originally defined for constant
concentration boundary conditions (B),
for which 
the result
is well-known \cite{Acrivos1960,Acrivos1962,Acrivos1965,Leal},
\begin{align}
{\rm Nu} = -\partial_\rho\bar{c}_0(\rho=1)&=
\begin{cases}
   1+\frac{1}{2} \bar{U} + ... & \mbox{for}~\bar{U}\ll 1\\
   0.6245\, \bar{U}^{1/3}
   & \mbox{for}~\bar{U}\gg 1
 \end{cases}
 \label{eq:NuUB}
\end{align}
with a prefactor that can be calculated analytically \cite{Acrivos1960,Leal}.

 \begin{figure}
  \begin{center}
   \includegraphics[width=0.99\linewidth]{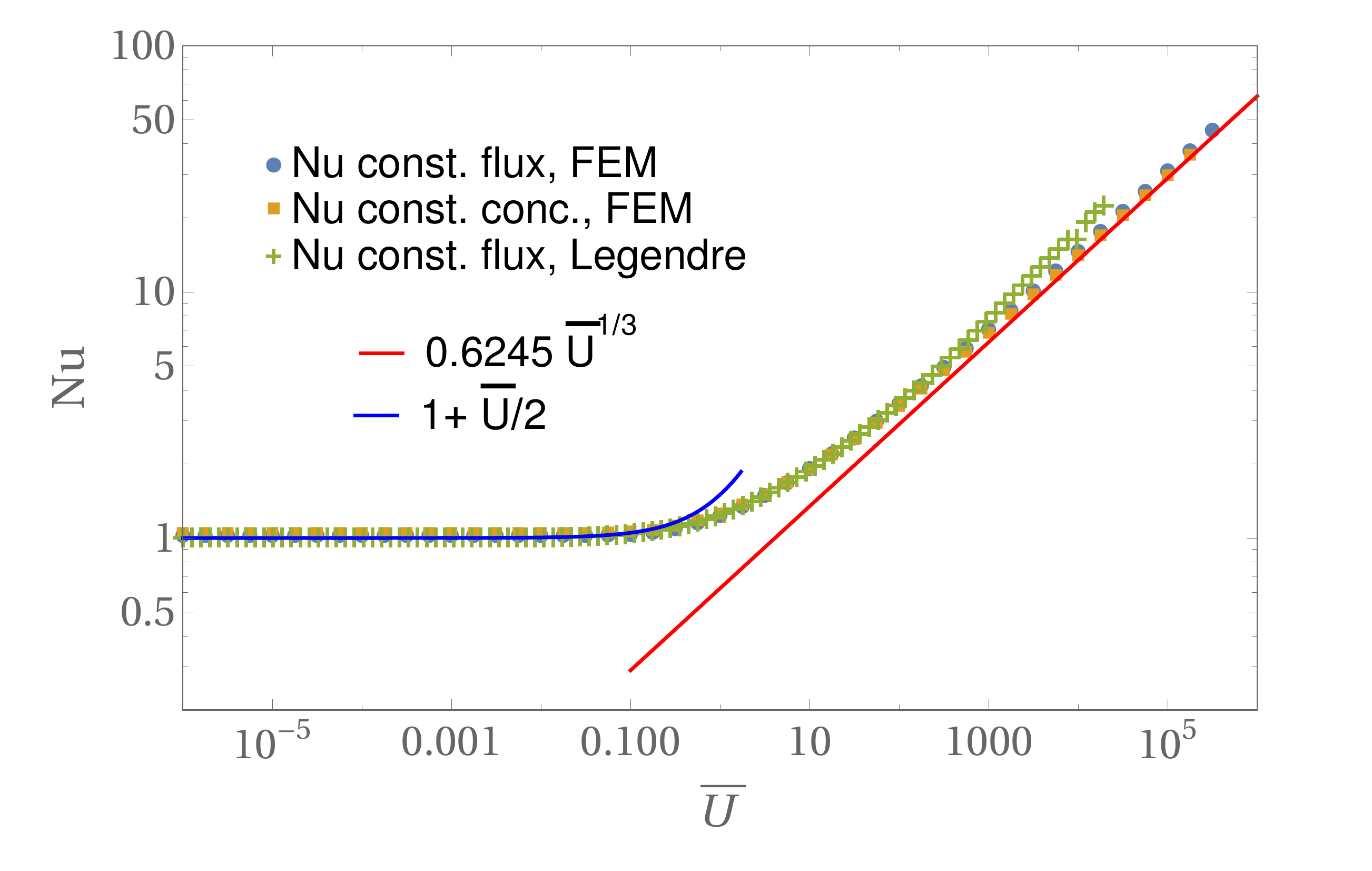}
   \caption{\label{fig:Nusselt}
     Average Nusselt number as a function of $\bar{U}$
     for constant flux and constant
     concentration boundary conditions.
     We compare results from numerical FEM solutions of the axisymmetric 
   diffusion-advection
   equation in two-dimensional
   angular representation with $\rho<\bar{R}=30$ and
   from numerical solutions in Legendre representation with Legendre
   components up to $n=61$ on a larger domain  $\rho<\bar{R}=300$.
 }
 \end{center}
 \end{figure}

We address the Nusselt number also 
for constant flux boundary conditions (A) and find
a very similar result (see Fig.\ \ref{fig:Nusselt})
\begin{align}
   {\rm Nu} = \frac{1}{\bar{c}_0(\rho=1)}&=
 \begin{cases}
    1+\frac{1}{2} \bar{U} & \mbox{for}~\bar{U}\ll 1\\
    0.65\, \bar{U}^{1/3}
    & \mbox{for}~\bar{U}\gg 1
  \end{cases},
      \label{eq:NuU}
\end{align}
where 
 the  prefactor $0.65$ is determined numerically from the
 data in Fig.\ \ref{fig:Nusselt}.
 This result will be derived below. As opposed to the
 case of a  constant concentration boundary condition, it is
 not possible to obtain an analytical result for the prefactor $0.65$. 
 Interestingly, the difference between both types of boundary
 conditions is small. We conclude  that the Nusselt number
 characterizes the mass transport mechanism by the advecting fluid 
 itself and  is rather robust with respect to the emission mechanism
 (diffusive emission, dissolution or production by a
 chemical reaction at the surface) by which
 the transported  molecules enter the advecting fluid.
 This is an important conclusion, which does not only apply
 to the microswimmer at hand,  but to laminar advective mass transport
 phenomena in general.

\begin{figure*}
  \begin{center}
   \includegraphics[width=0.75\linewidth]{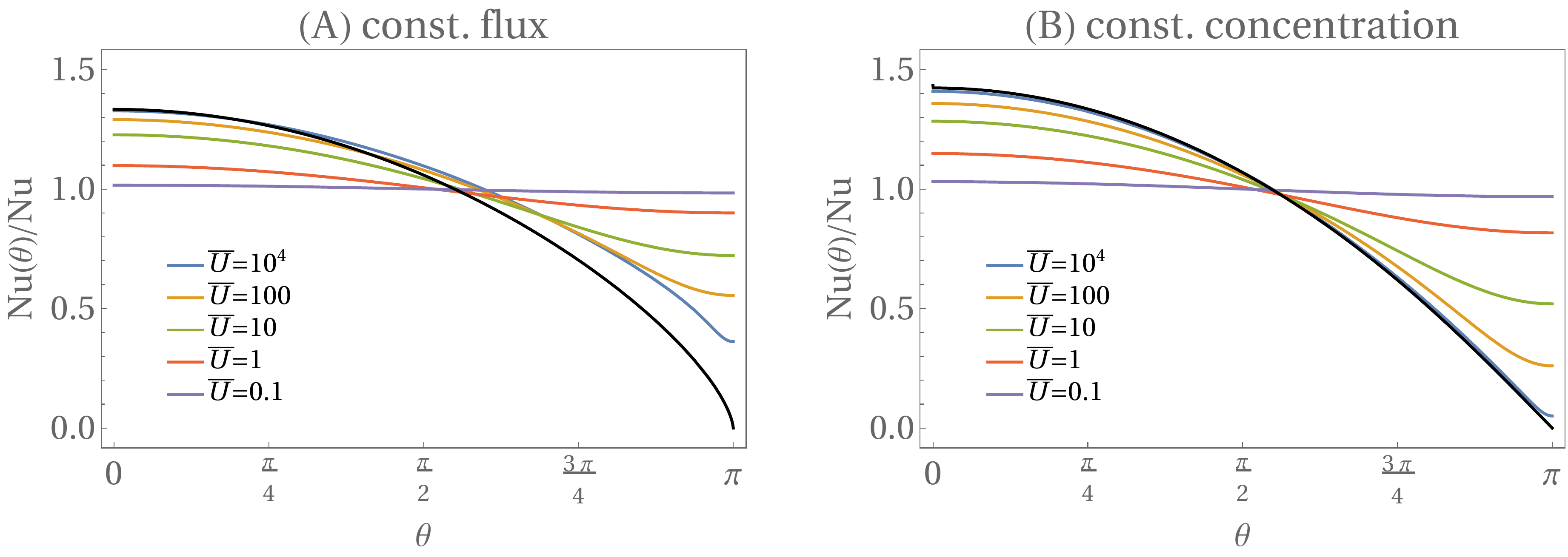}
   \caption{\label{fig:Nusseltlocal}
     Normalized local Nusselt number ${\rm Nu}(\theta)/{\rm Nu}$ (see text) 
     as a function of $\theta$
     for constant flux (left) and constant
     concentration (right) boundary conditions.
    Colored results are from numerical FEM solutions of the axisymmetric 
   diffusion-advection
   equation in two-dimensional
   angular representation with $\rho<\bar{R}=30$.
   Black lines are the exact analytical result (\ref{eq:Nuc}) and
    the approximate analytical result
   (\ref{eq:Nuflux}).
 }
 \end{center}
 \end{figure*}

\begin{figure*}
  \begin{center}
   \includegraphics[width=0.99\linewidth]{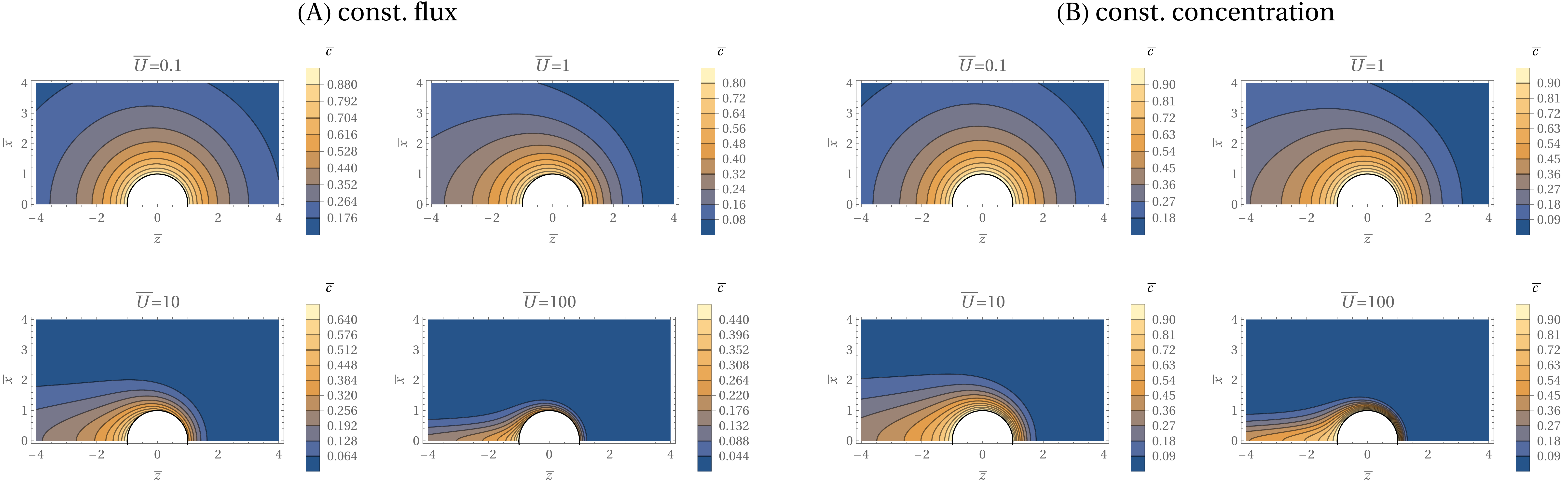}
   \caption{\label{fig:snaps}
     Concentration profiles  in the $\bar{z}\bar{x}$-plane
     for (A) constant flux and (B) constant concentration boundary
       conditions.
 }
 \end{center}
 \end{figure*}

 We can also define a local, i.e., angularly resolved
 Nusselt number via
 \begin{align}
 & {\rm Nu}(\theta) = \frac{-D\partial_r c(r=a,\theta)}{Dc(r=a,\theta)/a}
   = \frac{-\partial_\rho\bar{c}(\rho=1,\theta)}{\bar{c}(\rho=1,\theta)}
   \nonumber\\
  &  = -\partial_\rho (\ln \bar{c})(\rho=1,\theta),
                          \nonumber\\
 & \mbox{(A)~constant flux:}~~
  {\rm Nu}(\theta)  = \frac{1}{\bar{c}(\rho=1,\theta)},
                          \nonumber\\
&  \mbox{(B)~constant conc.:}~~
  {\rm Nu}(\theta)  ={-\partial_\rho\bar{c}(\rho=1,\theta)},
  \label{eq:Nulocal}
 \end{align}
 which is related to the average Nusselt number by 
 ${\rm Nu} = (\int_S{\rm Nu}(\theta)\,dA)/A_S$ for constant concentration
 boundary conditions (B) 
 and ${\rm Nu}^{-1} = (\int_S {\rm Nu}^{-1}(\theta)\,dA) /A_S$
 for constant flux boundary conditions (A).
 The local Nusselt number characterizes the symmetry breaking
 by advection; $ {\rm Nu}^{-1}(\theta)$
 gives the concentration profile   for constant flux (A), while
 ${\rm Nu}(\theta)$ gives the emission profile for constant
 concentration (B).
 Because ${\rm Nu}$ and ${\rm Nu}(\theta)$ are
 still $\bar{U}$-dependent (see  Eqs.\ (\ref{eq:NuU}) and  (\ref{eq:NuUB})),
 the angular dependence  in the Nusselt number profiles
 become more clear in the normalized local Nusselt number
 ${\rm Nu}(\theta)/{\rm Nu}$, which is  shown for both types
 of boundary conditions in Fig.\
\ref{fig:Nusseltlocal}. Again, the differences between constant flux
 (A) and constant concentration
 boundary conditions (B) are surprisingly
 small, at least for $\theta<\pi/2$.
 This becomes also evident by comparing the  snapshots of 
concentration profiles for constant concentration  (B) 
 and for constant flux boundary conditions (A) 
in Fig.\ \ref{fig:snaps}.

\subsection{Main results for Marangoni forces}

\begin{figure}
  \begin{center}
   \includegraphics[width=0.99\linewidth]{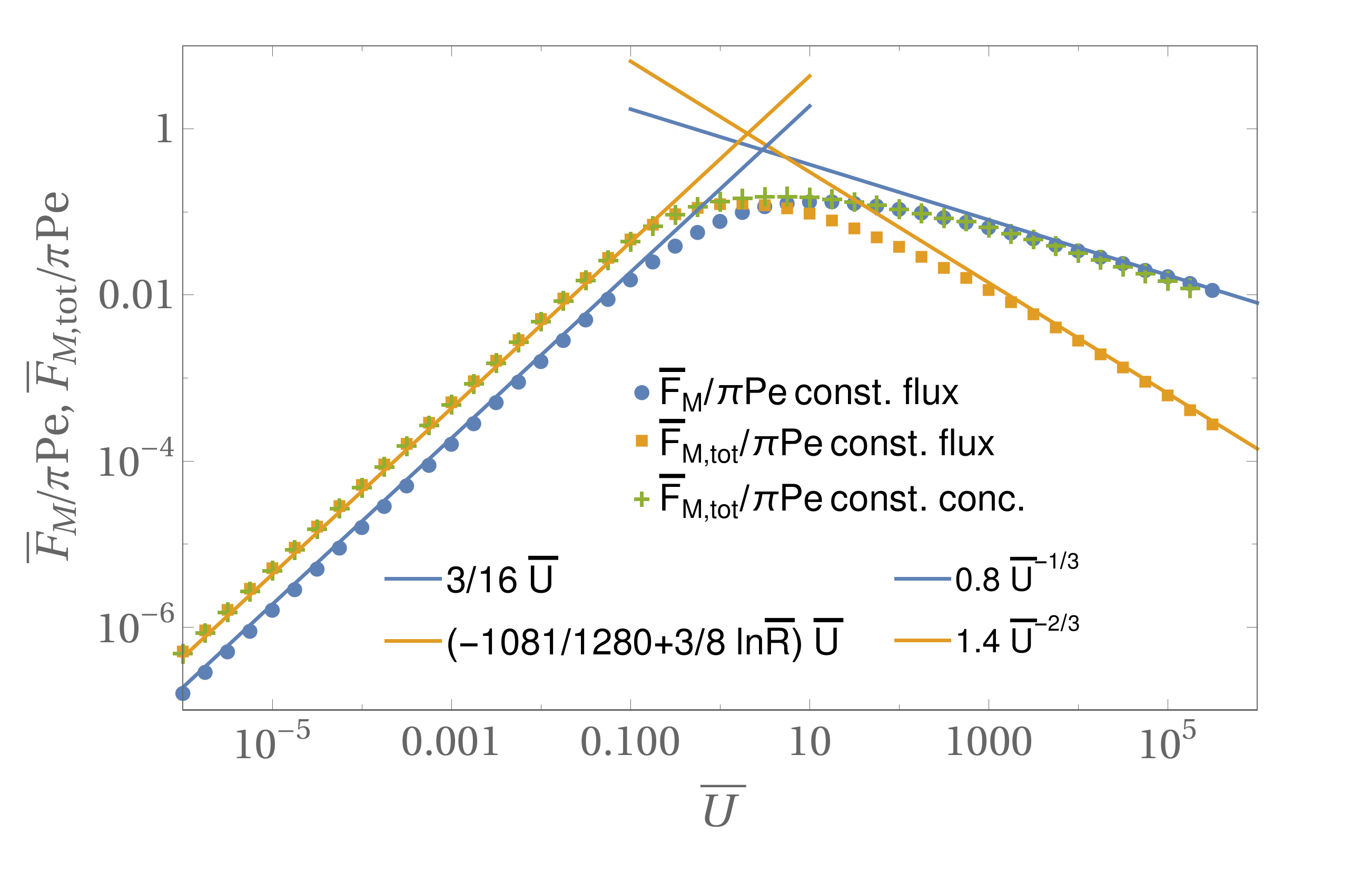}
   \caption{\label{fig:FM}
     Marangoni forces ${\bar{F}_{\rm M}}/{\pi \rm Pe}$ and
     ${\bar{F}_{\rm M, tot}}/{\pi \rm Pe}$ for constant flux boundary
     conditions and ${\bar{F}_{\rm M, tot}}/{\pi \rm Pe}$ for constant
     concentration boundary conditions  as a function of $\bar{U}$
     in the decoupled limit  ${\rm Pe} \ll \bar{U}$.
   All results are from numerical FEM solutions of the axisymmetric 
   diffusion-advection
   equation in two-dimensional
   angular representation with $\rho<\bar{R}=30$.
 }
 \end{center}
\end{figure}

For constant flux boundary conditions (A),  the main results for the 
 Marangoni forces as a function of a prescribed 
 velocity $\bar{U}$ are 
\begin{align}
 \frac{\bar{F}_{\rm M} }{\pi {\rm Pe}}  &=
\begin{cases}
   \frac{3}{16} \bar{U} & \mbox{for}~\bar{U}\ll 1\\
   d_{\rm M} \bar{U}^{-1/3}~\mbox{with}~d_{\rm M}\simeq 0.8
   & \mbox{for}~\bar{U}\gg 1
 \end{cases},
     \label{eq:c1Unew}
  \\
  \frac{\bar{F}_{\rm M, tot}}{\pi \rm Pe}
                    &=
       \begin{cases}
   -  \frac{1081}{1280}  \bar{U}
          +\frac{3}{8}  \bar{U}\ln \bar{R} & \mbox{for}~\bar{U}\ll 1\\
   d_{\rm M,fl} \bar{U}^{-2/3}~\mbox{with}~d_{\rm M,fl}\simeq 1.4
   & \mbox{for}~\bar{U}\gg 1
 \end{cases},
     \label{eq:c1Mflnew}
\end{align}
where numerical constants $d_{\rm M}$ and $d_{\rm M, tot}$ are
obtained from the numerical results, see Fig.\ \ref{fig:FM}.

For constant concentration boundary conditions (B), there is no direct
Marangoni force $\bar{F}_{\rm M}=0$ by definition
because there are no concentration and, thus, surface tension gradients
  along the  contact line $L$. 
Then, the total Marangoni force equals the Marangoni flow force
and is given by 
\begin{align}
  \frac{\bar{F}_{\rm M,tot}}{{\rm Pe}}
                    &=
       \begin{cases}
   -  \frac{563}{320}  \bar{U}
          +\frac{3}{8}  \bar{U}\ln \bar{R} & \mbox{for}~\bar{U}\ll 1\\
   d_{\rm M,B} \bar{U}^{-1/3}~\mbox{with}~d_{\rm M,B}\simeq 0.8
   & \mbox{for}~\bar{U}\gg 1
 \end{cases},
     \label{eq:c1MflnewB}
\end{align}
where the numerical constant $d_{\rm M,B}$ is
obtained from the numerical results.
Numerical results for these boundary conditions are also shown in Fig.\
\ref{fig:FM}.

The numerical result in Fig.\
  \ref{fig:FM}
  clearly confirms the existence of just two
  regimes for both types of boundary conditions.
At small $\bar{U}\ll 1$, the Marangoni forces are linear in $\bar{U}$
for both types of boundary conditions 
and can be calculated as linear response  in a perturbative approach.
In this limit, diffusion dominates.
For $\bar{U}\gg 1$, on the other hand, advection dominates, and a
concentration boundary layer forms around the half-sphere.
There is a markedly different
scaling for the total Marangoni force comparing both types
of boundary conditions, which we will explain below.
Figure \ref{fig:FM} shows that
direct and total Marangoni force reach maximal values 
$\bar{F}_{\rm M},\bar{F}_{\rm M, tot}\sim 0.15\,\pi{\rm Pe}$
in the crossover region 
 $\bar{U} \sim 1$ between diffusive and advective transport.

\subsection{Small velocity $\bar{U}$, perturbation theory}

At small $\bar{U}\ll 1$, there is a linear response of the
concentration field, which leads to a linear response
of the Nusselt number and Marangoni forces.
The coefficients can be calculated by perturbation theory
about the concentration field $\bar{c}^{(0)}(\vec{\rho}) = 1/\rho$ at
$\bar{U}=0$ in powers of $\bar{U}$.
A first approach is a naive perturbation series Ansatz
\begin{equation}
  \bar{c}_n(\rho) = \sum_{m=0}^\infty \bar{U}^m \bar{c}_n^{(m)}(\rho)
  \label{eq:Ansatz}
\end{equation}
for each Legendre coefficient starting with
$\bar{c}_0^{(0)}(\vec{\rho}) = 1/\rho$ and
$\bar{c}_{n>0}^{(0)}(\vec{\rho}) = 0$.
It turns out that this 
  will work only in the ``inner region'' $\rho < 1/\bar{U}$
  of a solution, because in the ``outer region'' 
  $\rho \gg 1/\bar{U}$, the  convection term can no longer
  be treated perturbatively, regardless how small $\bar{U}$ is
  \cite{Acrivos1962}.
   The problem that arises in performing such a naive expansion is that
  already $\bar{c}_1^{(1)}(\rho)$ does not vanish at infinity as required
  by the boundary conditions.
  What can be done, however, is to treat a finite
  system $\rho < \bar{R}$ and apply the boundary conditions
  $\bar{c}_0(\bar{R}) = 0$ and $\bar{c}_{n>0}(\bar{R})=0$
  as in the numerical approach.
  The above results (\ref{eq:c1Unew}), (\ref{eq:c1Mflnew})
  and (\ref{eq:c1MflnewB}) are obtained by this approach.
  We find excellent agreement between numerics and naive perturbation
  theory for such finite systems.

  In an infinite system, the situation differs
  because in the ``outer region'' 
  $\rho \gg 1/\bar{U}$ the  convection term can no longer
  be treated perturbatively  \cite{Acrivos1962}.
  These effects will only occur for system sizes $\bar{R}\gg 1/\bar{U}$,
  which become extremely large in the perturbative regime $\bar{U}\to 0$
  of interest.
  To address this problem, 
 in Ref.\ \cite{Acrivos1962},
  a systematic  expansion in inner and outer region and
  a matching procedure were performed for the 
  constant concentration  boundary condition  (B), which is posed in
  typical heat and mass transport problems in laminar flow
\cite{Acrivos1960,Acrivos1962,Leal}.
For the constant flux boundary condition (A), such 
calculations do  not exist at the moment.
We also adapt this  more advanced matching
procedure  to  the constant flux boundary condition (A).
 In Appendix \ref{app:perturbation}, we present the details of the
perturbative approach, both the naive perturbation theory and
the matching procedure. We find that in linear order in $\bar{U}$,
both approaches still agree in the inner region.
For the total Marangoni force, there is  a
 contribution $\propto \bar{U}\ln\bar{R}$
stemming from a $\rho$-integration of a  $\rho$-independent contribution
to $\bar{c}_1^{(1)}(\rho)$ in naive perturbation theory, see Eq.\
(\ref{eq:c1Mflnew}),  and (\ref{eq:c1MflnewB}).
In the framework of the matching procedure, this contribution
becomes $\propto -\bar{U}\ln\bar{U}$ as matching
provides an upper cutoff $\bar{R}\sim 1/\bar{U}$
to the otherwise unchanged inner region.

Regardless of whether this contribution is regularized by system
size $\bar{R}$ or by the boundary $\rho \sim 1/\bar{U}$ of the
inner region, the log-divergence of this
linear contribution  in  the total Marangoni force is a remarkable result
of these calculations.
Because  the linear term for the
direct Marangoni force stays finite,  its existence  means that the Marangoni
flow forces strongly \emph{increase} the direct force for $\bar{U}\ll 1$.

\subsection{Large velocity $\bar{U}$, concentration boundary layer}

\subsubsection{Scaling arguments}

  For large $\bar{U}\gg 1$, advection is strong and
a concentration boundary layer of  width $\Delta r$ develops around the
half-sphere.
The width $\Delta r$ is determined by the distance that
a surfactant molecule can diffuse during the time
$\Delta t\sim a/v(\Delta r/a)$ (see Eq.\ (\ref{v}))
that it takes to be transported along the
sphere by advection: $\Delta r^2 \sim D \Delta t$.
Because
$v(\Delta r/a) \sim  U \Delta r/a$
for $\Delta r/a \ll 1$ because of the
  no-slip boundary condition (see Eq.\ (\ref{v})), we find
\begin{equation}
  \Delta \rho = \Delta r/a  \sim  \bar{U}^{-1/3}.
  \label{eq:Deltarho}
\end{equation}
This is a classic result for the diffusion-advection problem
for constant concentration boundary conditions \cite{Acrivos1960,Leal},
but also holds for constant flux boundary conditions.

Because  the concentration will drop within the concentration
boundary layer
from its surface value to zero, we also have
$ -\partial_\rho \bar{c}(\rho=1,\theta)
\sim \bar{c}(\rho=1,\theta)/\Delta \rho $.
For constant flux boundary conditions (A) with $1=
-\partial_\rho \bar{c}(\rho=1)$, 
this  leads to a scaling
\begin{equation}
  {\rm Nu}^{-1}(\theta)=\frac{1}{\bar{c}(\rho=1,\theta)} \sim \Delta \rho \sim
  \bar{U}^{-1/3}~~\mbox{const.~flux~(A)} 
  \label{eq:cDeltarho}
\end{equation}
of the Nusselt number and the
symmetry-breaking concentration level at the sphere.
These scaling properties directly explain the results (\ref{eq:NuU}),
${\rm Nu}  \sim \bar{U}^{1/3}$,
for the Nusselt number and (\ref{eq:c1Unew}),
$ \bar{F}_{\rm M}/{\rm Pe}\sim  \bar{c}(\rho=1,\theta) \sim \bar{U}^{-1/3}$,
for the
direct Marangoni force in the limit $\bar{U}\gg 1$.

The result for the total Marangoni force (\ref{eq:c1Mflnew})
with constant flux boundary conditions 
 deviates from this scaling.
 Here, the expected boundary layer
 scaling is $ \bar{F}_{\rm M,tot}/{\rm Pe}\sim \Delta \rho^2 
 \bar{c}(\rho=1) \sim \bar{U}^{-1}$ (see Eq.\ (\ref{FMtotaldim}));
 this contribution is, however,
 only sub-dominant. The leading contribution comes from
 the advective tail in this limit of angular
 width $\Delta \theta \sim
 \bar{U}^{1/3}$,
  as follows from inspection of the 
stream function.
The (dimensionless) stream function for a sphere in Stokes flow is 
$\bar{\psi} = (\bar{U}/2) ( \rho^2 + 1/2\rho- 3\rho/2) \sin^2\theta$;
in the advection dominated regime $\bar{U}\gg 1$ fluid particles 
move along stream lines $\psi={\rm const}$, and 
the  fluid particles emerging from the boundary 
layer of width $\Delta \rho \sim \bar{U}^{-1/3}$
 around the sphere are transported into the advective tail
of angular width $\Delta \theta$ along a stream line. 
Therefore, $\Delta\theta$  follows by equating  the
respective scaling forms of the 
 stream function $\bar{\psi} \propto \rho^2\Delta \theta^2$ 
in the tail and $\bar{\psi }
\propto  3\Delta \rho^2\sin^2\theta/2$ in the boundary layer,
which gives $\rho \Delta\theta \sim \Delta \rho \sim \bar{U}^{-1/3}$. 
Therefore, the dominant contributions in Eq.\ (\ref{FMtotaldim})
are  $ \bar{F}_{\rm M,tot}\sim {\rm Pe}  \Delta \theta
\bar{c}(\rho=1,\theta) \sim \bar{U}^{-2/3}$
in agreement with the numerical results in Fig.\ \ref{fig:FM}.
 This also means that the Marangoni
 flow forces strongly \emph{decrease} the direct force
 (or effectively increase the drag) for $\bar{U}\gg 1$.

For constant concentration boundary conditions (B), the 
  drop of the concentration within the 
boundary layer
from its surface value $\bar{c}(\rho=1)=1$ to zero means that 
\begin{equation}
  {\rm Nu}(\theta)= -\partial_\rho \bar{c}(\rho=1,\theta)\sim
    \frac{1}{\Delta \rho} \sim
  \bar{U}^{1/3}~~\mbox{const.~conc.~(B)} 
  \label{eq:cDeltarhoB}
\end{equation}
Again,  these scaling properties directly explain the results (\ref{eq:NuU}),
${\rm Nu}  \sim \bar{U}^{1/3}$,
for the Nusselt number in the limit $\bar{U}\gg 1$.
The total Marangoni force  should scale 
$ \bar{F}_{\rm M,tot}/{\rm Pe}\sim \Delta \rho^2 
\bar{c}(\rho=1) \sim \bar{U}^{-2/3}$ from the boundary
layer contribution (see Eq.\ (\ref{FMtotaldim})),
which is again only subdominant.
As for constant flux boundary conditions,
the dominant contribution comes from the tail with
 $ \bar{F}_{\rm M,tot}/{\rm Pe}\sim  \Delta \theta
\bar{c}(\rho=1) \sim \bar{U}^{-1/3}$
which is in agreement with (\ref{eq:c1MflnewB}).

We also stress that, for both types of boundary conditions, we find
\begin{equation}
  {\rm Nu}(\theta) \sim \frac{1}{\Delta \rho(\theta)},
  \label{eq:NuDeltarho}
\end{equation}
i.e., the local Nusselt number can be interpreted as
the inverse local boundary layer width, which is also
evident from its  definition (\ref{eq:Nulocal}) as
an inverse decay length if the concentration profile
drops exponentially as a function of $\rho$.

\subsubsection{Rescaling  and similarity transformation}

More  stringent arguments are based on a corresponding scale
transformation of the entire diffusion-advection equation  (\ref{Dadvdim})
 in the decoupled limit $\vec{v}_{\rm M}\approx 0$.
Expecting a boundary layer of thickness $\Delta \rho \ll 1$, we can expand
(\ref{u}) and (\ref{v}) to obtain 
$v(\rho) \approx 3(\rho-1)/2$  and $u(\rho)\approx -3(\rho -1)^2/2$
to leading order.
Then we expand around the  surface of the sphere $\rho=1$ by 
introducing a rescaled distance 
$\xi \equiv (\rho-1) \bar{U}^m$. For $\bar{U}\gg 1$,
the leading  diffusion term is radial diffusion, which 
scales as $\bar{U}^{2m}$, while the  advection term scales as
$\bar{U}^{1-m}$. If advection and diffusion are both retained in the
boundary layer solution $m=1/3$ follows, which implies
a boundary layer $\rho-1 \sim \bar{U}^{-1/3}$ as in Eq.\ (\ref{eq:Deltarho}). 
If we also scale $\tilde{c} \equiv \bar{c} \bar{U}^{1/3}$,
the constant flux
 boundary condition (A) $\bar{U}^{1/3} \partial_\xi \bar{c}_0(0) = -1$
 becomes  $\bar{U}$-independent again, and 
 we end up with  $\bar{U}$-independent leading order equations
 in the rescaled variables $\xi$ and $\tilde{c}$,
 For constant concentration boundary conditions (B), no additional
 rescaling of $\bar{c}$ is necessary,  $\bar{c} = \tilde{c}$.

 We obtain in the rescaled variables for $\tilde{c}=\tilde{c}(\xi,\theta)$
\begin{align}
 \partial_\xi^2 \tilde{c}
  &= - \frac{3}{2} \xi^2  \cos\theta  \partial_\xi \tilde{c}
    + \frac{3}{2}\xi \sin\theta  \partial_\theta \tilde{c}
    \nonumber\\
  &= -\frac{1}{2} \xi^2 A'(\eta)  \partial_\xi \tilde{c}
    + \xi  A(\eta)  \partial_\eta \tilde{c}
    \nonumber\\
  & ~\mbox{with}~~~A(\eta) \equiv -\frac{3}{2} (1-\eta^2),
    ~~\eta \equiv \cos\theta
         \nonumber\\
  &\tilde{c}(\infty,\theta)  = 0, \nonumber\\
   &    \mbox{(A)~constant flux:}~~ \partial_\xi \tilde{c}(0,\theta) = -1,
     \nonumber\\
   &  \mbox{(B)~constant conc.:}~~
                \tilde{c}(0,\theta) =  1,
     \label{Dadvd}
\end{align}
i.e., a parameter-free equation confirming all  boundary layer scaling
results (\ref{eq:Deltarho}), (\ref{eq:cDeltarho}) and
(\ref{eq:cDeltarhoB}).

For the constant concentration boundary condition (B)
the equations (\ref{Dadvd})
can actually be solved analytically by a similarity transformation
\cite{Acrivos1960,Leal}, i.e., with an Ansatz
$\tilde{c}(\xi,\theta) = f(\xi g(\cos\theta))$
because this boundary
condition is compatible to a boundary condition $f(0)=1$
for the function $f(x)$.  Exact  results can be obtained
for the functions $g(\eta)$ and $f(x)$.
An immediate consequence of the existence of such a solution is that
the local Nusselt number is the inverse of the function $g(\cos\theta)$,
and that $g(\cos\theta)$ is 
identical  to the boundary layer width at angle $\theta$
because the function $f(\eta)$ is exponentially decaying on
a scale of order unity, 
\begin{equation}
  {\rm Nu}(\theta) = \frac{1}{g(\cos(\theta))} = \frac{1}{\Delta\rho (\theta)}.
  \label{eq:NuDeltarhoB}
\end{equation}
This confirms the scaling (\ref{eq:cDeltarhoB}) and (\ref{eq:NuDeltarho}).
The   exact results
for the functions $g(\eta)$ and $f(x)$ also give the  exact
asymptotics of 
the Nusselt number in Eq.\  (\ref{eq:NuU}),
${\rm Nu} = -\partial_\rho\bar{c}_0(\xi=0) \sim
c_0  \bar{U}^{1/3}$  with $c_0 = 3^{5/3}\pi^{2/3}/8\Gamma(1/3)
\simeq 0.624572$  \cite{Acrivos1962,Leal}.

A similarity transformation is, however, not possible
for the constant flux 
boundary conditions (A) $\partial_\xi \tilde{c}(0,\theta) = -1$,
which is  incompatible with the similarity Ansatz
$ \tilde{c}(\xi,\cos\theta) = f(\xi g(\cos\theta))$.
It turns out that we can reformulate the results for constant concentration
boundary conditions in terms of a flux balance argument, which can also
apply to the constant flux boundary conditions in order to obtain
an approximative result for the local Nusselt number.

\subsubsection{Flux balance argument for local Nusselt number}

Here, we consider the balance of the diffusive flux
out of the sphere at $\rho=1$ with the advective flux
assuming that a boundary layer $\Delta\rho\ll 1$ exists
to which the advective flux is constrained. 
We also assume that by its definition (\ref{eq:Nulocal}),
the local Nusselt number can be interpreted as
an inverse decay length, which is to be identified with the
boundary layer width
${\rm Nu}(\theta) \sim {1}/{\Delta \rho(\theta)}$, see
Eq.\ (\ref{eq:NuDeltarho}). 

For  the flux balance, we consider a volume
from $\theta=0$ up to an angle $\theta$ around the sphere $\rho=1$.
The diffusive outflux from the sphere gives the  particle influx
into this volume. For $\bar{U}\gg 1$, outflux from this volume
is dominated by advection in $\theta$-direction,
which is limited to the boundary layer
of thickness $\Delta \rho(\theta)={\rm Nu}^{-1}(\theta)$.
Both influx and outflux have to balance in a stationary state. 
In order to show the flux balance explicitly, we integrate on both sides
of equation (\ref{Dadvd}) (for the unrescaled $\bar{c}$ rather than
$\tilde{c}$).   The integrated diffusive term on the
left hand side gives the diffusive influx 
\begin{align*}
   I_{\rm in} &=   2\pi\int_{0}^{\infty} d\xi \int_\eta^1 d\tilde{\eta}
                \partial_\xi^2 \bar{c} 
  = - 2\pi \int_\eta^1 d\tilde{\eta}
    \partial_\xi \bar{c}(\xi=0,\eta)\\
  &= 2\pi  \int_0^\theta d\tilde{\theta}\sin\tilde{\theta}
                {\rm Nu}(\theta) \bar{c}(\xi=0,\theta).
\end{align*}
The integrated advective term on the right hand side gives
the advective outflux
\begin{align*}
  I_{\rm out}
  & =  \bar{U} 2\pi \int_{0}^\infty d\xi\bar{c}(\xi,\eta) \frac{3}{2} \xi
    (1-\eta^2)\\
  &\sim 2\pi \bar{U} \sin^2\theta \frac{3}{2}
               {\rm Nu}^{-2}(\theta) \bar{c}(\xi=0,\theta),
\end{align*}
where we used that outflux is confined to
a boundary layer of size $\Delta\rho (\theta) = {\rm
  Nu}^{-1}(\theta)$, see
Eq.\ (\ref{eq:NuDeltarho}), in the last equality. 
The integrated  Eq.\ (\ref{Dadvd})  thus transforms
into the flux balance $I_{\rm in}(\theta)=I_{\rm out}(\theta)$.

For constant concentration boundary conditions (B), we obtain
after differentiating with respect to $\theta$
\begin{align*}
  \frac{1}{3{\rm const}\bar{U}}
  &=
    \cos\theta \frac{1}{{\rm Nu}^3(\theta)}
    -\sin\theta  \frac{{\rm Nu}'(\theta)}{{\rm Nu}^4(\theta)}.
\end{align*}
Apart from the undetermined constant, this is exactly the differential
equation governing the scaling function $g(\cos\theta)$ in the similarity
solution \cite{Leal}, which confirms ${\rm Nu}(\theta) = g(\cos\theta)$.
The differential equation can be solved to give the well-known exact result
\cite{Leal}
\begin{align}
{\rm Nu}(\theta) &= \left(2{\rm const}\bar{U}\right)^{1/3}
             \frac{\sin\theta}{(\theta-\frac{1}{2} \sin(2\theta))^{1/3}}.
\label{eq:Nuc}
 \end{align}
For constant flux  boundary conditions (A),
$-\partial_\xi \bar{c}(\xi=0,\theta)=1$, we have
${\rm Nu}(\theta) = 1/\bar{c}(\xi=0,\theta)$ and 
flux balance gives 
\begin{align*}
  \int_0^\theta d\tilde{\theta}\sin\tilde{\theta}
  &={\rm const}  \bar{U} \sin^2\theta \frac{3}{2}
    {\rm Nu}^{-3}(\theta).
\end{align*}
This can be directly integrated to give
a new approximative result for the angular dependence of the
  Nusselt number,
\begin{align}
  {\rm Nu}(\theta) &=
                {\rm const}   \bar{U}^{1/3}
        \left( 1+\cos\theta \right)^{1/3}.   
                     \label{eq:Nuflux}
\end{align}
The normalized local Nusselt numbers ${\rm Nu}(\theta)/{\rm Nu}$
are plotted as black lines in Fig.\ \ref{fig:Nusseltlocal}. 
The agreement for large $\bar{U}$ is excellent for
constant concentration boundary conditions (B) and approximate
for  constant flux boundary conditions (A), as expected. 
The flux balance approach also
confirms the scaling ${\rm Nu}(\theta) \sim \bar{U}^{1/3}$,
see Eqs.\ (\ref{eq:cDeltarho}) and (\ref{eq:cDeltarhoB}).

\subsection{Anisotropic emission}

\begin{figure}
  \begin{center}
   \includegraphics[width=0.99\linewidth]{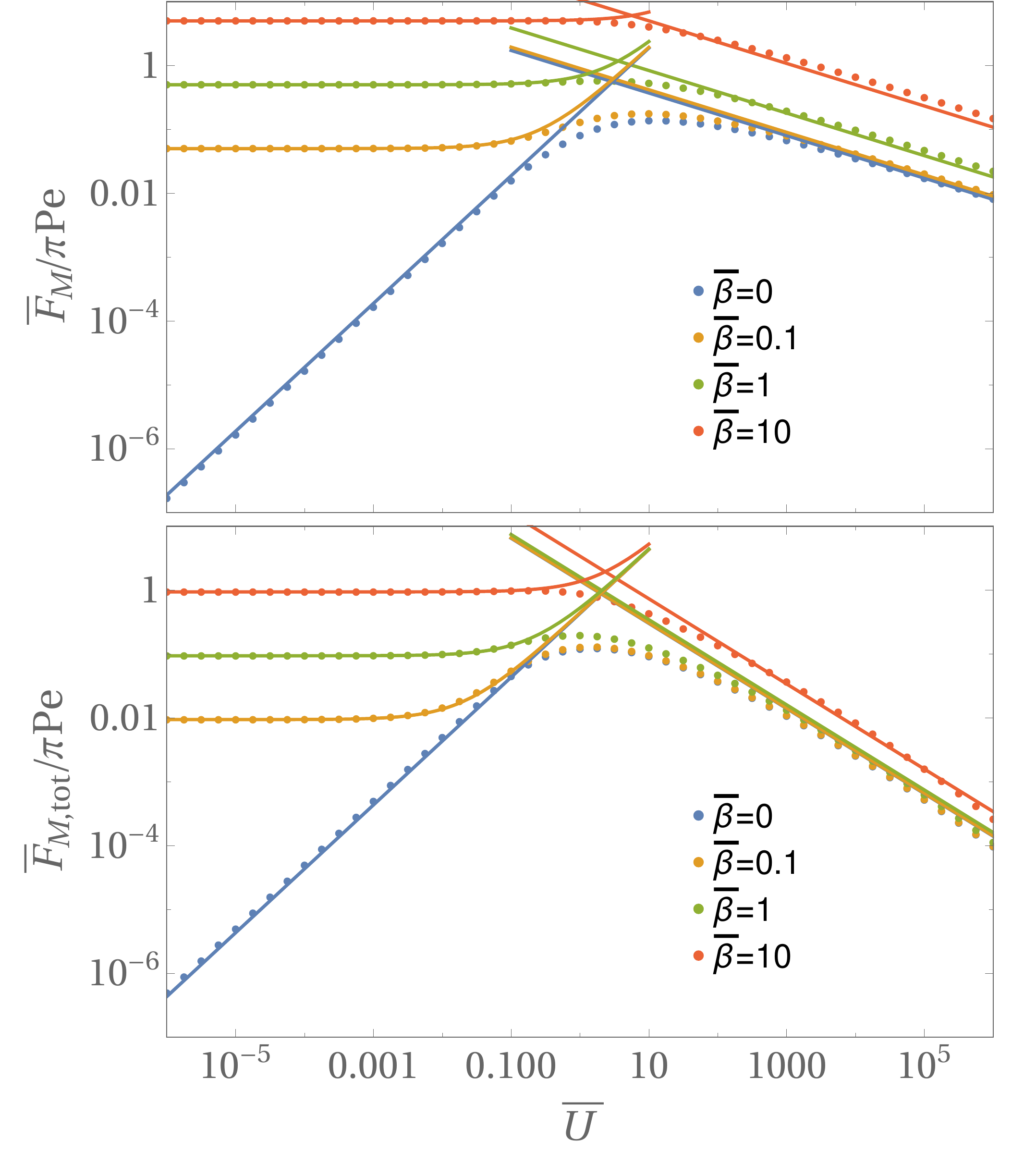}
   \caption{\label{fig:FMan}
     Marangoni forces ${\bar{F}_{\rm M}}/{\pi \rm Pe}$ and
     ${\bar{F}_{\rm M, tot}}/{\pi \rm Pe}$ for constant flux boundary
     conditions   as a function of $\bar{U}$
     in the presence of an anisotropy $\bar{\beta}$ in the emission.
   All results are from numerical FEM solutions of the axisymmetric 
   diffusion-advection
   equation in two-dimensional
   angular representation with $\rho<\bar{R}=30$.
   The solid lines for $\bar{U}<1$ are the analytical
   perturbative results (\ref{eq:c1Ubetanew})
   and (\ref{eq:c1Mflbetanew}).
   The solid lines for $\bar{U}>1$ are the scaling results
   (\ref{eq:c1Ubetanew2}) and (\ref{eq:c1Mflbetanew2}).
   For $\bar{\beta}=0$, we recover the results from Fig.\ \ref{fig:FM}. 
 }
 \end{center}
\end{figure}

Finally, we want to discuss the effect of an anisotropic
emission boundary condition using the example of an
anisotropic diffusive flux as characterized by a parameter
$\bar{\beta}>0$ in Eq.\ (\ref{eq:beta}).
In general, we expect higher Marangoni forces, because these
forces are caused by anisotropies in the concentration profile
around the half-sphere. If anisotropies are present 
without the need to create them by advection, this increases
the Marangoni forces as can also be seen in 
the numerical results in Fig.\ \ref{fig:FMan}.
These numerical results also show that
increasing the  anisotropic emission parameter $\bar{\beta}$
beyond $\bar{\beta}\sim 1$ erases the maximum in the
Marangoni forces in the crossover region 
 $\bar{U} \sim 1$ between diffusive and advective transport. 

In the diffusive limit $\bar{U}\ll 1$, the anisotropy
leads to an additional zeroth order term
$\bar{c}_1^{(0)}(\rho)=  -\bar{\beta}/{2\rho^2}$
in the concentration field, which results in
\begin{align}
  \frac{\bar{F}_{\rm M} }{\pi {\rm Pe}}
  &\approx 
               \frac{3}{16} \bar{U}  + \frac{1}{2}\bar{\beta},
                                           \label{eq:c1Ubetanew} \\
    \frac{\bar{F}_{\rm M, tot}}{\pi \rm Pe}
      &\approx  -  \frac{1081}{1280}  \bar{U}
        +\frac{3}{8}  \bar{U}\ln \bar{R} + \frac{3}{32} \bar{\beta},
       \label{eq:c1Mflbetanew}  
\end{align}
as derived in  Appendix  \ref{app:perturbation} (see Eqs.\
(\ref{eq:FMlinear}) and  (\ref{eq:FMtotlinear})).
These perturbative results are in excellent agreement with numerical
FEM results as can be seen in Fig.\ \ref{fig:FMan}.
For sufficiently small $\bar{U}$ the zeroth order term dominates.
If this term dominates,  Marangoni
flow forces \emph{decrease} the direct force because
$3 \bar{\beta}/32<\bar{\beta}/2$; this  is
similar to  the results  of Ref.\ \cite{Lauga2012}, where also
an explicitly asymmetric situation was considered.

In the  advective limit $\bar{U}\gg 1$, a boundary
layer of width $\Delta \rho \sim \bar{U}^{-1/3}$ determines the physics.
On the scale of the boundary layer thickness, the
concentration drops   from its surface value $\bar{c}(\rho,\theta)$
to zero. 
For constant flux boundary conditions (A), this led to
a concentration level 
$\bar{c}(\rho,\theta)\sim \Delta \rho\sim \bar{U}^{-1/3}$
(see Eq.\ (\ref{eq:cDeltarho})) at the sphere. 
In the presence of an explicitly  symmetry-breaking
emission $\partial_\rho\bar{c}_1(\rho=1) = \bar{\beta}$, this
contribution will also decay on the scale of the boundary layer
$\Delta \rho$, and  we expect a corresponding
contribution 
$\bar{\beta} \bar{U}^{-1/3}$ to the  concentration level  at the sphere,
$\bar{c}(\rho=1,\theta) \sim  ({\rm const}+\bar{\beta})     \bar{U}^{-1/3}$.
Because the direct Marangoni force scales as
$\bar{F}_{\rm M}/{\rm Pe}\sim  \bar{c}(\rho=1,\theta)$,
this leads to 
\begin{align}
  \frac{\bar{F}_{\rm M} }{\pi {\rm Pe}}
  &\approx 
           (d_{\rm M}+\bar{\beta})     \bar{U}^{-1/3},
                  \label{eq:c1Ubetanew2} 
\end{align}
which is in good agreement with numerical results as shown
in Fig.\ \ref{fig:FMan}.
The total Marangoni force scaling is dominated by the advective
tail, which led to $\bar{F}_{\rm M,tot}\sim {\rm Pe}  \Delta \theta
\bar{c}(\rho=1,\theta)$; we find
\begin{equation}
  \frac{\bar{F}_{\rm M,tot} }{\pi {\rm Pe}}
  \approx 
    d_{\rm M,\beta} \left(\frac{d_{\rm M,fl}}{d_{\rm M,\beta}} +\bar{\beta}\right)
    \bar{U}^{-2/3}~\mbox{with}~d_{\rm M,\beta}\simeq 0.2.
         \label{eq:c1Mflbetanew2} 
\end{equation}
This result is also in good agreement with numerical results  as shown
in Fig.\ \ref{fig:FMan}.

\section{Diffusion-advection with strong
  Marangoni flow ${\rm Pe} \gg \bar{U}$}

For a strong Marangoni flow, ${\rm Pe} \gg \bar{U}$, the linear
response regime $\bar{U}\ll 1$ becomes modified.
We first have to address the dominant Marangoni flow problem (iib),
which determines the Marangoni flow $\vec{v}_{\rm M}$.
For  ${\rm Pe} \gg \bar{U}$, this is the dominant
contribution to the fluid flow in the diffusion-advection problem (iii). 
The Marangoni  flow pattern is a  stationary
 Marangoni vortex ring 
 around the spherical swimmer  below and parallel to
 the fluid interface $S_{\rm Int}$ as can be seen in Fig.\ \ref{fig:flows}.
 Because this solution lacks axisymmetry a
 complete and analytical solution is no longer
 possible.

   Applying  mass conservation $\bar{J} \sim
   2\pi \bar{c} \bar{v}_{\rm M}\rho  \bar{l}_c = {\rm const}$
   and the Marangoni boundary condition
 to  concentration profile and Marangoni flow field
 in a concentration boundary layer of width
 $\bar{l}_c \sim  (\rho/\bar{v}_{\rm M})^{1/2}$ below
 the fluid interface $S_{\rm Int}$, we find a scaling \cite{Ender2020}
 \begin{align}
  \bar{c}(\rho) &= \bar{c}(1) \rho^{-2/3}
  ~~\mbox{with}~~ \bar{c}(1) \sim {\rm Pe}^{-1/3},
  \label{eq:cMarangoni}\\
  \bar{v}_{\rm M} &\sim \bar{c}^{-2} \rho^{-3}
  \sim \bar{c}^{-2}(1) \rho^{-5/3} \sim {\rm Pe}^{2/3} \rho^{-5/3}.
  \label{eq:vMarangoni}
 \end{align}
for strong Marangoni flows. 
Here, we will further test this result  in numerical FEM solutions,

We see that the advective current
$\bar{j} \sim \bar{c} \bar{v}_{\rm M}\sim {\rm Pe}^{1/3} \rho^{-7/3}$
becomes smaller than the corresponding diffusive current
$\bar{j}_D \sim -\partial_\rho \bar{c} \sim {\rm Pe}^{-1/3} \rho^{-5/3}$
for $\rho > {\rm Pe}$. Then our assumption of advective transport
breaks down, and this should mark the boundary of the
Marangoni advection dominated region. Therefore
\begin{equation}
  \rho_{\rm M} \sim  {\rm Pe}
  \label{eq:rhoM}
\end{equation}
should be the scaling of the size of the Marangoni vortex
around the sphere for low Reynolds numbers.
At larger distances, a crossover to diffusive transport
with  $\bar{c} \propto \rho^{-1}$  sets in.

We can also introduce the dimensionless Marangoni number
for the radial Marangoni flow, which  exactly compares
advective Marangoni current and diffusive current by definition, 
\begin{equation}
  {\rm Ma} = \frac{j_{\rm M}}{j} = \frac{v_{\rm M} r}{D} 
  = \bar{v}_{\rm M} \rho =  {\rm Pe}^{2/3} \rho^{-2/3},
  \label{eq:Ma}
\end{equation}
and see that $\rho_{\rm M}$ is determined by the condition that
the regime ${\rm Ma} >1$ determines the size of the Marangoni vortex.

\begin{figure}
  \begin{center}
     \includegraphics[width=0.99\linewidth]{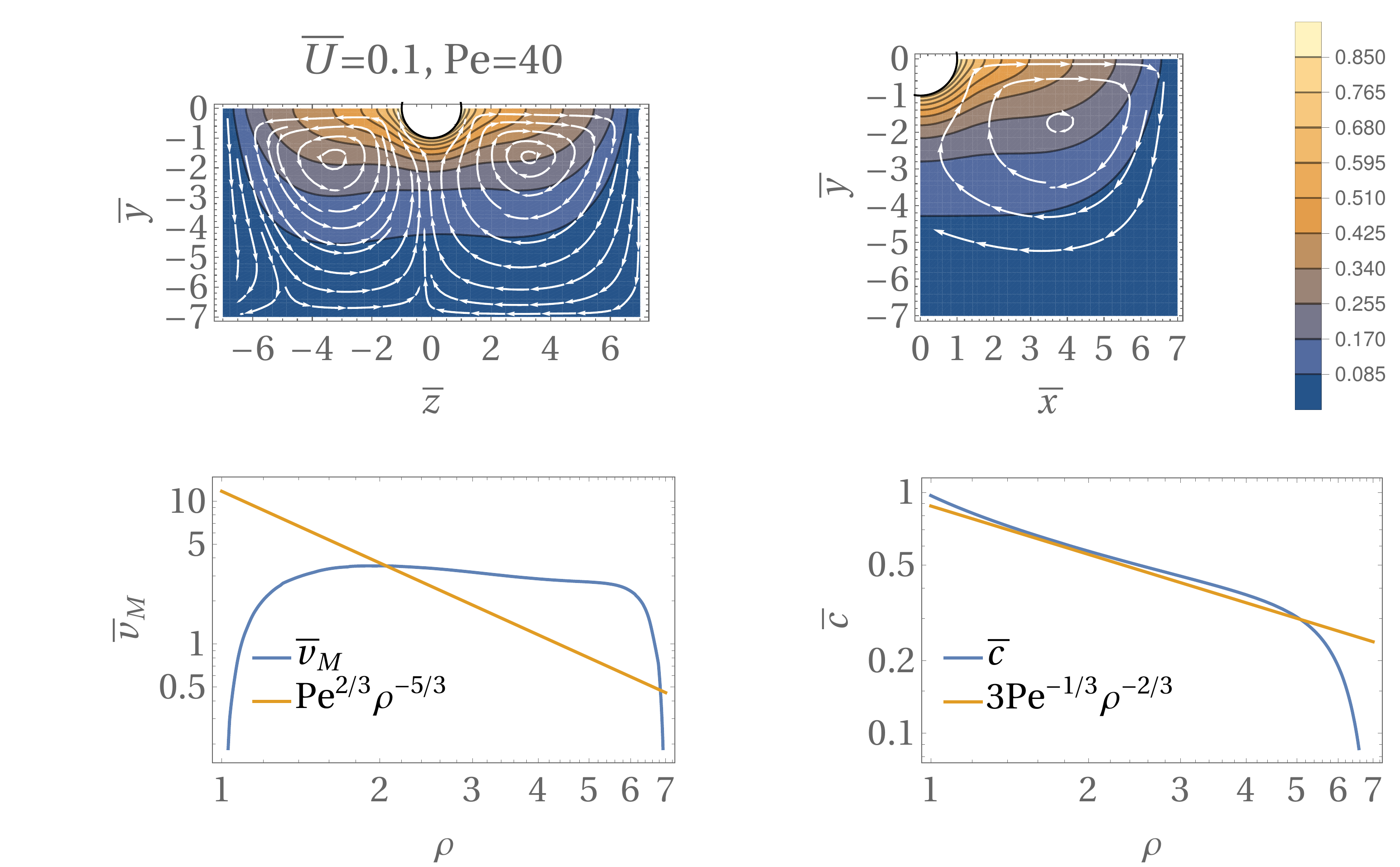}
    \caption{\label{fig:vMc}
      Iterative three-dimensional  FEM results for ${\rm Pe}=40$
      and $\bar{U}=0.1$
      in  a  cubic system with  $-7 < \bar{y} <0$,
    $0<\bar{x}<7$, $-7<\bar{z}<7$.
    Top:  Contour plots of the concentration $\bar{c}(\vec{\rho})$
   and the stream lines of the Marangoni flow field
   $\vec{v}_{\rm M}(\vec{\rho})/{\rm Pe}$.
   Bottom: $\vec{v}_{\rm M}$ as a function of $\rho$ and $\bar{c}$ as a function
   of $\rho$ at the interface $\bar{y}=0$ along with the
    predictions (\ref{eq:vMarangoni}) and (\ref{eq:cMarangoni}).
  }
\end{center}
\end{figure}

We can test the predictions (\ref{eq:cMarangoni}) and (\ref{eq:vMarangoni})
  in numerical FEM solutions, see Fig.\ \ref{fig:vMc}.
  One problem is that, for large Peclet numbers,
  the finite size of the numerical system becomes too small
  to accommodate the Marangoni vortex of size $\rho_{\rm M} \sim  {\rm Pe}$
  properly. This results in deviations of the interfacial 
  Marangoni flow field from Eq.\ (\ref{eq:vMarangoni}).
  The numerical results  for the interfacial concentration
   field show excellent agreement with (\ref{eq:cMarangoni}).

\begin{figure}
  \begin{center}
    \includegraphics[width=0.99\linewidth]{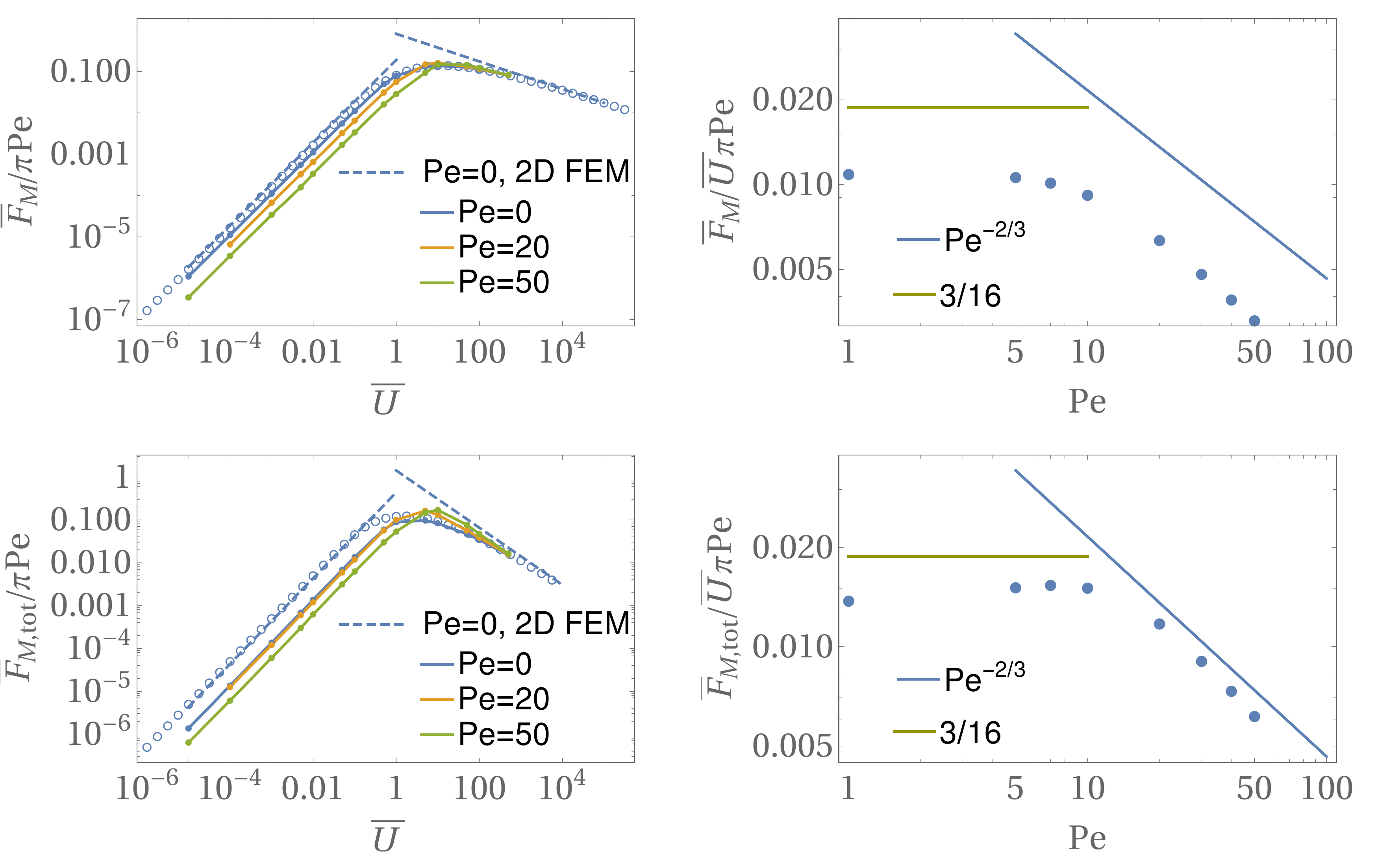}
    \caption{\label{fig:FMtotalPe}
  Left: Iterative three-dimensional  FEM results for
      ${ \bar{F}_{\rm M}}/\pi{\rm Pe}$ (top) and ${\bar{F}_{\rm
        M,tot}}/\pi {\rm Pe}$ (bottom) as a function of $\bar{U}$
    for ${\rm Pe} = 0-50$
    for a  cubic system with  $-7 < \bar{y} <0$,
    $0<\bar{x}<7$, $-7<\bar{z}<7$.
    Blue open circles are results for ${\rm Pe}=0$
    from FEM solutions to the axisymmetric diffusion-advection
    equation in two-dimensional angular
    representation with $\bar{R}=30$.
    The slope in the linear response regime for $\bar{U}\ll 1$
    is reduced according to Eq.\ 
     (\ref{eq:FMtotalMarangoni}). Results for
    $\bar{U}\gg 1$ are essentially not
    affected by strong Marangoni flows ${\rm Pe}\gg {\bar U}$.
 Right: Corresponding  slopes
      ${ \bar{F}_{\rm M}}/\bar{U}\pi{\rm Pe}$ and ${\bar{F}_{\rm
        M,tot}}/\bar{U}\pi {\rm Pe}$  as a function of ${\rm Pe}$
    calculated from the results for  $\bar{U}=0.1$.
  }
\end{center}
\end{figure}

So far, we considered the leading order of our problem by setting
$\bar{U}\approx 0$; going one order further, 
we get the linear response for small $\bar{U}$
with the ansatz $\bar{c}
=  \bar{c}^{(0)}   + \bar{U} \bar{c}^{(1)}$
with $\bar{c}^{(0)}(\rho)$ given by (\ref{eq:cMarangoni}).
In the total flow $\vec{v} + \vec{v}_{\rm M}$, the
Marangoni flow (\ref{eq:vMarangoni}) is the zeroth-order result,
$\vec{v}_{\rm M}= \vec{v}_{\rm M}^{(0)}$, 
while the Stokes swimming flow $\vec{v}= \vec{v}^{(1)}$
is linear in $\bar{U}$.
In an advection dominated situation, mass conservation
in the boundary layer still holds in the
presence of Stokes flow,
\begin{equation*}
  1 \sim  (\bar{c}^{(0)} + \bar{U} \bar{c}^{(1)})
  (\bar{U}\bar{u}\cos\theta + \bar{v}_{\rm M})^{1/2}\rho^{3/2},
\end{equation*}
 where the radial component $\bar{u}$ of the Stokes flow
 is considered. 
Expanding up to first order in $\bar{U}$ we find a scaling 
 \begin{equation*}
  \bar{c}^{(1)}(\rho) \sim   \frac{1}{ \bar{v}_{\rm M}^{1/2}(\rho)}
                        \bar{c}^{(0)}(\rho)  \bar{u}(\rho)
                        \sim {\rm Pe}^{-2/3}  \rho^{1/6} \bar{u}(\rho),
 \end{equation*}
 which will give rise to a 
  Marangoni force scaling 
 \begin{align}
   \frac{ \bar{F}_{\rm M}}{\pi{\rm Pe}}
   &\sim \bar{U}  {\rm Pe}^{-2/3},&
   \frac{ \bar{F}_{\rm M,tot}}{\pi {\rm Pe}}
   &\sim \bar{U}  {\rm Pe}^{-2/3}.
   \label{eq:FMtotalMarangoni}
\end{align}
Numerical FEM results show that both prefactors are of order unity
 (but hard to quantify because of finite size effects), see
  Fig.\ \ref{fig:FMtotalPe}.
This shows that Marangoni flows depress the total driving force
 in the linear response regime by a factor ${\rm Pe}^{-2/3}$
 because 
 it is harder to break
 the symmetry in the presence of the strong Marangoni flow advection.
 Numerical results in Fig.\ \ref{fig:FMtotalPe}
 also show that the total Marangoni force is somewhat larger
 than the direct Marangoni force,
 $ \bar{F}_{\rm M,tot} >  \bar{F}_{\rm M}$.
 In this respect,
 our previous results for linear response regime for ${\rm Pe}\ll \bar{U}$
 remain unchanged:
 The Marangoni flow force {\em increases} the direct force.

In the advection-dominated regime 
    $\bar{U}\gg 1$, on the other hand, results  are essentially not
    affected by strong Marangoni flows ${\rm Pe}\gg {\bar U}$
    as the numerical results in Fig.\ \ref{fig:FMtotalPe} show.
     The flow field $\vec{v}$ will still give
 rise to a concentration boundary layer of thickness
 $\Delta \rho \sim \bar{U}^{-1/3}$
 around the sphere. On the scale of the boundary layer, the Marangoni
 flows $\vec{v}_{\rm M}$ are not yet developed; they develop only 
 further away
 at $1\ll \rho < \rho_{\rm M} \sim  {\rm Pe}$ because of the no-slip
 boundary condition for the Marangoni flow in (iib). 
 Therefore, the results for $\bar{U}\gg 1$ are essentially 
 unaffected by a strong Marangoni flow for ${\rm Pe}\gg {\bar U}$.

\section{Diffusion-advection in the presence of evaporation}

 In the presence of evaporation, we have
 a convective (Robin)  boundary condition (\ref{eq:bcevap}),
 which is governed by the dimensionless Biot number (\ref{eq:Biot}),
 instead of the Neumann condition (\ref{eq:bcnoevap}),
 which is recovered for vanishing Biot number $\bar{k}=0$.
 In general, evaporation of surfactant
 depletes the interface of surfactant and, thus, decreases
 the Marangoni driving forces (both direct and flow forces).
 For volatile camphor,  we find a Biot number
 $\bar{k} = ak/D \approx 550$ using results from Ref.\ \cite{Soh2008},
 whereas other surfactants such as PEG are non-volatile and have a very
 small Biot number \cite{Ender2020}. 

 The Biot number can also be interpreted as an extrapolation
 length scale.
The concentration profile will fall off exponentially
perpendicular to the interface in the outward direction on a
dimensionless extrapolation length scale $\Delta \bar{y}\sim 1/\bar{k}$
given by the inverse of the Biot number.

In Ref.\ \cite{Ender2020},  we  developed a qualitative scaling theory
 based on the assumption that
 the total evaporation flux balances the total emission flux
 of surfactant in a stationary state.
 In the diffusive regime  $\bar{U}\ll 1 $, this leads to 
\begin{align}
  \bar{F}_{\rm M}  &\sim
 \left.  \bar{F}_{\rm M}\right|_{\bar{k}=0} \frac{1}{\bar{k}+1},
&   \bar{F}_{\rm M,tot}
    &\sim  \left. \bar{F}_{\rm M,tot}\right|_{\bar{k}=0}
      \frac{1}{\bar{k}+1}.
      \label{eq:FMtotalkUsmall}  
 \end{align}
 In  the advection-dominated limit $\bar{U}\gg 1$, we find
 \begin{align}
  \bar{F}_{\rm M}  &\sim
                     \left.  \bar{F}_{\rm M}\right|_{\bar{k}=0}
                     \frac{\bar{U}^{1/3}}{\bar{k}+\bar{U}^{1/3}},
& \bar{F}_{\rm M,tot}
    &\sim  \left. \bar{F}_{\rm M,tot}\right|_{\bar{k}=0}
      \frac{\bar{U}^{1/3}}{\bar{k}+\bar{U}^{1/3}}.
      \label{eq:FMtotalkU}
 \end{align}
 In both limits, Marangoni forces are reduced by evaporation,
 because it reduces the surfactant concentration.

\section{Swimming condition, symmetry breaking, and speed}

Now, we have a rather
 complete picture of the solution of
problems (i)-(iii), i.e., diffusion-advection coupled to
hydrodynamics 
for a prescribed swimmer velocity $\bar{U}$ at low Reynolds numbers.
In particular, we know the  Marangoni forces as a function of the
prescribed velocity  $\bar{U}$.

\subsection{Swimming condition}

The swimming condition  (\ref{eq:swimcond})
gives an additional force balance
relation between Marangoni forces and $\bar{U}$,
which has to be satisfied in the swimming state and determines
the selected swimming speed $\bar{U}=\bar{U}_{\rm swim}$ as a function
of Peclet number ${\rm Pe}$  and Biot number $\bar{k}$. 
In general, the swimming velocity increases with ${\rm Pe}$ and
decreases with $\bar{k}$.

The force balance condition can be interpreted such that
intersections of the linear Stokes friction relation
$-\bar{F}_{\rm D}= 3\pi  \bar{U}$
and the total Marangoni force 
$\bar{F}_{\rm M, tot} = \bar{F}_{\rm M, tot}(\bar{U})$ relation 
give the swimming speed  $\bar{U}=\bar{U}_{\rm swim}$.
The resulting swimming state can only be stable  if the Marangoni force
curve  $\bar{F}_{\rm M, tot}(\bar{U})$  intersects the
straight Stokes friction line $3\pi  \bar{U}$ from \emph{above}.
Then, a speed fluctuation $\delta \bar{U}>0$ will
give rise to $\bar{F}_{\rm M, tot}<-\bar{F}_{\rm D}$ such that
friction dominates, and the swimming speed is decreased again.

All curves  $(\bar{F}_{\rm M, tot}/\pi{\rm Pe})(\bar{U})$ in Figs.\
\ref{fig:FM},  \ref{fig:FMan}  and \ref{fig:FMtotalPe}
start linearly $\propto \bar{U}$
in the diffusive regime $\bar{U}\ll 1$ and then cross over
to sublinear growth and finally decrease in the advective regime $\bar{U}>1$.
Therefore, all intersection points with the linear Stokes friction function
will represent \emph{stable} swimming states, also
if an anisotropic emission is included. These results remain
unchanged if  evaporation is included.

\begin{figure}
  \centerline{\includegraphics[width=0.99\linewidth]{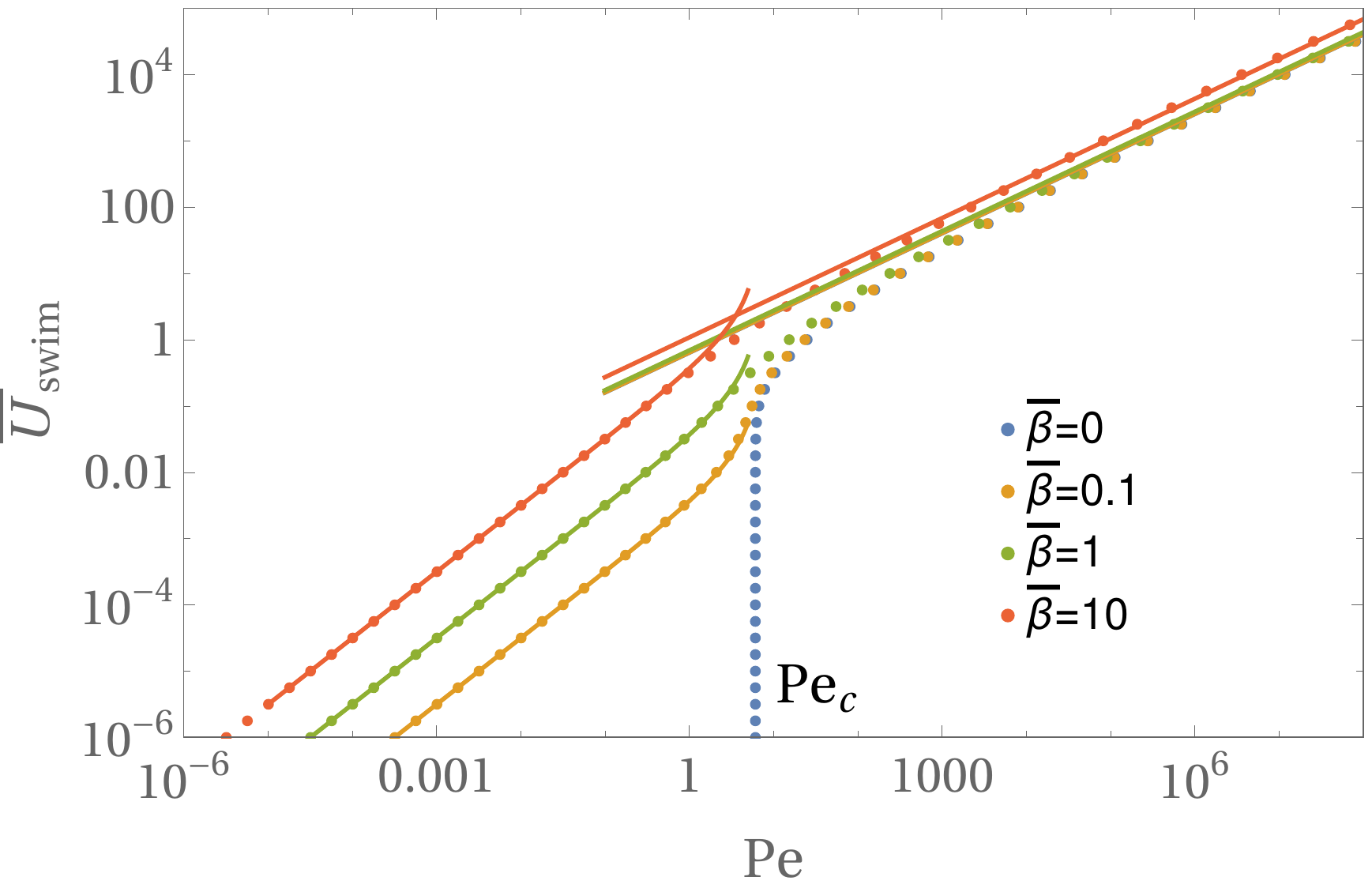}}
  \caption{Swimming speed $\bar{U}_{\rm swim}$ as a function of the
    Peclet number ${\rm Pe}$ (representing emission strength)
    based on Marangoni forces from the FEM solutions
    in Fig.\ \ref{fig:FMan} in the absence
    of evaporation and in the decoupled limit.
    For $\bar{\beta}=0$, the blue vertical line of data points
    ends in the critical Peclet number ${\rm Pe}_c$ at zero
     swimming speed. 
    In the presence of an anisotropic emission $\bar{\beta}>0$, the
    swimming bifurcation in the diffusive regime
    becomes avoided resulting in an initial
    linear relation $\bar{U}_{\rm swim}\propto {\rm Pe}$, crossing
    over to $\bar{U}_{\rm swim}\propto {\rm Pe}^{3/5}$ in the advective
    regime.  The solid lines for $\bar{U}<1$ are derived from the analytical
   perturbative results (\ref{eq:c1Ubetanew})
   and (\ref{eq:c1Mflbetanew}).
   The solid lines for $\bar{U}>1$ are derived from the scaling results
   (\ref{eq:c1Ubetanew2}) and (\ref{eq:c1Mflbetanew2}).
     }
\label{fig:UPe}
\end{figure}

In the decoupled limit ${\rm Pe}\ll \bar{U}$, the total Marangoni
force is always trivially linear in ${\rm Pe}$. Then, we can directly
obtain
the swimming condition in the form 
\begin{align}
    {\rm Pe} &= \frac{3\bar{U}_{\rm swim}}{(\bar{F}_{\rm M, tot}/\pi{\rm
               Pe})(\bar{U}_{\rm swim})}.
               \label{eq:PeU}
\end{align}
Using the  Marangoni forces from  Fig.\ \ref{fig:FMan}
in the decoupled limit, which include an  asymmetric emission $\bar{\beta}$
and reduce to Eq.\ (\ref{eq:c1Mflnew}) for  $\bar{\beta}=0$
and 
inverting this relation, we obtain the swimming relation in Fig.\
\ref{fig:UPe}.

\subsection{Swimming bifurcation}

For $\bar{\beta}=0$, i.e., a symmetrically emitting swimmer,
we see a sharp spontaneous symmetry breaking
above a critical Peclet number ${\rm Pe}_c$ in  the swimming relation
in Fig.\ \ref{fig:UPe} (blue vertical line of data points).
From  Eq.\ (\ref{eq:c1Mflnew}), we obtain the existence
of a symmetry-broken swimming state for 
${\rm Pe}>  {\rm Pe}_c \sim  8/\ln \bar{R} \to 0$,
which approaches zero for large system sizes. Therefore, the
symmetry is essentially always spontaneously
broken in  a large  swimming vessel.

The swimming bifurcation in the force balance
is governed by the leading order linear
terms $\propto \bar{U}$ (from the drag force and the linear
response regime of the Marangoni forces) and the next order correction
$\propto \bar{U}^3$ in the Marangoni force. Therefore,
we expect a supercritical pitchfork bifurcation analogously
to a $\phi^4$-theory for a second order phase transition.
In the presence of the additional symmetry-breaking
emission rate $\bar{\beta}>0$, which contributes a constant $\bar{U}^0$-term
to the force balance. This corresponds to an additional
symmetry-breaking field in the $\phi^4$-theory and gives rise
to an avoided bifurcation. This bifurcation scenario is
clearly reflected in Fig.\ \ref{fig:UPe}.

Figure \ref{fig:UPe} and Eq.\  (\ref{eq:c1Mflnew})
were, however, derived for
the decoupled limit ${\rm Pe}\ll \bar{U}$. 
At the swimming bifurcation, we have ${\rm Pe} ={\rm Pe}_c \gg \bar{U}\approx
0$, such that the feedback of  Marangoni flows
onto the diffusion-advection problem has to be taken into account,
 and the decoupling approximation
 should not   be used. 
 Then, Eq.\ (\ref{eq:FMtotalMarangoni}) describes the Marangoni forces
 in the linear response regime, which further reduces the critical Peclet
 number to ${\rm Pe}_c \sim 1/(\ln \bar{R})^{3}\to 0$.
 In the presence of evaporation with $\bar{k}\gg 1$, as
 appropriate for surfactants such as camphor, the total Marangoni force is
 further
 depressed according to Eq.\ (\ref{eq:FMtotalkUsmall}) resulting
 in an increased ${\rm Pe}_c =  \bar{k}^3/(\ln \bar{R})^{3} \to 0$, which is,
 however, still approaching zero for large swimming vessel sizes $\bar{R}$.

 For strong Marangoni flows and in the presence of evaporation,
 we still have  a linear response of the Marangoni forces
 for small $\bar{U}$ (see Fig.\  \ref{fig:FMtotalPe} and
 Eq.\ (\ref{eq:FMtotalkUsmall})) with higher order correction terms
 competing with a linearly $\bar{U}$-dependent drag force.
 Therefore, the above supercritical bifurcation scenario
 should persist.

\subsection{Swimming relation}

 For ${\rm Pe}> {\rm Pe}_c$, a spontaneously symmetry-broken
 swimming state with $\bar{U}_{\rm swim}>0$ exists for a symmetrically
 emitting swimmer with $\bar{\beta}=0$.
 Because the Marangoni force  Eq.\ (\ref{eq:c1Mflnew})
 remains approximately linear up to $\bar{U}\sim O(1)$, as can also
 be seen in Fig.\ \ref{fig:FM}, the swimming velocity
 rises steeply for ${\rm Pe}\gtrsim {\rm Pe}_c$ and quickly
 enters the asymptotics for the advection-dominated regime $\bar{U}_{\rm swim}
 \gg 1$ as can be clearly seen in Fig.\ \ref{fig:UPe}.

In the advective regime, 
we find  the swimming relations
\begin{subequations}
    \label{eq:swimrelation}
 \begin{align}
   \bar{U}_{\rm swim} &\sim {\rm Pe}^{3/5} 
    &&\mbox{for}~ \bar{k}\ll {\rm Pe}^{1/5},
           \label{eq:UPe}\\
   \bar{U}_{\rm swim} &\sim  \bar{k}^{-3/4} {\rm Pe}^{3/4} 
    &&\mbox{for}~   \bar{k}\gg {\rm Pe}^{1/5}.
           \label{eq:UPek}
\end{align}
\end{subequations}
Also in this regime, we have 
${\rm Pe} \gg \bar{U}_{\rm swim}$ such that
Marangoni flows are strong, but this
has little influence on the swimming speed because of the
concentration boundary layer that forms in this regime.
Evaporation is significant  for $\bar{k} \gg {\rm Pe}^{1/5}$ and
reduces the swimming speed 
because it reduces  the driving Marangoni forces.

For an anisotropically emitting swimmer with
$\bar{\beta}>0$, the bifurcation is avoided, and we find a
 linear swimming relation for small ${\rm Pe}$.
 In the vicinity of the bifurcation, the force balance
 can be written as
 $\bar{U}_{\rm swim} = ({\rm Pe}/{\rm Pe}_c) \bar{U}_{\rm swim}
 +{\rm Pe}\bar{\beta}/32$,
 which results in the linear
 swimming relation
 \begin{align}
      \bar{U}_{\rm swim}&= \frac{{\rm Pe}}{1-{\rm Pe}/{\rm Pe}_c}
                          \frac{\bar{\beta}}{32}.
                          \label{eq:UPelinear}
 \end{align}
 This describes the linear   relations 
$\bar{U}_{\rm swim}\propto {\rm Pe}$  for small ${\rm Pe}$ 
in the swimming relation in Fig.\ \ref{fig:UPe}.
 In the advective regime, we still have a 
crossover to the above swimming relations
  (\ref{eq:swimrelation}), but  with
  a slightly increased prefactor, i.e.,
  \begin{subequations}
    \label{eq:swimrelationbeta}
    \begin{align}
      \bar{U}_{\rm swim} &\sim ({\rm const} + \bar{\beta}) {\rm Pe}^{3/5}
                         \label{eq:UPebeta}, \\
      \bar{U}_{\rm swim} &\sim  \bar{k}^{-3/4}
                           ({\rm const} + \bar{\beta}) {\rm Pe}^{3/4}.
                           \label{eq:UPekbeta}
    \end{align}
 \end{subequations}

\section{Discussion and conclusion}

At low Reynolds numbers, 
we developed a complete theory for Marangoni  boat propulsion for 
a completely symmetric,  half-spherical, surfactant emitting
swimmer.
Symmetric PEG-alginate Marangoni surfactant  boats can be produced
down to  radii $a\sim 150\,{\rm \mu m}$
\cite{Ender2020} with  Reynolds numbers
${\rm Re} \sim 1-10$ such that the low Reynolds number regime
becomes accessible for surfactant-loaded boats.
Recently, asymmetric  thermal Marangoni surfers propelled
 by the {thermal} Marangoni effect were  successfully
 realized \cite{Dietrich2020}.
 Here, the thermal diffusion constant replaces the
 surfactant diffusion constant and is  by a factor $O(10^3)$ larger.
 Moreover, radii $a \sim 3\,{\rm \mu m}$ could be reached.
 At the same time, swimming velocities are still in the range
 above $10^3-10^5 \, {\rm \mu m/s}$. These parameters correspond
 to dimensionless velocities
 $\bar{U}_{\rm swim} \sim 2\times 10^{-2}-2$, which is
 mostly in the diffusive regime $\bar{U}\ll 1$
 and at low Reynolds numbers
 ${\rm Re} \sim 6\times 10^{-6}$.
 These swimmers were asymmetrically heated with a temperature
 difference $\Delta T$ across the swimmer corresponding to
 a constant concentration asymmetry
 $ \bar{c}_{S,1} \propto  \Delta T$. We therefore expect to be
 in a constant  concentration situation, which is
 analogous to the linear regime in the constant
 flux  swimming relation in  Fig.\ \ref{fig:UPe}.
 This is in accordance with the theoretical results of W{\"u}rger
 \cite{Wurger2014}, because advection plays no role in this regime
 and agrees with the experimental observations in Ref.\ \cite{Dietrich2020}.

Our theoretical description comprises the coupled
problems of surface tension reduction by surfactant adsorption
at the air-water interface including the possibility of surfactant
evaporation, fluid flow (both Marangoni flow and
flow induced by swimmer motion), diffusion and advection of the
surfactant.
Conceptually, there is no difference for a thermal Marangoni
surfer as realized in Ref.\ \cite{Dietrich2020}.
In previous theoretical approaches to surfactant \cite{Lauga2012} or thermal
\cite{Wurger2014} Marangoni boats, advection
has been neglected. For surfactant driven Marangoni boats, this
is typically a bad approximation as estimates in Ref.\ \cite{Ender2020}
show; for thermal Marangoni boats, this is typically justified as our
above estimates show.

The three coupled problems  of surfactant adsorption, low Reynolds number
fluid flow and diffusion-advection of surfactant
are first solved for prescribed
swimmer velocity $U$; the actual swimming velocity
$U_{\rm swim}$ is determined by force balance between the
drag force, the direct Marangoni force from the surface tension
contribution at the  air-water-swimmer   contact line and
the Marangoni flow force.
We employ the reciprocal theorem, which we could reinterpret
in terms of energy transduction, to calculate the Marangoni forces.

Non-dimensionalization reveals that 
two dimensionless control parameters exist, the Peclet number (\ref{eq:Pe}),
which is the dimensionless emission rate of surfactant, and
the Biot number (\ref{eq:Biot}), which is the dimensionless
evaporation rate. Evaporation is practically absent for PEG (Biot number
 $\bar{k}\ll 1$), but strong
for other frequently studied soap boat swimmers such as camphor
boats (Biot numbers $\bar{k}\approx 550$ \cite{Soh2008}).
In Ref.\ \cite{Ender2020}, it is shown that
  evaporation is relevant to quantitatively
  understand the large  differences in the swimming relation
   $\bar{U}_{\rm swim}=\bar{U}_{\rm swim}({\rm Pe})$
  between PEG-alginate swimmers and camphor boats from Ref.\
  \cite{Boniface2019}, but  these Marangoni
  boats operate at moderate Reynolds numbers.  
  For thermal Marangoni surfers
  \cite{Dietrich2020}, evaporation corresponds to a convective boundary
  condition for heat
  transfer from the  water surface to the air; the corresponding convection
  coefficient will depend on the nature of the air flow that is applied to
  transfer heat, which is difficult to quantify.
  It also depends on the temperature difference to the surrounding air that
   can be established. 
  Because the thermal
  Marangoni surfers from Ref.\ \cite{Dietrich2020}
  mostly operate in the diffusive regime $\bar{U}\ll 1$, we expect
  convection to reduce the Marangoni force according to Eq.\
  (\ref{eq:FMtotalkUsmall}) if the corresponding Biot number $\bar{k}$ is
  sufficiently high.

Moreover, the dimensionless swimmer velocity $\bar{U}$ plays
an important role as it controls the transition from
a diffusive regime $\bar{U}\ll 1$ to an advective regime
$\bar{U}\gg 1$. 
Non-dimensionalization of the coupled equations also
shows a decoupling
of the Marangoni flow problem for weak Marangoni flows
${\rm Pe}\ll \bar{U}$. Then, the concentration field around the
interfacial Marangoni swimmer with velocity $U$
is essentially equivalent to the concentration field
around a mass emitting sphere moving with velocity $U$ through
a bulk viscous fluid, which is a classical diffusion-advection problem.
We developed solutions for this diffusion-advection problem
for two types of boundary conditions which seem
  most important for applications:
 constant flux boundary conditions (A) for 
diffusive emission of surfactant from the swimmer and 
constant concentration boundary conditions (B) if the surfactant dissolves
from the surface or is produced by a chemical reaction on the
surface. We could obtain novel results for
constant flux boundary conditions, which are  unusual in the
diffusion-advection literature. In particular, we could obtain
qualitative results for the local Nusselt number by a novel
flux balance argument. 
All theoretical results are supported by numerical FEM simulations.

Apart from extensive results for the decoupled limit ${\rm Pe}\ll \bar{U}$,
we also addressed strong Marangoni flow in the limit
${\rm Pe}\gg \bar{U}$ and evaporation on the basis of scaling
arguments and numerical FEM simulations. 
This allowed us to obtain the Marangoni forces as a function of
a prescribed swimmer speed $\bar{U}$ for all relevant
situations, also including a possible anisotropic emission.
For all cases, our theoretical results agree well with the
numerical FEM calculations.
Knowledge of the Marangoni forces is the basis to discuss
the swimming bifurcation and swimming speed
as a function of the Peclet number as main control parameter via
the force balance condition.

We showed that a spontaneous symmetry breaking, i.e., a spontaneous
transition into a swimming state, is possible also for a completely
symmetric swimmer above a critical Peclet number.
The swimming bifurcation is a supercritical pitchfork bifurcation
analogous to a second order  symmetry-breaking phase transition,
and the presence of an explicitly symmetry-breaking emission
gives rise to an avoided bifurcation. 
Spontaneous symmetry breaking resulting in propulsion
  is  possible    by establishing
    an asymmetric surfactant concentration profile that is maintained
    by advection.
The symmetry breaking mechanism is similar to what  has been
proposed  for  autophoretic swimmers
\cite{michelin2013,Michelin2014}
and liquid Marangoni swimmers \cite{Izri2014} before.

In Eq.\ (\ref{eq:swimrelation}), we obtain the power-laws
governing the swimming velocity as a function of Peclet and Biot number,
which are $\bar{U}_{\rm swim} \propto  {\rm Pe}^{3/5}$,
without evaporation (PEG)  and
$\bar{U}_{\rm swim} \propto \bar{k}^{-3/4} {\rm Pe}^{3/4}$,
in the presence of strong evaporation (camphor).
In Eq.\ (\ref{eq:swimrelationbeta}), the result is extended in the
presence of an explicitly symmetry-breaking emission.
Then, a linear regime emerges in the diffusive limit $\bar{U}\ll 1$,
which is caused by the avoided bifurcation. This regime
is observed for the thermal Marangoni surfers in Ref.\ \cite{Dietrich2020}.

\begin{acknowledgement}
We acknowledge financial support by the Deutsche Forschungsgemeinschaft 
via SPP 1726 ``Microswimmers'' (KI 662/7-1 and KI 662/7-2).  
\end{acknowledgement}

\section{Authors contributions}
J.K. and H.E.
developed the theoretical model and 
performed the analytic calculations and  numerical
simulations. J.K. wrote the manuscript with support from H.E. 
%

\appendix
\numberwithin{equation}{section}

\section{Energy transduction and reciprocal theorem}
\label{app:energy}

In this Appendix, we discuss the reciprocal
theorem in terms of  energy transduction, in order to see
how power input from Marangoni stresses via the flow fields
is transduced to the sphere for propulsion.
This will also provide an alternative derivation of the
results obtained by Masoud and Stone \cite{Masoud2014} via
the reciprocal theorem.

For this, we switch to the laboratory frame in the following. 
We first consider the 
dissipation rate $\Phi \equiv 2\mu \int_V dV  e_{ij} e_{ij}$
of a solution of the Stokes equation
in an arbitrary volume $V$. Here,
$ e_{ij} =  \frac{1}{2}\left( \partial_i v_j +  \partial_j v_i \right)$ is the
strain tensor, and $\sigma_{ij} = -p \delta_{ij} + 2\mu e_{ij}$ is the stress
tensor. 
The kinetic energy $K = \frac{1}{2} \rho \int_V dV  \vec{v}^2$
of the fluid changes
according to $dK/dt = \int_{\partial V} da_i v_j \sigma_{ij} -
\Phi$, i.e., by the external power input $P_{\partial V}$ across the surface
of the volume and by dissipation.
In a stationary state, $dK/dt=0$ and dissipation and
power input are equal,
\begin{equation}
    \Phi = P_{\partial V} = \int_{\partial V} da_i v_j \sigma_{ij}
\label{eq:dissipation}
\end{equation}
($d\vec{a}$ is the outward normal to volume $V$).
Applying this equality to the Stokes flow field $\vec{v}$ and
the liquid volume with the boundary $\partial V=S+S_{\rm Int}$ and
using $\sigma_{ij}=0$  at the liquid-air interface
$S_{\rm Int}$,   we find
\begin{align}
 \Phi &=   \int_{S} da_i v_{j} \sigma_{ij}
        +  \int_{S_{\rm Int}} da_i v_{j} \sigma_{ij}
        \nonumber\\
          &= P_S =  U_j \int_{S} da_i \sigma_{ij}
            = -U F_D >0
      \label{eq:Phiv}
\end{align}
with the Stokes drag force  $F_{D} \equiv - \int_{S} da_i \sigma_{\rm iz}$
from Eq.\ (\ref{FD}).
Applying the same dissipation relation (\ref{eq:dissipation}) 
to the Marangoni flow field $\vec{v}_{\rm M}$
and the liquid volume with the boundary $\partial V = S+S_{\rm Int}$ and
using the  no-slip condition  $\vec{v}_{\rm M}=0$ at the half-sphere $S$,
we find
\begin{align}
  \Phi_{\rm M}  
   &=       \int_{S_{\rm Int}} da_i v_{\rm M,j} \sigma_{M,ij}
     = P_{M, S_{\rm Int}}.
     \label{eq:PhiM}
 \end{align}
This means that the power input by Marangoni stresses $\sigma_{M,ij}$
 on the interface
 $S_{\rm Int}$ via the Marangoni flows $v_{\rm M,j}$ (right hand side)
 is dissipated entirely within the fluid without transmitting
 any mechanical power onto the half-sphere because
 $\vec{v}_{\rm M}=0$ on $S$.

 Now, we  consider the  mutual
dissipation rate 
for  {\em two} solutions $\vec{v}^{(1)}(\vec{r})$ and
$\vec{v}^{(2)}(\vec{r})$ to the Stokes equation.
The mutual
dissipation can be shown to be
given by the mutual power input in a stationary state,
\begin{align}
  \Phi^{(12)}  &=  2\mu \int_V dV  e^{(1)}_{ij} e^{(2)}_{ij}
               = \int_{\partial V} da_i v^{(2)}_j \sigma^{(1)}_{ij}.
                 \label{eq:mutual}
\end{align}
The symmetry  $\Phi^{(12)}=\Phi^{(21)}$  leads directly to the reciprocal
theorem
\begin{equation}
  \int_{\partial V} da_i v^{(2)}_j \sigma^{(1)}_{ij}
  =  \int_{\partial V} da_i v^{(1)}_j \sigma^{(2)}_{ij}.
   \label{eq:reciprocal}
 \end{equation}
Now, we can apply this finding to the Stokes and Marangoni flow fields,
 which both satisfy the Stokes equation,
 and to  the liquid volume with the boundary $S+S_{\rm Int}$
 resulting in a mutual  dissipation
 \begin{align}
  \Phi_{\rm mut} &=   \int_{S+S_{\rm Int}} da_i v_{j} \sigma_{\rm M, ij} + 
                   \int_{S+S_{\rm Int}} da_i v_{\rm M,j } \sigma_{ij}
   \nonumber\\
  &= 2\int_{S+S_{\rm Int}} da_i v_{j} \sigma_{\rm M, ij}  =
    2\int_{S+S_{\rm Int}} da_i v_{\rm M,j } \sigma_{ij} = 0,
    \label{qe:Phimut}
\end{align}
 because $v_{\rm M,j} =0$ on the  surface $S$ of the half-sphere and
 $\sigma_{ij}=0$ on the interface $S_{\rm Int}$ such that
 the last equality holds. 
 The reciprocal theorem is thus equivalent to a vanishing 
 mutual dissipation between Stokes and Marangoni flow.

This has consequences for the 
 dissipation relation for the total  flow field $ \vec{v}_{\rm tot} =
 \vec{v}+ \vec{v}_{\rm M}$, which
 also includes the mutual dissipation $\Phi_{\rm mut}$
 of both contributions.
 The total power transmission onto the fluid
 volume with the boundary $\partial V = S+S_{\rm Int}$ is
 $P_{\rm tot, S} +P_{\rm tot, S_{\rm Int}}$ with 
 \begin{align}
   P_{\rm tot, S_{\rm Int}}
   &=
     \int_{S_{\rm Int}} da_i (v_j  + v_{\rm M,j}) \sigma_{\rm M,ij}
     \nonumber\\
    &= \int_{S_{\rm Int}} da_i v_j\sigma_{\rm M,ij} +  P_{M, S_{\rm Int}},
    \nonumber\\
   P_{\rm tot, S}
   &=
      \int_{S} da_i v_j  (\sigma_{ij}+ \sigma_{\rm M,ij}) =
     P_S - UF_{\rm M,fl},
     \label{eq:Ptot}
 \end{align}
 by employing boundary conditions on $S$ and $S_{\rm Int}$, Eqs.\
 (\ref{eq:PhiM}) and (\ref{eq:Phiv}), 
 and by introducing the Marangoni flow force
 \begin{equation}
   F_{\rm M, fl} \equiv - \int_{S} da_i \sigma_{\rm M, iz}
   \label{eq:FMfldef}
 \end{equation}
 as the  drag  force exerted
by the Marangoni flow field onto the sphere. ($d\vec{a}$ is the
inward normal to the sphere.) $F_{\rm M, fl} <0$ signals
additional Marangoni drag, while $F_{\rm M, fl} >0$ signals
an additional Marangoni flow propulsion force. 
Because of the vanishing mutual dissipation, we
obtain for the total dissipation 
\begin{align}
   \Phi_{\rm tot} &=  \Phi + \Phi_{\rm M} + 2\Phi_{\rm mut}
                    \nonumber\\
                    &=\Phi + \Phi_{\rm M}
                      =  P_S + P_{M, S_{\rm Int}}.
       \label{eq:Phitot}
 \end{align}
 The equality $\Phi_{\rm tot} = P_{\rm tot, S} +P_{\rm tot, S_{\rm Int}}$
 and Eq.\ (\ref{eq:Ptot})
 finally gives the energy transduction for the fluid, 
 \begin{align}
   P_S + P_{M, S_{\rm Int}}&= \Phi_{\rm tot}
   =  P_{\rm tot, S} +P_{\rm tot, S_{\rm Int}}
     \nonumber\\
    &=P_S - UF_{\rm M,fl}
          +\int_{S_{\rm Int}} da_i v_j\sigma_{\rm M,ij} +
      P_{M, S_{\rm Int}}
      \nonumber\\
\mbox{or}~~~   0 &=  - UF_{\rm M,fl} +\int_{S_{\rm Int}} da_i v_j\sigma_{\rm M,ij}.
  \label{eq:transduction}
 \end{align}
 This states that the mutual power input by Marangoni stresses via the
 Stokes flow field is \emph{completely}
 transduced via the  Marangoni flow force
 onto the sphere,
 while the power input  by Marangoni stresses via the
 Marangoni flow field itself is completely dissipated (see Eq.\
 (\ref{eq:PhiM})).

\section{Legendre  polynomial decomposition}
\label{app:Legendre}

In the axisymmetric decoupling approximation,
we can  employ a decomposition of the 
diffusion-advection  equation into Legendre polynomials
\begin{align}
  \bar{c}(\rho,\theta) &= \sum_{n=0}^\infty \bar{c}_n(\rho) P_n(\cos \theta),
                \nonumber\\
   \bar{c}_n(\rho) &= \frac{2n+1}{2}
                     \int_0^\pi d\theta \sin\theta P_n(\cos\theta)
                     \bar{c}(\rho,\theta).
                \label{Ldecomp}
\end{align}
with Legendre polynomials $P_n(\cos\theta)$.
Derivatives of Legendre polynomials are associated Legendre
polynomials 
$\partial_\theta P_n(\cos\theta)   = P_n^1(\cos\theta)$. 
The decomposition into Legendre polynomials is
advantageous as the Stokes  velocity field (\ref{u}) and (\ref{v})
can be written
in terms of $n=1$ polynomials only, $P_1(\cos\theta) = \cos\theta$ and
$P_1^1(\cos\theta) = -\sin\theta$, 
\begin{align}
     \bar{\hat{u}}(\rho,\theta) &= \bar{U} P_1(\cos\theta) u(\rho)=
                                         \bar{U} \cos\theta u(\rho),
                \label{u1}\\
  \bar{\hat{v}}(\rho,\theta) &= -\bar{U} P_1^1(\cos\theta) v(\rho)
                         =\bar{U} \sin\theta v(\rho).
                         \label{v1}
\end{align}
Therefore, the diffusion-advection equation (\ref{Dadvdim}) only couples
 coefficients $\bar{c}_n(\rho)$ to coefficients $\bar{c}_{n\pm 1}(\rho)$.

Both 
the direct  Marangoni force $F_{\rm M}$ (see (\ref{FMdim})) and
the total Marangoni force $ F_{\rm M,tot}$ (see (\ref{FMtotaldim}))
can also be written in terms
 of Legendre components of the concentration field at $r=a$:
 \begin{align}
   \frac{\bar{F}_{\rm M}}{\rm Pe}
   &= - 2 \int_0^\pi d\theta \cos\theta \bar{c}(1,\theta)
         \nonumber\\
       &= -2 \sum_{n=1}^\infty f_n \bar{c}_n(1)
            = -\pi   \bar{c}_M(1)
         ~~~\mbox{with}
         \label{FM2}\\  
 \bar{c}_M(\rho) & \equiv \frac{2}{\pi}\int_0^\pi d\theta \cos\theta
                   \bar{c}(\rho,\theta)
                   \nonumber\\
                &= \frac{2}{\pi}\sum_{n=1}^\infty f_n \bar{c}_n(\rho)
               \approx  \bar{c}_1(\rho) + ....,
                     \label{eq:cM_2}\\
   \bar{c}_1(\rho) &= \frac{3}{2}
                     \int_0^\pi d\theta \sin\theta\cos\theta
                     \bar{c}(\rho,\theta),
                   \label{eq:c1}  \\
 f_n &\equiv \int_0^\pi d\theta P_1(\cos\theta)  P_n(\cos\theta)
         \nonumber\\
   &=  (-1)^{(1-n)/2}\pi \frac{\Gamma\left(1+\frac{n}{2}\right)}
     {\Gamma\left(\frac{n+1}{2}\right)
         \Gamma\left((1-\frac{n}{2}\right))\Gamma\left(\frac{n+3}{2}\right)}
         \nonumber\\
       &= \begin{cases}   n=2k: & 0 \\
         n=2k+1 & \pi \frac{k+1/2}{k+1} \frac{(2k)!^2}{k!^416^k}
          =
   \frac{\pi}{2},~\frac{3\pi}{16},~\frac{15\pi}{128},...
         \end{cases},
                  \label{eq:fn}
   \\
   \frac{F_{\rm M,tot}}{\rm Pe} 
       &=  -\pi \int_1^\infty d\rho
         \frac{3}{4}\left(\rho^{-1} - \rho^{-3} \right) \bar{c}_M(\rho)
    \nonumber\\
       &= -2\sum_{n=1}^\infty f_n  \int_1^\infty d\rho
         \frac{3}{4}\left(\rho^{-1} - \rho^{-3} \right) \bar{c}_n(\rho),
        \label{FMtotal2}
 \end{align}
 where we used results from Ref.\ \cite{Rashid1986} to calculate the $f_n$
 in (\ref{eq:fn}).
 Advection always gives rise to an asymmetry where $\bar{c}(\rho,\theta)$ is an
 increasing function of $\theta$; it follows that $\bar{c}_1(\rho)<0$
 and $\bar{c}_M(\rho)<0$
 and $\bar{F}_{\rm M}>0$ and $\bar{F}_{\rm M,tot}>0$. 
The coefficients are decreasing and fall off as $f_n \sim {2}/{n}$
 for large $n$. This motivates to neglect all but the first
 $n=1$ component.
 For small $\bar{U}\ll 1$, the Legendre coefficients will scale as
 $\bar{c}_n(\rho) \sim \bar{U}^n$ and this becomes
 an excellent approximation; for $\bar{U}\gg 1$,
 this approximation becomes worse. Therefore,
 Legendre decomposition is useful for  $\bar{U}\ll 1$. 

Using the decomposition (\ref{Ldecomp}) and the
orthogonality relations
\begin{equation}
  \int_0^\pi d\theta \sin\theta P_n(\cos\theta) P_m(\cos\theta)
  = \delta_{nm} \frac{2}{2n+1},
\end{equation}
  the diffusion-advection  equation (\ref{Dadvdim}) becomes
\begin{align*}
 & \frac{2}{2n+1}  \left[ \frac{1}{\rho} \partial_\rho^2(\rho\bar{c}_n)
   - \frac{n(n+1)}{\rho^2}  \bar{c}_n\right]
   \\
  &=  \bar{U} u(\rho) \sum_m \left( \int d\theta \sin\theta P_1 P_n P_m \right)
    \partial_\rho \bar{c}_m \\
  &~~
    - \bar{U} \frac{v(\rho)}{\rho}
    \sum_m \left( \int d\theta \sin\theta P_1^1 P_n P_m^1 \right)
    \bar{c}_m.
 \end{align*}
The integrals on the right-hand side can be evaluated in closed form
using Wigner 3-j symbols \cite{Mavromatis1999}:
\begin{align*}
  &\int d\theta \sin\theta P_1 P_n P_m
    =
        2\begin{pmatrix} 1 & n & m \\ 0 & 0& 0 \end{pmatrix}^2,
  \\
  & \int d\theta \sin\theta P_1^1 P_n P_m^1 =
    \\
 &  =  -2 \begin{pmatrix} 1 & n & m \\ 0 & 0& 0 \end{pmatrix}
   \begin{pmatrix}  1 & n & m \\ 1 & 0& -1 \end{pmatrix}
                                       \left( 2(m+1)m\right)^{1/2},
\end{align*}
which gives non-vanishing contributions only for $m=n\pm 1$.
For these values, we find
\begin{align*}
 &    2\begin{pmatrix} 1 & n & n-1 \\ 0 & 0& 0 \end{pmatrix}^2
           = \frac{2n}{(2n-1)(2n+1)} \\
 &  2\begin{pmatrix} 1 & n & n+1 \\ 0 & 0& 0 \end{pmatrix}^2
                                       = \frac{2(n+1)}{(2n+1)(2n+3)},
  \\
 &   -2 \begin{pmatrix} 1 & n & n-1 \\ 0 & 0& 0 \end{pmatrix}
   \begin{pmatrix}  1 & n & n-1 \\ 1 & 0& -1 \end{pmatrix}
                 \left( 2n(n-1)\right)^{1/2}\\
                     & =  -\frac{2n(n-1)}{(2n-1)(2n+1)},
  \\
 &     -2 \begin{pmatrix} 1 & n & n+1 \\ 0 & 0& 0 \end{pmatrix}
   \begin{pmatrix}  1 & n & n+1 \\ 1 & 0& -1 \end{pmatrix}
                                     \left( 2(n+1)(n+2)\right)^{1/2}=
                                          \\
  &     =  \frac{2(n+1)(n+2)}{(2n+1)(2n+3)}.    
\end{align*}
Finally, we obtain
the diffusion-advection  equation in Legendre representation
Eq.\ (\ref{DadvLd}).

\section{Diffusion-advection equation, perturbation theory for $\bar{U}\ll 1$}
\label{app:perturbation}

For small fluid velocities,  $\bar{U}\ll 1$, we can  expand about the
isotropic undisturbed  diffusion solution at $\bar{U}=0$, which is given by
\begin{equation}
  \bar{c}_0^{(0)}(\rho)  = \frac{1}{\rho}~,~~\bar{c}_n^{(0)}(\rho)  = 0~
  \text{for}~n>0
     \label{eq:U=0}
\end{equation}
in Legendre representation.
A first approach is a naive perturbation series Ansatz (\ref{eq:Ansatz})
\begin{equation}
  \bar{c}_n(\rho) = \sum_{m=0}^\infty\bar{U}^m c_n^{(m)}(\rho)
\end{equation}
for each Legendre coefficient.

It turns out that this 
  will work only in the ``inner region'' $\rho < 1/\bar{U}$
  of a solution, because in the ``outer region'' 
  $\rho \gg 1/\bar{U}$, the  convection term can no longer
  be treated perturbatively.
  In Ref.\ \cite{Acrivos1962},
  a systematic  expansion in inner and outer region and
  a matching procedure were performed for the 
  constant concentration  boundary condition  (B),
  which we will adapt also to the  constant flux boundary
  conditions (A) in this Appendix.

\subsection{Naive perturbation theory}

We are most interested  in  the $n=1$ Legendre coefficient
$\bar{c}_1(\rho)$, which will give access to both direct and
total Marangoni forces at small $\bar{U}$  because
the Legendre coefficients will scale as
 $\bar{c}_n(\rho) \sim \bar{U}^n$ and the Legendre series 
 (\ref{FM2}) and (\ref{FMtotal2}) for the Marangoni forces can be terminated
 after $n=1$ if we are interested in the linear response for $\bar{U}\ll 1$.
Because of the boundary conditions (A) $\partial_\rho \bar{c}_{n>0}(1)=0$
or (B) $\bar{c}_{n>0}(1)=0$,
all $n>0$ modes are ``generated'' in Eq.\ (\ref{DadvLd})
from lower-order terms $\bar{c}_{n-1}$ on the right hand sides:
 \begin{align}
&  \frac{1}{\rho} \partial_\rho^2(\rho \bar{c}_0)
  =  \bar{U} \left( u(\rho) 
       \frac{1}{3} \partial_\rho \bar{c}_{1}-
    \frac{v(\rho)}{\rho} \frac{2}{3} \bar{c}_{1}  \right),
\nonumber\\
 &   \left[ \frac{1}{\rho} \partial_\rho^2(\rho \bar{c}_1)
    - \frac{2}{\rho^2}  \bar{c}_1  \right]
  =  \bar{U} u(\rho) \left(    \partial_\rho \bar{c}_{0}
    + \frac{2}{5} \partial_\rho \bar{c}_{2}
   \right)
   \nonumber\\
 &  ~~~~~~~~~~~~+\bar{U} \frac{v(\rho)}{\rho}
    \left(
    -  \frac{6}{5} \bar{c}_{2}  \right),
\nonumber\\
 & ...
\nonumber\\
 & \bar{c}_0(\infty)  = \bar{c}_\infty~,~~ \bar{c}_{n>0}(\infty)=0,
  \nonumber\\
 & \mbox{(A)~constant flux:}~~
  \partial_\rho \bar{c}_0(1) = -1~,~~\partial_\rho \bar{c}_{n>0}(1) = 0,
                               \nonumber\\
&  \mbox{(B)~constant concentration:}~~
  \bar{c}_0(1) = 1~,~~\bar{c}_{n>0}(1) = 0.
                 \label{DadvLd01}
\end{align}
This hierarchy results in
\begin{equation}
  \bar{c}_n(\rho) = \sum_{m=n}^\infty \bar{U}^m \bar{c}_n^{(m)}(\rho),
\end{equation}
i.e., the leading orders are $\bar{c}_n(\rho) \propto \bar{U}^n$.

In a naive perturbative approach, we start with
\begin{equation*}
  \bar{c}_0^{(0)}(\rho) = \frac{1}{\rho}~,~~ \bar{c}_n^{(0)} =0 ~~n\ge 1.
\end{equation*}
for both boundary conditions. 
We  obtain in first order
\begin{align*}
& \left[ \frac{1}{\rho} \partial_\rho^2(\rho \bar{c}_1^{(1)})
    - \frac{2}{\rho^2}  \bar{c}_1^{(1)}  \right]
  =   u(\rho) \left(    \partial_\rho \bar{c}_{0}^{(0)} \right).
\end{align*}
Solving for $\bar{c}_1^{(1)}$ and
regularizing using a finite system $\rho<R$ (i.e., the boundary
  conditions $\bar{c}_0(\bar{R}) = 0$ and $\bar{c}_{n>0}(\bar{R})=0$),
we find up to linear order in $\bar{U}$
\begin{align}
  & \mbox{(A)~constant flux:}~~\nonumber\\
  &\bar{c}_1(\rho)
  =   -\frac{1}{2}\bar{\beta}  \frac{1}{\rho^2}+
    \bar{U}\left( \frac{1}{8}\frac{1}{\rho^3}
           -  \frac{9}{16} \frac{1}{\rho^2} 
    +\frac{3}{4} \frac{1}{\rho} - \frac{1}{2} \right)
    + O\!\left(\frac{1}{\bar{R}}\right),
   \nonumber  \\
  & \mbox{(B)~constant concentration:}~~\nonumber\\
   & \bar{c}_1(\rho)
  =  \bar{c}_{S,1} \frac{1}{\rho^2}+
    \bar{U} \left(-  \frac{3}{8}\frac{1}{\rho^2} 
     +\frac{1}{8}\frac{1}{\rho^3} 
     + \frac{3}{4} \frac{1}{\rho} - \frac{1}{2} \right)+
     O\!\left(\frac{1}{\bar{R}}\right),
\nonumber
 \end{align}
 which fulfills the more general explicitly symmetry-breaking
 flux boundary
 condition $\partial_\rho\bar{c}_1(1)= \bar{\beta}$ (see Eq.\ (\ref{eq:beta}))
 with $\bar{\beta}=0$  as constant flux (A)
 or concentration boundary condition
 $\bar{c}_1(1) = \bar{c}_{S,1}$ see Eq.\ (\ref{eq:c1S}))
 with $\bar{c}_{S,1}=0$ as constant
concentration (B). These more general boundary conditions
just add a zeroth order $U^0$-term to the $n=1$ component
$\bar{c}_1(\rho)$. 

We obtain for  flux boundary conditions (A)
 \begin{align}
   \frac{\bar{F}_{\rm M}}{\pi {\rm Pe}} &\approx
   -\bar{c}_1(1)    =\frac{1}{2}\bar{\beta}+  \frac{3}{16}  \bar{U},
   \label{eq:FMlinear}\\
   \frac{\bar{F}_{\rm M,tot}}{\pi {\rm Pe}}
    &=   -\int_1^\infty d\rho
      \frac{3}{4}\left(\rho^{-1} - \rho^{-3} \right) c_1(\rho)
      \nonumber\\
          &= \frac{3}{32}\bar{\beta}-  \frac{1081}{1280}  \bar{U}
            +\frac{3}{8}  \bar{U}\ln \bar{R}.
 \label{eq:FMtotlinear}
 \end{align}
 This means that the direct Marangoni force  $\bar{F}_{\rm M}$ 
 is linear with a finite linear coefficient for large $\bar{R}$, whereas 
 the total Marangoni force $\bar{F}_{\rm M,tot}$ 
 has a logarithmically diverging linear  coefficient for large $\bar{R}$
 (stemming from the  $\rho$-independent contribution
 $c_1^{(1)}(\rho)= -1/2+...$).
 The contribution from explicit symmetry breaking ($\bar{\beta}$)
 to the direct Marangoni force is always \emph{weakened}
 by the presence of Marangoni flows. 

 For  concentration boundary conditions (B), we have
  \begin{align}
    \frac{\bar{F}_{\rm M}}{\pi {\rm Pe}} &\approx
          -\bar{c}_1(1)=-\bar{c}_{S,1} + O(\bar{U}^3),
              \label{eq:FMlinearB}       \\
   \frac{\bar{F}_{\rm M,tot}}{\pi {\rm Pe}}
    &= -\int_1^\infty d\rho
      \frac{3}{4}\left(\rho^{-1} - \rho^{-3} \right) c_1(\rho)
      \nonumber\\
  &= -\frac{3}{8}\bar{c}_{S,1}
    -  \frac{563}{320}  \bar{U}
    +\frac{3}{4}  \bar{U}\ln \bar{R}.
    \label{eq:FMtotallinearB}
 \end{align}
 This means that the direct Marangoni force
 $\bar{F}_{\rm M}$ [see (\ref{FMdim})]
 is only present for explicit symmetry breaking
 ($\bar{c}_{S,1}\neq 0$), whereas 
 the total Marangoni force $\bar{F}_{\rm M,tot}$ (see(\ref{FMtotaldim}))
 has a logarithmically diverging linear  coefficient for large $\bar{R}$,
 which is identical to the constant flux case. 
The contribution from explicit symmetry breaking ($\bar{c}_{S,1}$)
 to the direct Marangoni force is always \emph{weakened}
 by the presence of Marangoni flows.

\subsection{Matching procedure}

In order to go beyond naive perturbation theory for the
constant flux boundary condition (A) in the fluid
velocity window  $1/R <\bar{U}\ll 1$, we must adapt the matching
procedure that was developed for constant-$c$ boundary
conditions (B) in Ref.\ \cite{Acrivos1962}.
This matching procedure employs both the  real space
 representation and a Legendre decomposition.
 We start from the diffusion-advection equation  (\ref{Dadvdim})
 (neglecting  $\bar{\vec{v}}_{\rm M}$ in the decoupled limit)
in the original angular representation but in dimensionless form for 
$\bar{c} = \bar{c}(\rho,\theta)$ or
$\bar{c} = \bar{c}(\rho,\eta)$ ($\eta \equiv \cos\theta$).
\footnote{
  Note that in Ref.\ \cite{Acrivos1962}, the angle $\theta$ is defined
  in the opposite sense such that the flow toward the sphere
  is in negative $\theta$-direction.
  This corresponds to $\eta \to -\eta$.}

A perturbation expansion
\begin{align}
  \bar{c}(\rho,\eta) &= \sum_{n=0}^\infty f_n(\bar{U}) \bar{c}^{(n)}(\rho,\eta),
\nonumber\\
  &\lim_{\bar{U}\to 0} \frac{f_{n+1}}{f_n} =0 ~~\mbox{and}~~
  f_0(\bar{U})=1
  \label{eq:cexpansion}
\end{align}
with constant flux boundary conditions 
\begin{equation}
  \partial_\rho \bar{c}^{(0)}(1,\eta)  = -1+\bar{\beta}\eta ~,
    ~~\partial_\rho \bar{c}^{(n\ge 1)}(0,\eta)  = 0
\end{equation}
is used  in the ``inner region'' $\rho < 1/\bar{U}$.
In the   ``outer region'' 
  $\rho \gg 1/\bar{U}$, the  convection term can no longer
  be treated perturbatively, regardless how small $\bar{U}$ is
  \cite{Acrivos1962}.
  Here, we rescale  $\sigma \equiv \rho \bar{U}$ and
  $\bar{C}(\sigma,\eta) = \bar{c}(\sigma/\bar{U},\eta)$ to 
  obtain an equation
\begin{align}
  \vec{\nabla}_\sigma^2 \bar{C} 
        &= \eta u\left(\frac{\sigma}{\bar{U}}\right) \partial_\sigma \bar{C}
          - (1-\eta^2) \frac{v\left(\frac{\sigma}{\bar{U}}\right)}{\sigma}
          \partial_\eta \bar{C}
  \label{eq:outer}
\end{align}
 and use an expansion 
 \begin{equation}
   \bar{C}(\sigma,\eta) = \sum_{n=0}^\infty F_n(\bar{U})
                \bar{C}^{(n)}(\sigma,\eta)
  ~,~~ \lim_{\bar{U}\to 0} \frac{F_{n+1}}{F_n} =0
\end{equation}
with outer boundary conditions
\begin{equation}
  \bar{C}^{(n)}(\infty,\eta)  = 0.
\end{equation}
Both expansions have to match at a large but finite $\rho$, such that 
\begin{equation}
  \bar{c}(\rho\to \infty,\eta) = \bar{c}(\sigma/\bar{U}\to \infty,\eta)=
   \bar{C}(\sigma\to 0,\eta)
  \label{eq:match}
\end{equation}
in all orders in $\bar{U}$  and 
as a function of $\eta$ or in all Legendre coefficients.

Plugging the expansions into the  inner and outer equations
(\ref{Dadvdim})
and (\ref{eq:outer}), respectively,
and isolating the $\bar{U}^0$-terms, 
 the functions $\bar{c}^{(0)}(\rho,\eta)$
and $\bar{C}^{(0)}(\sigma,\eta)$ have 
 to fulfill
 \begin{align*}
   \bar{\vec{\nabla}}^2 \bar{c}^{(0)} &=0,
   \\
    \vec{\nabla}_\sigma^2 \bar{C}^{(0)} 
        &= -\eta  \partial_\sigma \bar{C}^{(0)}
          - (1-\eta^2) \frac{1}{\sigma} \partial_\eta \bar{C}^{(0)}
 \end{align*}
 and constant flux boundary conditions,
resulting in
 \begin{align*}
   \bar{c}^{(0)}(\rho,\eta) &= B_0 + \frac{1}{\rho} + \bar{\beta}\rho P_1(\eta) 
                   \\
                 &~~ + \sum_{k=1}^\infty B_k (\rho^{-k-1} +
              \frac{k+1}{k}\rho^k ) P_k(\eta) 
 \end{align*}
   and (unchanged from the constant concentration case \cite{Acrivos1962})
\begin{align*}
  \bar{C}^{(0)}(\sigma,\eta)
  &= e^{-\sigma\eta/2} \left(\frac{\pi}{\sigma}\right)^{1/2}
                     \sum_{k=0}^\infty C_k K_{k+1/2}(\sigma/2)P_k(\eta)
\end{align*}
with the modified Bessel functions
\begin{align*}
  K_{k+1/2}\left(\frac{\sigma}{2}\right)
  &=
      \left(\frac{\pi}{\sigma}\right)^{1/2}e^{-\sigma/2}
                  \sum_{m=0}^k\frac{(k+m)!}{(k-m)!m! \sigma^m}.
\end{align*}
 Matching the zeroth-order contributions according to (\ref{eq:match})
  means 
   \begin{align*}
     \bar{c}^{(0)}(\sigma/\bar{U},\eta)
     &=
       B_0 +\frac{\bar{U}}{\sigma}
       +   \bar{\beta}\frac{\sigma}{\bar{U}}  P_1(\eta)\\
     &~~ + \sum_{k=1}^\infty B_k 
       \frac{k+1}{k}\frac{\sigma^k}{\bar{U}^k}  P_k(\eta)
   \end{align*}
   has to match
   \begin{align*}
 F_0(\bar{U})  \bar{C}^{(0)}(\sigma,\eta)    
         &=  F_0(\bar{U})\frac{\pi}{\sigma}
         \left[ 1+ \frac{\sigma}{2}(-\eta-1) +
           ...\right]  \\
     &~~\times \sum_{k=0}^\infty C_k \frac{(2k)!}{k! \sigma^k}P_k(\eta)
   \end{align*}
   for small $\sigma$ and $\bar{U} \to 0$
   for all Legendre coefficients $k$. This yields
\begin{align*}
       F_0(\bar{U}) &= \bar{U}~,~~C_0 = \frac{1}{\pi}~,~~
                      B_0 = -\bar{U}/2 \to 0~,~~B_1 = -\bar{\beta}/2
       \\
          B_{k\ge 2}&=0~,~~ C_{k\ge 1}=0.
\end{align*}
$B_0 = -\bar{U}/2$ gives an $O(\bar{U})$-contribution to $\bar{c}^{(0)}$,
which should be attributed to  the next order $\bar{c}^{(1)}$. 
 The final result for the zeroth order contributions is
 \begin{align*}
   \bar{c}^{(0)}(\rho,\eta) &= \frac{1}{\rho}
                              -\frac{\bar{\beta}}{2\rho^2}\eta, \\
   \bar{C}^{(0)}(\sigma,\eta)
     &= \frac{1}{\sigma} \exp\left( \frac{\rho}{2}(-\eta-1) \right),
 \end{align*}
 which is  unchanged from the constant $c$ case \cite{Acrivos1962}
 (apart from the different $\theta$-convention leading to $\eta\to -\eta$).

Assuming $f_1(\bar{U}) = \bar{U}$ and $F_1(\bar{U}) = \bar{U}^2$,
plugging the expansions including  the
already calculated zeroth order
contributions into the inner and outer equations  (\ref{Dadvdim})
and (\ref{eq:outer}), respectively,  and isolating the  $\bar{U}^1$-terms,
we find that 
 the  first  order contributions  $\bar{c}^{(1)}(\rho,\eta)$
and $\bar{C}^{(1)}(\sigma,\eta)$ have 
to fulfill the same equations as for the
constant $c$ case \cite{Acrivos1962},
  \begin{align*}
   \bar{\vec{\nabla}}^2 \bar{c}^{(1)}
   &=   u(\rho) P_1(\eta) \partial_\rho \bar{c}^{(0)}
              =\left[1 - \frac{3}{2} \frac{1}{\rho}
                    + \frac{1}{2} \frac{1}{\rho^3} \right]
                    \frac{\eta}{\rho^2},
   \\
    \vec{\nabla}_\sigma^2 \bar{C}^{(1)} 
         &= -\eta  \partial_\sigma \bar{C}^{(1)}
           - (1-\eta^2) \frac{1}{\sigma} \partial_\eta \bar{C}^{(1)}\\
        &~~   +\frac{3}{2}\frac{\eta}{\sigma}\partial_\sigma \bar{C}^{(0)}
           + \frac{3}{4}\frac{1-\eta^2}{\sigma} \partial_\eta \bar{C}^{(0)}.
 \end{align*}
A particular solution for $\bar{c}^{(1)}$ is
 \begin{equation*}
     \bar{c}_p^{(1)} = -\left(\frac{1}{2} - \frac{3}{4\rho} -
       \frac{1}{8\rho^3}\right) \eta;
 \end{equation*}
 the full solution that fulfills the flux boundary condition
 $\partial_\rho \bar{c}^{(1)}(0,\eta)  = 0$ is
 \begin{align*}
   \bar{c}^{(1)}
     &=  B_0 + \left[ (2B_1 - \frac{9}{8} ) \rho +
                 \frac{B_1}{\rho^2} - \left( \frac{1}{2} - \frac{3}{4\rho} -
       \frac{1}{8\rho^3}\right)\right] \eta
       \\
     &~~   + \sum_{k=2}^\infty B_k  (r^{-k-1} +  \frac{k+1}{k}r^k ) P_k(\eta).
\end{align*}
Now we can determine all $B_k$ by matching
all contributions up  to $O(\bar{U})$
according to (\ref{eq:match}),
\begin{align*}
 & \bar{c}^{(0)} (\sigma/\bar{U}\to \infty,\eta)+
  \bar{U} \bar{c}^{(1)} (\sigma/\bar{U}\to \infty,\eta) \\
 &~~ =
    \bar{U} \bar{C}^{(0)}(\sigma\to 0,\eta)
 \end{align*}
 resulting in
\begin{equation*}
    B_0 = -1/2~,~~ 
      B_1 = -9/16~,~~
          B_{k\ge 1}=0.
\end{equation*}
$B_0=-1/2$ for $\bar{c}^{(1)}$ is indeed equivalent to our above
$B_0 =  -\bar{U}/2$ for $\bar{c}^{(0)}$; now this term is
consistently attributed to $\bar{c}^{(1)}$.
All in all, we have up to $O(\bar{U})$
 \begin{align}
 \bar{c}^{(0)}+ \bar{U}  \bar{c}^{(1)}
   &=  \frac{1}{\rho}-\frac{\bar{\beta}}{2\rho^2}\eta
     \nonumber\\
      &~~  +\bar{U}\left[ -\frac{1}{2} -
     \left( \frac{1}{2} - \frac{3}{4\rho} +
     \frac{9}{16\rho^2}-\frac{1}{8\rho^3} \right) \eta \right].
     \label{eq:c01}
 \end{align}
We can use this result to calculate  the first Legendre coefficient
at the boundary  $-\bar{c}_1(\rho=1)$, which gives
\begin{equation}
  -\bar{c}_1(1) = -\frac{3}{2} \int_{-1}^1 d\eta \eta
  \left.(\bar{c}^{(0)}+ \bar{U}  \bar{c}^{(1)})\right|_{\rho=1}
    = \frac{\bar{\beta}}{2} + \frac{3}{16} \bar{U}
 \label{eq:c1linear2}
\end{equation}
in complete agreement with our above result  (\ref{eq:FMlinear})
from naive perturbation theory. 

We continue with the next order $C^{(1)}(\sigma,\eta)$ of the outer solution. 
Here we obtain the same result as for the constant concentration
case \cite{Acrivos1962},
\begin{align*}
  \bar{C}^{(1)} &= e^{-\sigma\eta/2} \left[
                  \left(\frac{\pi}{\sigma}\right)^{1/2}
                  \sum_{k=0}^\infty C_k^* K_{k+1/2}(\sigma/2)P_k(\eta)\right.
                  \\
                &~~~~~~~~\left.
                  + \sum_{i=0}^2 \tilde{R}_i(\sigma) P_i(\eta)\right]
\end{align*}
with  functions $\tilde{R}_i(\sigma) = (-1)^iR_i(\sigma)$ with
the functions  $R_i(\sigma)$ from 
\cite{Acrivos1962}
(because of  the different $\theta$-convention leading to $\eta\to -\eta$).

In the variable  $\sigma \equiv \rho \bar{U}$, in which the matching
to the outer solution is performed, the inner solution (\ref{eq:c01})
becomes
 \begin{align}
 \bar{c}^{(0)}+ \bar{U}  \bar{c}^{(1)}
   &=  \bar{U} \left[ \frac{1}{\sigma} +\frac{1}{2}(-\eta-1) \right]
        +\bar{U}^2 \frac{1}{\sigma} \frac{3}{4} \eta + ...
     \label{eq:c01sigma}
\end{align}
All terms $\sigma^{-m}$ ($m\ge 1$) of the inner solution from
$\bar{c}^{(n\ge 2)}$ are of
higher order and, thus, at least $O(\bar{U}^3)$; only an additional
constant term $O(\bar{U}^2)$ from $\bar{c}^{(2)}$ is possible.
Therefore, in order to match (\ref{eq:c01sigma}),
all terms $\sigma^{-m}$ for $m\ge 2$ of the outer solution
$\bar{C}^{(1)}$ have to be zero and the $\sigma^{-1}$-term has to equal
$\sigma^{-1}\frac{3}{4}\eta$.
We conclude that 
\begin{align*}
     C_{k\ge 3}^*  &=0,\\
  C_2^* &= -\frac{1}{4\pi}(3-\ln \gamma)~,~~
          C_1^* = -\frac{3}{4\pi}(1-\ln\gamma),\\
  C_0^* &= \frac{1}{2\pi} \ln\gamma
\end{align*}
(where $\ln\gamma=0.577216$ is the Euler constant),
i.e., only $C_0^*$ is different from the constant concentration
results of
\cite{Acrivos1962} ($C_1^*$ has an additional minus sign because
of  $\eta \to -\eta$).
The resulting outer solution
\begin{align}
 & \bar{U} \bar{C}^{(0)} + \bar{U}^2 \bar{C}^{(1)}
  =\nonumber\\
 & =  \frac{\bar{U}}{\sigma}\exp\left( \frac{\sigma}{2}(-\eta-1) \right)
     +\bar{U}^2\left( -\frac{\ln\sigma}{2} +
  \left(\frac{1}{2}-\frac{\ln\gamma}{4} \right) \right)
  \nonumber\\
 &~~   - \bar{U}^2\left( -\frac{3}{4\sigma}\ln\gamma -\frac{3}{16}(\ln\gamma
    -1)\right) \eta
  \nonumber\\
 &\approx
    \frac{\bar{U}}{\sigma}\left[1-\frac{\sigma}{2}+\frac{\sigma^2}{6} -
   \left( \frac{\sigma}{2}-\frac{\sigma^2}{4}\right) \eta \right]
   \nonumber\\
  &~~   +\bar{U}^2\left( -\frac{\ln\sigma}{2} +
    \left(\frac{1}{2}-\frac{\ln\gamma}{4} \right) \right)P_0(\eta)
    \nonumber\\
  &~~
    + \bar{U}^2\left( \frac{3}{4\sigma}\ln\gamma -\frac{3}{16}(\ln\gamma
    -1)\right) P_1(\eta)
 \label{eq:C12}
\end{align}
contains terms $\bar{U}^2 \sigma^0P_0(\eta)$ and $\bar{U}^2\ln \sigma P_0(\eta)$,
which suggests that  the second-order contribution
$\bar{c}^{(2)}$ to the inner solution should also contain
constant terms $\bar{U}^2$ and $\bar{U}^2\ln \bar{U}$ in order
to match the outer solution.

Now, we turn to this contribution $\bar{c}^{(2)}(\rho,\eta)$.
Plugging the expansion  (\ref{eq:cexpansion}) including the
already calculated (\ref{eq:c01}) up to the first order
into the inner equation  (\ref{Dadvdim})
and isolating the  $\bar{U}^2$-terms,  we find 
 \begin{align*}
   \bar{\vec{\nabla}}^2 \bar{c}^{(2)}
   &=
         u(\rho) P_1(\eta) \partial_\rho \bar{c}^{(1)} -
     \frac{v(\rho)}{\rho} (1-\eta^2) \partial_\eta  \bar{c}^{(1)}.
 \end{align*}
 Inserting the first order part from (\ref{eq:c01}) on the right hand side,
 we finally obtain
 \begin{align*}
   \bar{\vec{\nabla}}^2 \bar{c}^{(2)}
   &= Z_0(\rho) P_0(\eta) + Z_2(\rho) P_2(\eta) ~~~\mbox{with}
   \\
   Z_0(\rho) &= \frac{1}{3\rho} -\frac{1}{2\rho^2} +\frac{23}{96\rho^4}
               +\frac{1}{16\rho^5}-\frac{9}{32\rho^6}+\frac{1}{12\rho^7},\\
   Z_2(\rho) &= -\frac{1}{3\rho} +\frac{7}{4\rho^2}-\frac{9}{4\rho^3}
               +\frac{175}{96\rho^4}\\
             &~~  -\frac{5}{16\rho^5}-\frac{9}{32\rho^6}+\frac{5}{48\rho^7}.
 \end{align*}
 A $P_1(\eta)$-term is absent on the right hand side
 because $\bar{c}^{(1)}$
from (\ref{eq:c01})
 contains no $P_0(\eta)$-component
 but is a pure $P_1(\eta)$-term.
  In order to obtain the first two Legendre components $k=0,1$ of 
 $\bar{c}^{(2)}$ we thus have to solve 
\begin{align*}
 \partial_\rho^2 \bar{c}^{(2)} + \frac{2}{\rho}\bar{c}^{(2)}
   &=
     Z_0(\rho)
\end{align*}
and find
\begin{align*}
  \bar{c}^{(2)} &=  B_0 + L_0(\rho) +\eta B_1\left(\rho +
                  \frac{1}{2\rho^2}\right) + P_2(\eta) \left(...\right) + ...,
                  \\
   L_0(\rho) &=  \frac{1}{240\rho^5}-\frac{3}{128\rho^4}
               +\frac{1}{48\rho^3}+\frac{23}{192}{\rho^2}\\
             &~~  -\frac{9}{16\rho}
                +\frac{527}{1920}+\frac{\rho}{6}-\frac{1}{2}\ln\rho.
\end{align*}
Using $\rho= \sigma /\bar{U}$ and matching with
$ \bar{U} \bar{C}^{(0)} + \bar{U}^2 \bar{C}^{(1)}$ from (\ref{eq:C12})
gives
\begin{align*}
  B_0&= \left(\frac{1}{2}-\frac{\ln\gamma}{4} \right)-\frac{527}{1920}
  ~,~~ B_1 = 1/4.
\end{align*}
Up to the second order, we obtain from (\ref{eq:c01})
and $\bar{c}^{(2)}$  a $P_1(\eta)$-contribution
\begin{align*}
  \bar{c}_1(\rho)
    &= -\frac{\bar{\beta}}{2\rho^2} -\bar{U} 
     \left( \frac{1}{2} - \frac{3}{4\rho} +
      \frac{9}{16\rho^2}-\frac{1}{8\rho^3} \right)\\
    &~~  +\bar{U}^2 \frac{1}{4}\left(\rho +
      \frac{1}{2\rho^2}\right)
      + O(\bar{U}^3).
\end{align*}
We can directly obtain the
value $\bar{c}_1(1)$ of the first Legendre coefficient
at the surface as
\begin{align}
  -\bar{c}_1(\rho=1)
  &=  \frac{\bar{\beta}}{2} +
    \frac{3}{16}\bar{U} - \frac{3}{4}\bar{U}^2 + O(\bar{U}^3).
 \label{eq:matchingc1}         
\end{align}
The leading $O(\bar{U})$-term agrees with the naive
perturbation expansion results (\ref{eq:FMlinear}).
The matching procedure gives, however, a non-vanishing second-order
contribution $O(\bar{U}^2)$, which is absent in the naive
perturbation expansion.
The matching to the outer solution, which features a $P_1(\eta)$-contribution
in second order, see (\ref{eq:C12}), enforces this term.
Matching is required for $\bar{R}\gg 1/\bar{U}$.
For $\bar{U}\ll 1/\bar{R}$, the completely symmetric outer boundary condition
$\bar{c}(\rho=\bar{R})=0$ suppresses this term and the naive perturbation
expansion for $\bar{c}_1(\rho)$
only contains odd powers of $\bar{U}$.

For constant concentration boundary conditions (B),
we can directly employ the results from  Ref.\ \cite{Acrivos1962}
to obtain
up to the second order   a $P_1(\eta)$-contribution
\begin{align*}
  \bar{c}_1(\rho)
    &= \frac{\bar{c}_{S,1}}{\rho^2} -\bar{U} 
     \left( \frac{1}{2} - \frac{3}{4\rho} +
      \frac{3}{8\rho^2}-\frac{1}{8\rho^3} \right)\\
  &~~    -\bar{U}^2 \left(\frac{7}{16\rho^2} -
      \frac{1}{4}\rho + \frac{1}{4} - \frac{3}{8\rho} -
          \frac{1}{16\rho^3} \right)
      + O(\bar{U}^3)
\end{align*}
leading to 
\begin{align*}
  -\bar{c}_1(1)
    &= -\bar{c}_{S,1}  + O(\bar{U}^3).
\end{align*}
This agrees with the naive
perturbation expansion result (\ref{eq:FMlinearB}).

There is, however, an important difference in evaluating
the $\rho$-integrals in Eq.\ (\ref{eq:FMtotlinear}) and
(\ref{eq:FMtotallinearB}) in order to  calculate the total Marangoni force
as compared to the naive perturbation theory. 
Also, these integrals have to be divided into inner and outer region in
the framework of the matching procedure,
which essentially provides an upper cutoff $\bar{R}\sim 1/\bar{U}$
to the otherwise unchanged inner region. Therefore, contribution
$\propto \bar{U}\ln\bar{R}$ are to be replaced by
corresponding contributions $\propto -\bar{U}\ln\bar{U}$.

  
\bibliographystyle{epj}
\bibliography{marangoni2}

\end{document}